\documentclass{emulateapj}

\usepackage{rotating}
\usepackage{subfigure}
\usepackage{captcont}
\usepackage{amsmath}
\usepackage{appendix}
\usepackage{morefloats}

\newcommand{\noprint}[1]{}

\newcommand\aastex{AAS\TeX}
\def\h2{H$_2$}
\def\-2e{$^{-2}$\ }
\def\-2{$^{-2}$}

\newcommand{\HI}{\ensuremath{\mbox{\rm \ion{H}{1}}}}
\newcommand{\HII}{\ensuremath{\mbox{\rm \ion{H}{2}}}}

\renewcommand{\t}[1]{\mathrm{#1}}
\renewcommand{\H}{\ensuremath{\mathrm{H}}}

\newcommand{\msun}{\ensuremath{M_\odot}}

\newcommand{\sunits}{\mbox{\msun ~pc$^{-2}$}}

\newcommand{\kms}{\mbox{km~s$^{-1}$}}

\newcommand{\co}[1]{\mbox{$^{#1}$CO}}

\newcommand{\vunits}{\mbox{km s$^{-1}$ pc$^{-1}$}}

\newcommand{\arc}{\mbox{$^{\prime\prime}$}}

\begin{document}

\slugcomment{Accepted for Publication in ApJ}

\title{Angular Momentum in Giant Molecular Clouds. II. M33}

\author{Nia Imara, Frank Bigiel, and Leo Blitz}
   \affil{Astronomy Department, University of California, Berkeley, CA 94720}
   \email{imaran@berkeley.edu}

  \begin{abstract}
We present an analysis comparing the properties of 45 giant molecular clouds (GMCs) in M33 and the atomic hydrogen (\HI) with which they are associated.  High-resolution VLA observations are used to measure the properties of \HI~in the vicinity of GMCs and in regions where GMCs have not been detected.  The majority of molecular clouds coincide with a local peak in the surface density of atomic gas, though $7\%$ of GMCs in the sample are not associated with high-surface density atomic gas. The mean \HI~surface density in the vicinity of GMCs is 10 \sunits~and tends to increase with GMC mass as $\Sigma_\t{HI}\propto M_\t{GMC}^{0.27}$.  Thirty-nine of the 45 \HI~regions surrounding GMCs have linear velocity gradients of $\sim 0.05$ \vunits.  If the linear gradients previously observed in the GMCs result from rotation, $53\%$ are counterrotating with respect to the local \HI.  And if the linear gradients in these local \HI~regions are also from rotation, 62\% are counterrotating with respect to the galaxy.  If magnetic braking reduced the angular momentum of GMCs early in their evolution, the angular velocity of GMCs would be roughly one order of magnitude \emph{lower} than what is observed. Based on our observations, we consider the possibility that GMCs may not be rotating.   Atomic gas not associated with GMCs has gradients closer to 0.03 \vunits, suggesting that events occur during the course of GMC evolution that may increase the shear in the atomic gas.

 \end{abstract}


\section{Introduction}\label{sec:intro}

Observations of molecular clouds at millimeter wavelengths show that they often have systematic velocity gradients.  If, as many authors have argued, these velocity gradients are indicative of cloud rotation (e.g., Kutner et al. 1977; Blitz 1993; Rosolowsky et al. 2003), then  angular momentum conservation should contribute to our understanding of the formation and evolution of molecular clouds.  Significant progress was made in this direction when Rosolowsky et al. (2003) studied the angular momentum properties of giant molecular clouds (GMCs) in M33.  Their measurements, based on high-resolution \co{12}($J=0\rightarrow 1$) observations, showed that M33 molecular clouds have velocity gradient magnitudes that are comparable to Milky Way GMCs ($\sim 0.1$ \vunits).  They found that typical gradients of the GMCs are 5 to 10 times smaller than would be expected from simple formation theories.  Furthermore, just as Galactic GMCs are sometimes observed with gradients that are not parallel to the sense of Galactic rotation (e.g., Blitz 1993; Phillips 1999; Imara \& Blitz 2010), Rosolowsky et al. (2003) found that the gradient directions of M33 GMCs are often not in alignment with the galaxy rotation.  This is the so-called ``angular momentum problem'' of GMC formation.
 
It is our goal in this paper to address the question of the origin of velocity gradients in GMCs by doing a detailed analysis of the gradients in the atomic gas associated with the M33 molecular clouds catalogued by Rosolowsky et al. (2003).  This paper is an extension of our study in the Milky Way (Imara \& Blitz 2010, henceforth Paper I) in which we compared the velocity fields of GMCs to those of the local atomic gas surrounding them.  Our study is prompted, in part, by observations in external galaxies showing that high-density \HI~appears to be a necesssary but not sufficient condition of GMC formation (e.g., Mizuno et al. 2001; Engargiola et al. 2003).  Furthermore, it has been suggested that the discrepancy between observations and theoretical expectations in GMC angular momentum arises because angular momentum has been redistributed or ``shed'' during the course of GMC evolution (Mouschovias \& Paleologou 1979; Mestel \& Paris 1984; Rosolowsky et al. 2003, Kim et al. 2003).  It seems reasonable, then, to look for evidence of this redistribution in the immediate environs of GMCs.    And thus, we are motivated to search for clues to the angular momentum problem by comparing the velocity fields in GMCs to those of the atomic gas with which they are associated.  Whereas Rosolowsky et al. (2003) make inferences about the \HI~velocity field from the M33 rotation curve, in this study, we take measurements of the gradients in the atomic gas associated with the GMCs directly from observations.

In the next section, we describe the observations used to perform this analysis.  In \S \ref{sec:properties}, we provide the methodology and discuss the properties of atomic gas surrounding molecular clouds in M33.  Implications for GMC formation and evolution are discussed in \S \ref{sec:discuss}, and the results are summarized in \S \ref{sec:summary}.

\section{Data}\label{sec:data}
The data for the M33 molecular clouds are obtained from the Rosolowsky et al. (2003) catalog.  Their work was a high-resolution continuation to the Engargiola et al. (2003) survey of 148 GMCs.   Rosolowsky et al. observed 17 fields of the highest-mass clouds from their previous study in \co{12}($J=0\rightarrow 1$) using the C configuration of the BIMA array (Welch et al. 1996) in the fall of 2000 and spring of 2001.  The $\sim 6\arc$ resolution of the observations, corresponding to a linear resolution of 20 pc, was sufficiently high to resolve most GMCs and measure their sizes, which was done using the prescription of Rosolowsky \& Leroy (2006).  The high spatial resolution, in combination with the velocity resolution of 2 \kms, also allowed them to determine velocity gradients as small as $\sim 0.01$ \vunits~in the GMCs.  The magnitudes and directions of the gradients measured by Rosolowsky et al. are listed in Table 1.

Observations of the 21-cm line in M33 were taken from VLA archives and reduced by Schruba (A. Schruba \& A. LeRoy, priv. comm.).  The data, taken by Thilker et al. in 1997, where first described by Thilker et al. (2002).  We converted the data cube from units of flux density ($S_\nu$) to brightness temperature ($T_b$) via the Rayleigh-Jeans relation,
\begin{equation}
T_b = 1360 \frac{S_\nu \lambda^2}{b_{maj}\times b_{min}}~\t{K} = 6.055 \times 10^5\frac{S_\nu}{b_{maj}\times b_{min}}~K,
\end{equation}
where $S_\nu$ is in Jy, $\lambda=21.1$ cm is the observing wavelength, and  $b_{maj}\times b_{min} = 8.3^{\prime\prime}\times 7.5^{\prime\prime}$ is the synthesized beam size in arcseconds.  The final map has $2^{\prime\prime}$ pixels.  The LSR velocity range covered was $-321$ to $-65$ \kms~with  $1.3$ \kms~channel spacing, which was sufficient to include all of the gas in the disk of M33.  

\section{\HI~Properties}\label{sec:properties}
In order to compare the velocity gradients of molecular clouds with the gradients of the material from which they formed, we develop a set of criteria for selecting regions of atomic gas that are associated with the GMCs in our sample.  The challenge is to determine \emph{physically meaningful}  boundaries of atomic gas surrounding a given molecular cloud.  Thus, for each GMC, we begin by isolating subcubes in the VLA 21-cm data set that coincide with the GMCs kinematically and spatially.  Properties of the molecular clouds and the regions surrounding them are listed in Table 1.  We use a distance to M33 of 850 kpc for this analysis (Kennicutt 1998), and we assume a galactic inclination of $51^\circ$ (Corbelli \& Salucci 2000) when calculating the gas surface densities.

\subsection{Selecting \HI}\label{subsec:hi}
Molecular clouds have well-defined boundaries at which a phase transition occurs between the molecular and atomic gas and, thus, they generally have well-defined dimensions and masses (e.g., Kutner et al. 1977; Blitz \& Thaddeus 1980; Blitz 1993). The neutral gas associated with GMCs, however, does not have such distinct edges but rather has slower, more gradual transitions from high- to low-surface density material.  In trying to determine appropriate boundaries for \HI~associated with GMCs, therefore, our challenge is to define regions that are large enough so that most of the \HI~belonging to a given system is included, and yet the regions must be centered close enough to the GMCs so that the associations are physically meaningful and we do not include too much unrelated material.

There are a number of properties of \HI~and GMCs observed both in the Milky Way and M33 that help us develop our selection criteria.  First, an \HI~peak is typically found to have a velocity distribution which may be approximated by a Gaussian and whose line center falls near the CO line center of the affiliated molecular cloud (Andersson et al. 1991; Engargiola et al. 2003).  We assume that the atomic gas associated with a given GMC will be at approximately the same velocity as the GMC, within the limits set by the velocity dispersion inherent to the atomic gas.  
Since the velocity dispersion of the \HI~gas is several times larger than that of the CO, we select velocities falling within  $\pm 25$ \kms~of the CO line center, so that we include as much emission as possible from the atomic gas associated with a given GMC.  We decide to select from a wider range of velocities than the previous study (i.e., Chapter 2) because the work of Engargiola et al. (2003) shows that the \HI~associated with GMCs in M33 has emission extending as much as $\pm 35$ \kms~from the line center of the CO emission.  And because M33 has a moderate inclination ($\sim 51^\circ$; Corbelli \& Salucci), the problem of source confusion is minimized, allowing us to select a more generous range of velocities.   Changing the range of velocities by a 5 -- 10 \kms~does not significantly change the results.  This is because the velocity maps we create are weighted by the intensity (see \S \ref{subsec:gradients}),  so low-level emission occurring near the wings of the spectra does not make significant contributions to the velocity maps nor, therefore, to the measurement of the velocity gradients.

\begin{figure*}[htbp]\centering
\includegraphics[scale=.6]{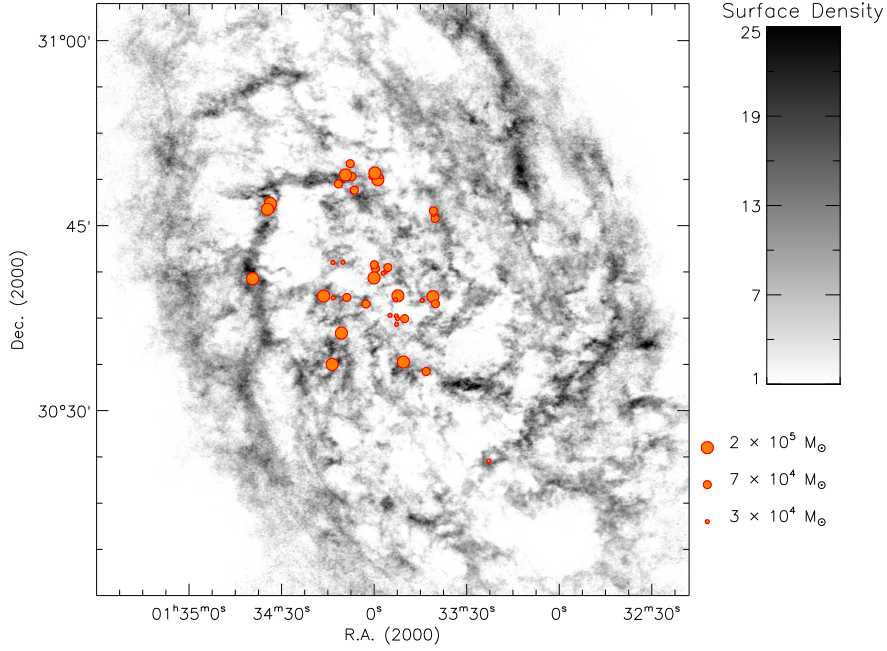}
\caption{M33: Grayscale image in units of \sunits~of the 21-cm emission of the central $45^\prime\times 45^\prime$ field.  
Molecular clouds are overlaid with area scaled to mass.  Nearly all (93\%) GMCs lay in regions of high-density \HI.
The galactic mean of $\Sigma_\t{HI}$ is roughly 4 \sunits, while the mean value in the vicinity of GMCs is 10 \sunits.}\label{fig:m33}
\end{figure*}

\begin{figure*}[htbp]\centering
\includegraphics[scale=0.5]{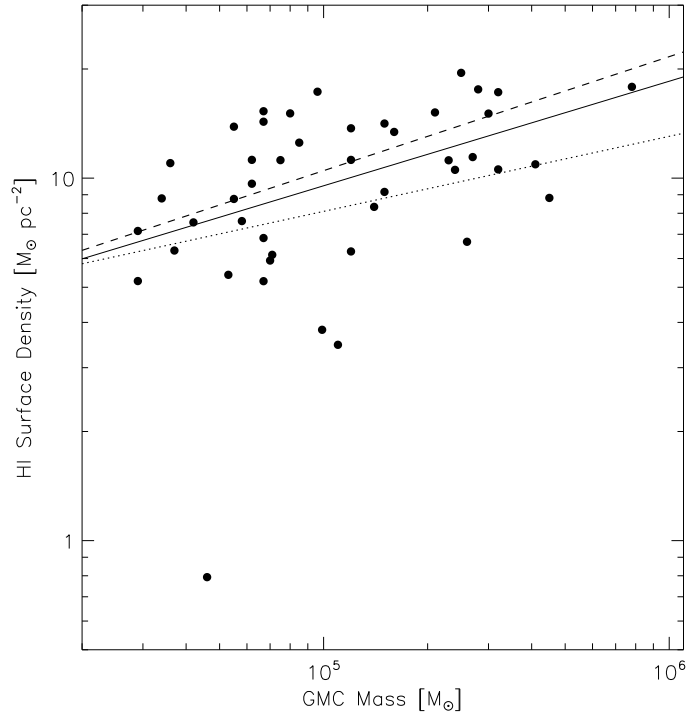}
\caption{Mean local $\Sigma_\t{HI}$ versus GMC mass.  The solid line is the least-squares fit for the plotted points, which show the mean \HI~surface density within 70 pc of GMCs. The dotted and dashed lines are the least-squares fits for  $\Sigma_\t{HI}$ within $R<150~\t{pc}$ and $\Sigma_\t{HI}$ within $R<50$ pc, respectively.  The average surface density of atomic gas surrounding GMCs is $10.2\pm 0.4$ \sunits. The \HI~surface density increases slowly with increasing GMC mass as $\Sigma_\t{HI} \propto M_\t{GMC}^{0.27 \pm 0.06}$, where the power-law exponent is the mean of the three least-squares fits and the uncertainty is the 1-$\sigma$ spread.  }\label{fig:surf_mass}
\end{figure*}

Identifying appropriate spatial boundaries is less clear-cut.  Figure \ref{fig:m33} shows that in M33, there is a spatial correspondence between GMCs and over-dense \HI~filaments.  The radial accumulation length, $R_A$, can give us an idea of the extent of the atomic gas associated with GMCs.  The size of an area which collapses to form a molecular cloud, $R_A$, can be estimated by requiring that the total \HI~mass in the region be equal to the observed mass of the cloud.  The minimum value of $R_A$ is determined by assuming that the initial configuration of atomic gas is a cylinder stretching to infinity in the direction perpendicular to the galactic disk:
\begin{equation}
\pi R_A^2\Sigma_\t{HI}=M_\t{GMC}, \label{eq:accum}
 \end{equation}
where $\Sigma_\t{HI}$ is the average local surface density of atomic hydrogen, measured from the 21-cm data.  We estimate $\Sigma_\t{HI}$ for different sized regions centered on the GMCs.  The mean value of $\Sigma_\t{HI}$ spans a narrow range of values for regions having a size of $\le 200$ pc.   Within 150 pc of molecular clouds, $\Sigma_\t{HI}$ is 8.8 \sunits; within 50 pc the average surface density is 11.5 \sunits.  Figure \ref{fig:surf_mass} plots the mean \HI~surface density within 70 pc of GMCs versus GMC mass.  The figure shows that as GMC mass increases, the  surface density of associated \HI~(and therefore \HI~mass) slowly increases as $\Sigma_\t{HI} \propto M_\t{GMC}^{0.27}$.  We find that the average \HI~surface density in the vicinity of GMCs is roughly 10 \sunits.   This is the saturation level of atomic gas, observed in other spiral galaxies including the Milky Way, above which gas becomes primarily molecular (Martin \& Kennicutt 2001; Wong \& Blitz 2002; Bigiel et al. 2008). We calculate $R_A$ for the slightly different estimates of $\Sigma_\t{HI}$ and find  a mean value of $\sim 70$ pc.  Thus, we define \HI~regions in the 21-cm data cube as 140 pc $\times$ 140 pc $\times$ 50 \kms~connected regions that are centered on the locations of GMCs in position and velocity space.  Note that the area defined by the accumulation radius of a GMC contains gas which has filled in the space previously occupied by that which formed the GMC.

We briefly note that defining \HI~associated with GMCs by identifying contiguous regions in position-velocity space above a given temperature threshold would not have been appropriate for our purposes. Given that the \HI~has such a diffuse distribution compared to the molecular gas bound in GMCs, it is unclear what such a threshold should be.  Even if we were to define \HI~clouds using this method, we would still be left with the problem of deciding which clouds in the vicinity of a given GMC were associated with the said GMC and which which were not.  Based on this rationale, we find it much more suitable to identify \HI~regions using the accumulation radius analysis described above.

\begin{figure*}[htbp]
\includegraphics[scale=0.75]{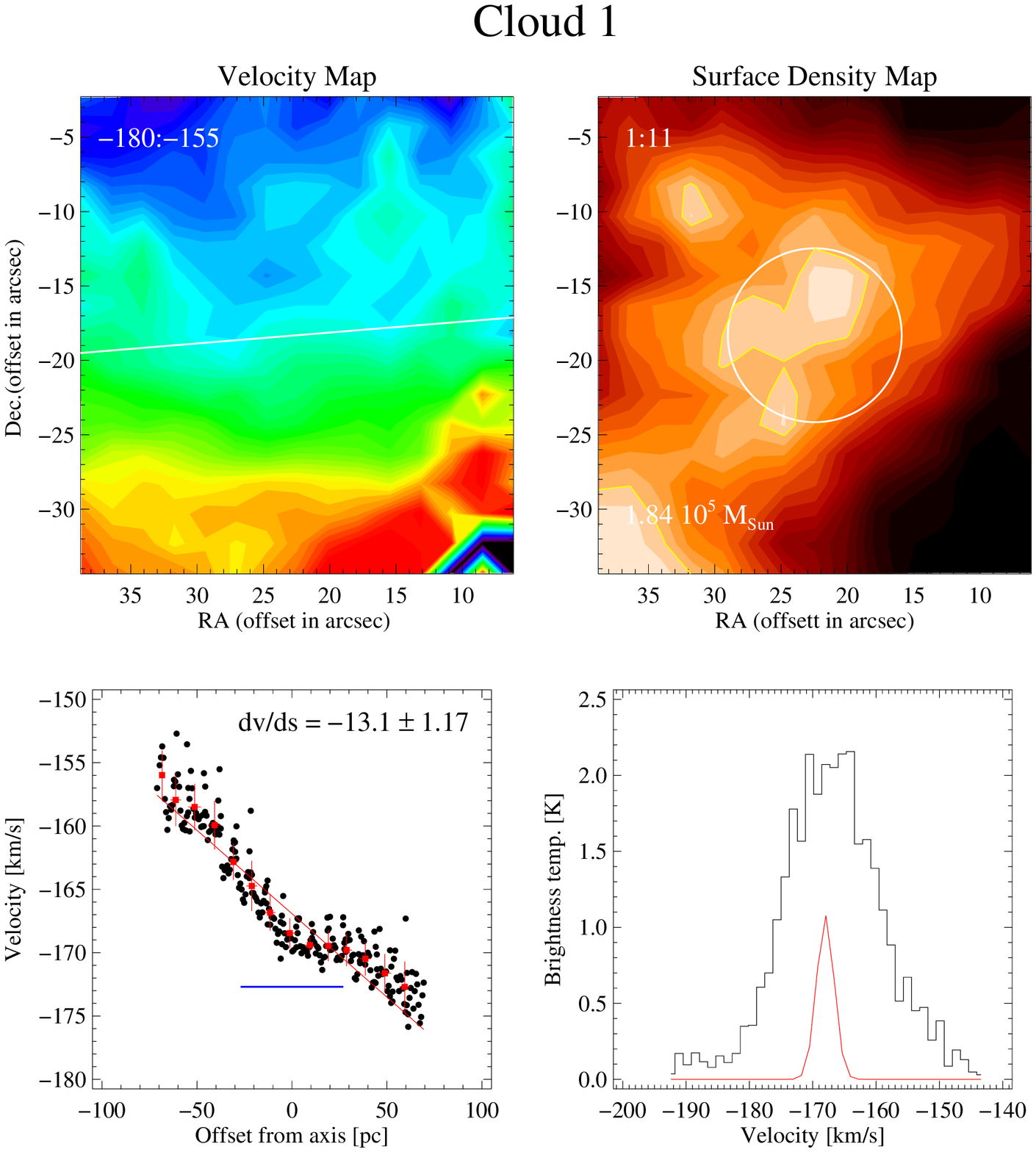}
\vspace{-1.0em}
\caption{Cloud 1: The top left figure shows the intensity-weighted first moment map of the \HI~ with the gradient axis overlaid.  The velocity range of the map is indicated in the top left corner in units of \kms; red represents the maximum speed.  Below this figure is a plot of the central velocity at a given location in the first-moment map versus perpendicular offset from the gradient axis; the linearity of the plot indicates that a plane is a good fit to the first-moment map; the radial extent of the GMC is demarcated by the horizontal blue line. The top right figure is a surface density map of the \HI~overlaid with a circle matching to the size of the associated GMC; the 10 $\sunits$ contour is overlaid in yellow.  The range of \HI~surface densities displayed in the map are in the top left corner in units of \sunits, and the total \HI~mass in the region is written in the bottom left corner.  Below is a plot of the average spectra of \HI~emission (black) and CO emission (red) toward the region.}
\label{fig:grad1}
\end{figure*}

\begin{figure*}[htbp]
\includegraphics[scale=0.8]{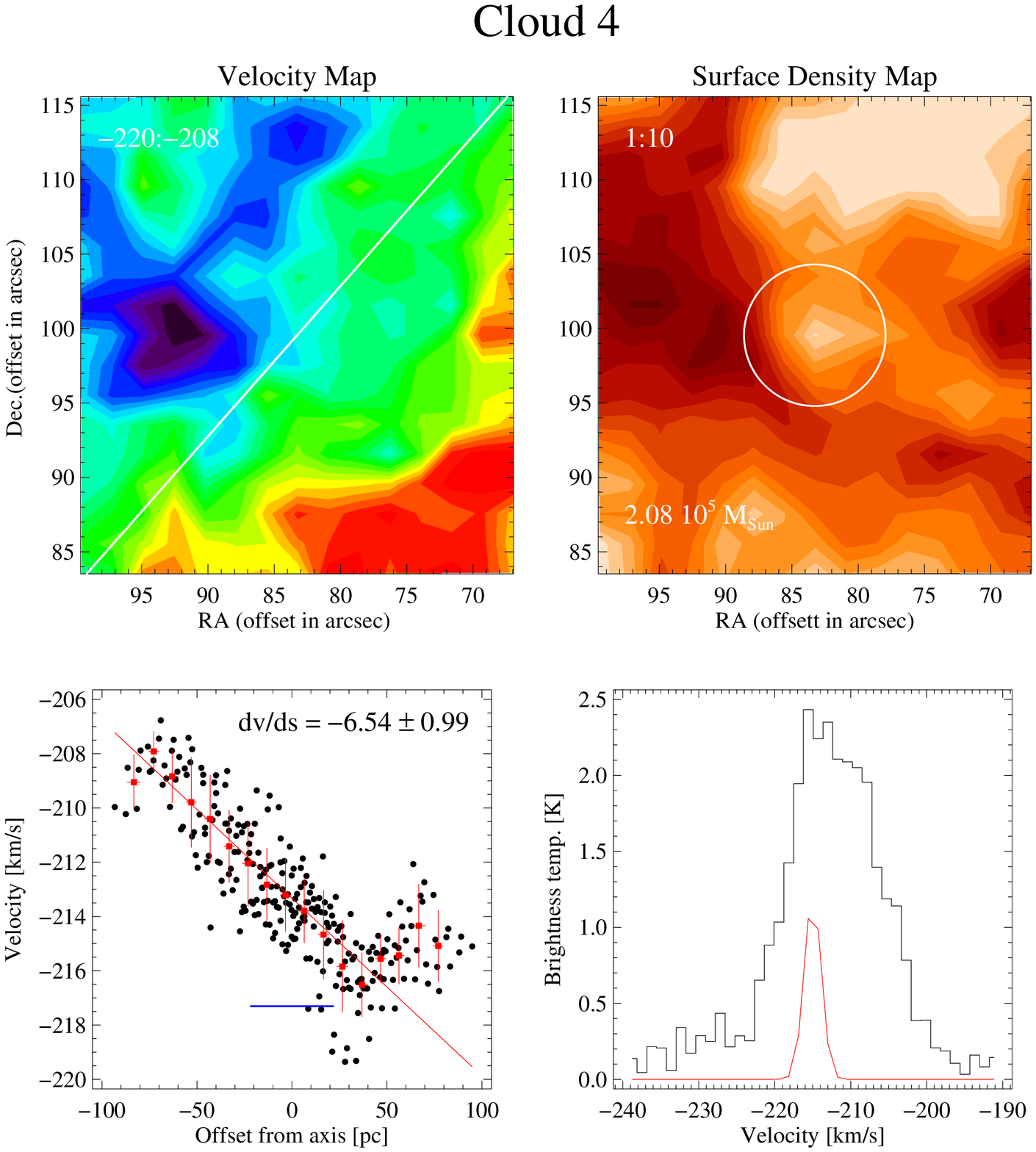}
\caption{Cloud 4: Same as Figure \ref{fig:grad1}.}\label{fig:grad2}
\end{figure*}

\begin{figure*}[htbp]
\includegraphics[scale=.8]{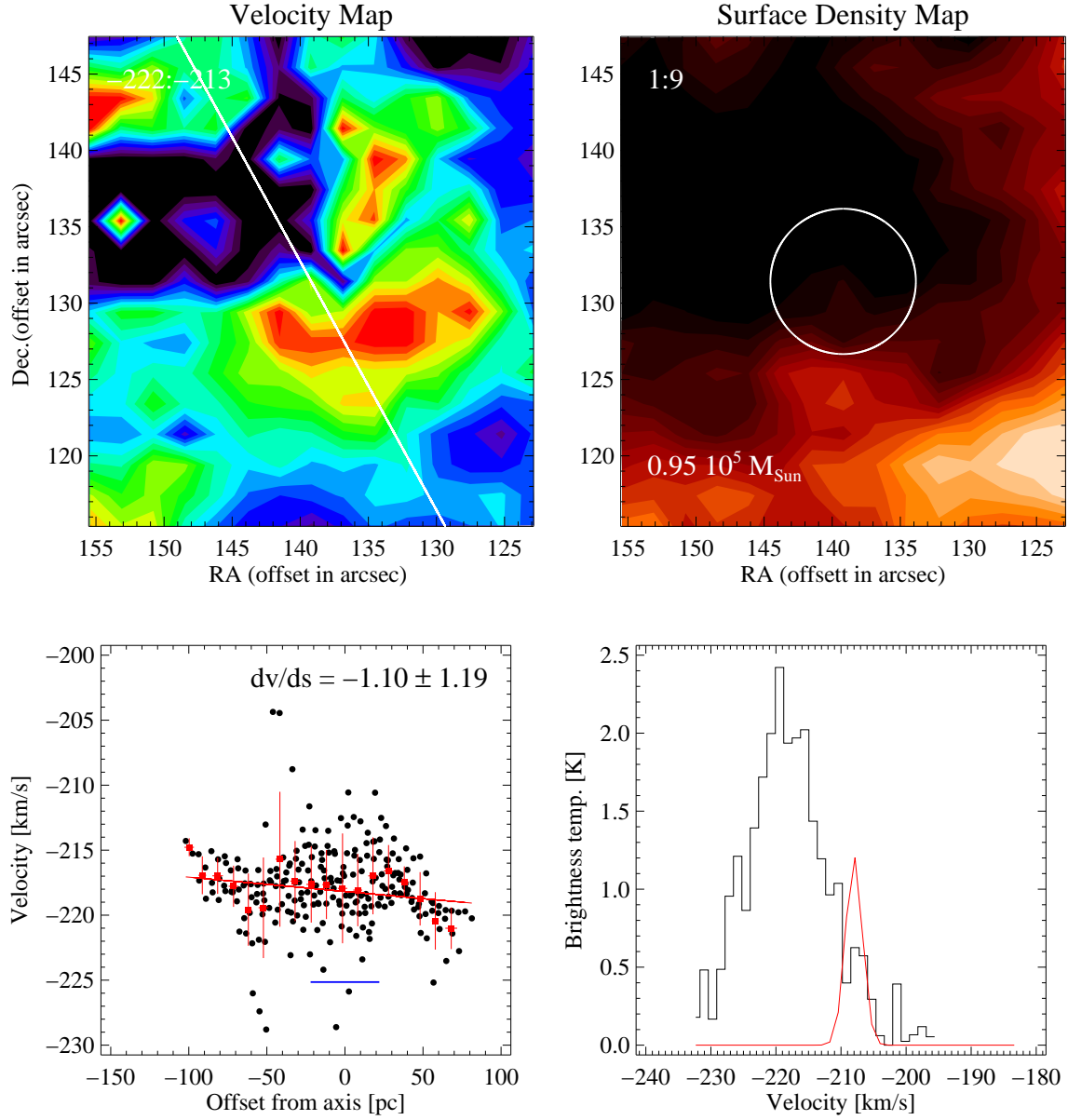}
\caption{Cloud 13: Same as Figure \ref{fig:grad1}, accept that the slope of the position-velocity plot is close to zero, indicating that a plane is not a good fit to the first-moment map.  Cloud 13 is an example of a region that does not exhibit a linear gradient and in which the GMC does not coincide with a local maximum in the \HI.}\label{fig:grad3}
\end{figure*}

\begin{figure*}[htbp]
\includegraphics[scale=0.8]{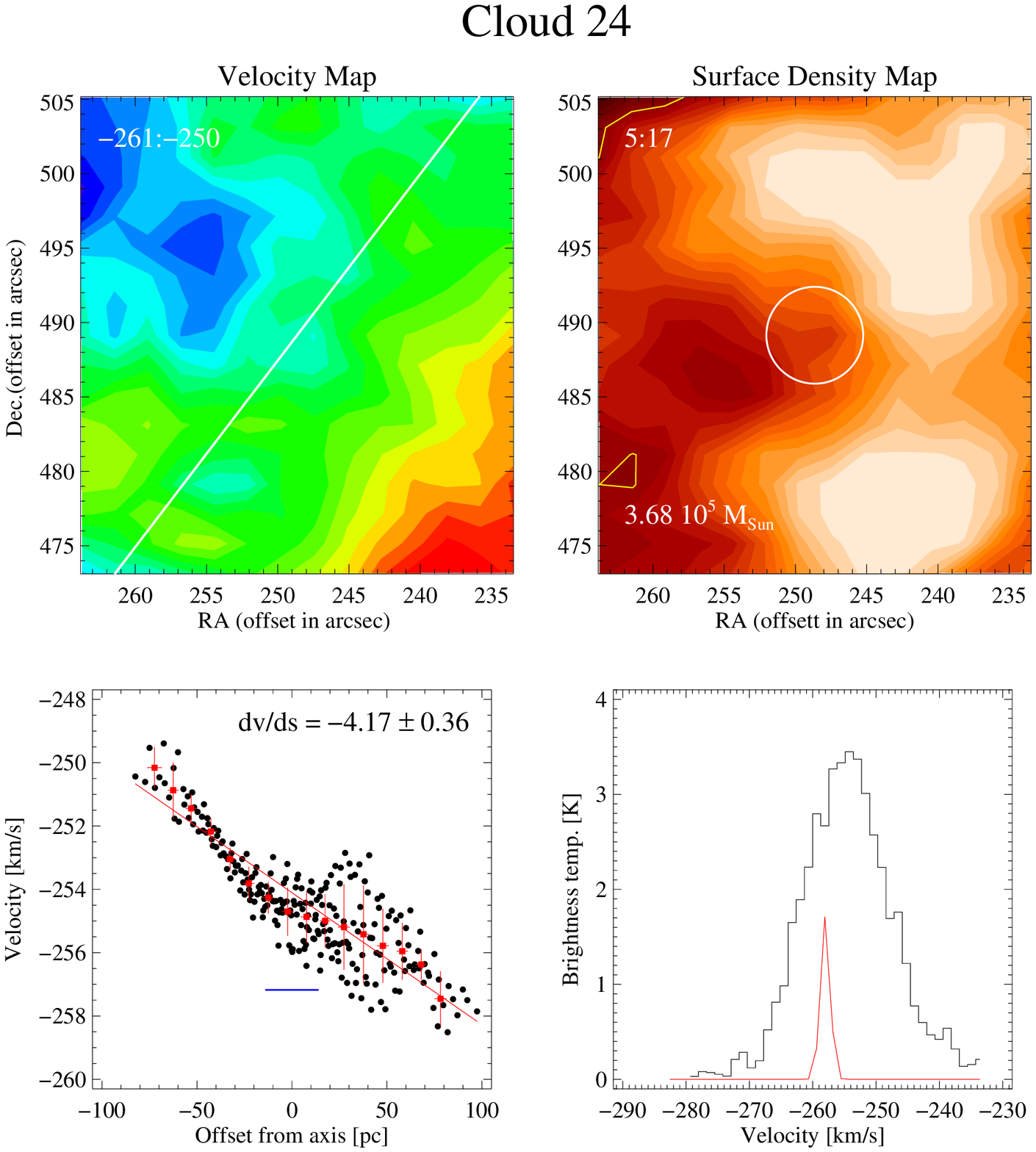}
\caption{Cloud 24: Same as Figure \ref{fig:grad1}.}\label{fig:grad4}
\end{figure*}

\begin{figure*}[htbp]
\includegraphics[scale=.8]{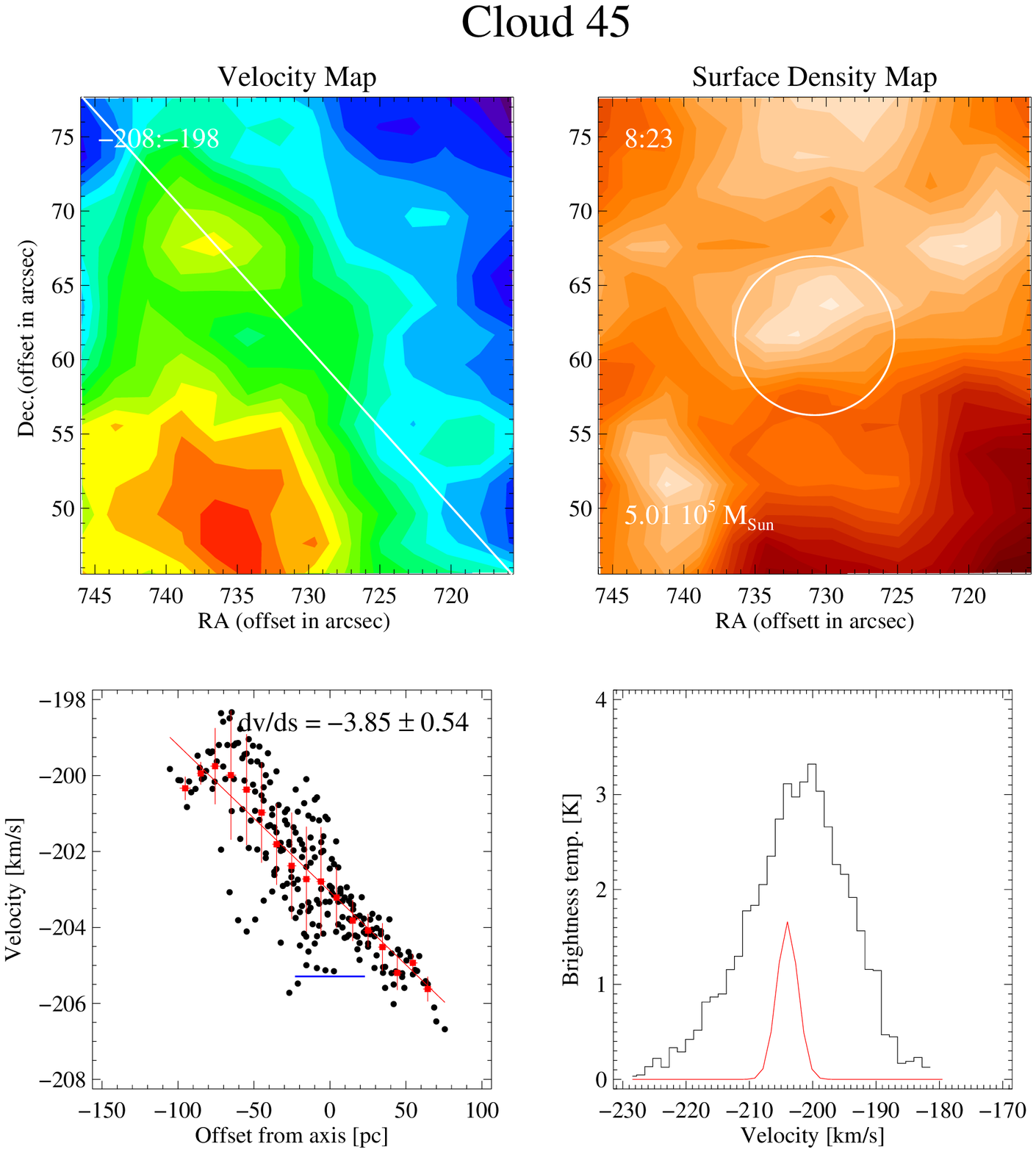}
\caption{Cloud 45: Same as Figure \ref{fig:grad1}.}\label{fig:grad6}
\end{figure*}

The top-right panels of Figures \ref{fig:grad1} -- \ref{fig:grad6} show surface density maps of the \HI~surrounding six of the 45 GMCs in this study. The  bottom-right plot displays the average \co{13} and \HI~spectra toward the clouds.  (The left-hand panels are described in \S \ref{subsec:gradients}.)  Based on the spatial and kinematic proximity of GMCs to local maxima in the atomic gas, we observe three rough ``classes'' of clouds.  Most GMCs in the catalog are spatially and kinematically coincident with local \HI~maxima.  Clouds 1, 4, 39, and 45 are representative examples---at least $1/3$ of the area of the molecular cloud coincides with a local peak in the atomic gas.  Twenty-nine out of 45 the GMCs, $64\%$, fall into this class.  The average GMC mass in this class is $2\times 10^5$ \msun, with a dispersion of 80\%.  Another group of clouds is kinematically but not quite spatially coincident with \HI~maxima.  GMCs in this class, including Clouds 16, 24 (Figure \ref{fig:grad4}), 35, and 44 are located near the edge of a bright \HI~peak or sit on a filament between two peaks. The thirteen  ($29\%$)  molecular clouds belonging to this class tend to have their mean LSR velocities offset from the mean \HI~velocity by a few \kms.   The typical GMC mass in this second class is nearly 3 times smaller than in the first:  $7\times 10^4$ \msun, with a dispersion of 30\%. 

At least two GMCs in the catalog, Clouds 13 (Figure \ref{fig:grad3}) and 38, are not associated with any local \HI~maximum.  For both of these clouds, the center of the GMC is roughly 60 pc from the nearest peak in \HI~surface density, and the central velocity of the molecular and atomic clouds differ by at least 10 \kms.  At $2.05\times 10^5$ \msun, the total mass (molecular and atomic) in the region surrounding Cloud 13 is among the lowest in this study; only four other regions have less total mass, including Clouds 7, 8, 9, and 43.  Cloud 38 does not appear to have any other characteristics, such as mass or  size, that distinguishes it from others in the catalog.  The one other case in which the there are no \HI~peaks in the vicinity of the molecular cloud may be Cloud 43. It has the lowest mass of the sample, and it is one of three GMCs farthest from the galactic center.  However, because of its anomalously high LSR velocity of $-74$ \kms, compared to a mean galactic value of $-111$ \kms~at that location, this molecular cloud---which was not observed in the original Engargiola et al. (2003) catalog---may be a false detection.  In their study of star formation in the outer parts of M33, Gardan et al. (2007) also found GMCs that were not associated with strong \HI~emission.  Neither of these two clouds, located at ($ 01^\t{h}34^\t{m}25^\t{s}$, $30^\circ 54^\prime 50^{\prime\prime}$) and ($ 01^\t{h}34^\t{m}16.9^\t{s}$, $30^\circ 59^\prime 31.4^{\prime\prime}$ ), were detected by Engargiola et al. (2003), whose study was confined to the central region of M33.

\subsection{\HI~Without GMCs}\label{subsec:non-gmc}
Since over-dense regions of atomic hydrogen are generally necessary but not sufficient for molecular cloud formation, we would like to know what the distinguishing features are between atomic gas harboring GMCs and regions of dense \HI~in which GMCs have not been detected.  One way to address this issue is to search for velocity gradients along ``non-GMC'' \HI~filaments in M33 and then compare the properties to regions of \HI~containing GMCs.  We would like to know, for example, if we can find regions of \HI~having low angular momentum---that is, angular momentum comparable to that which is observed in GMCs?

\begin{figure*}[htbp]
\includegraphics[scale=.9]{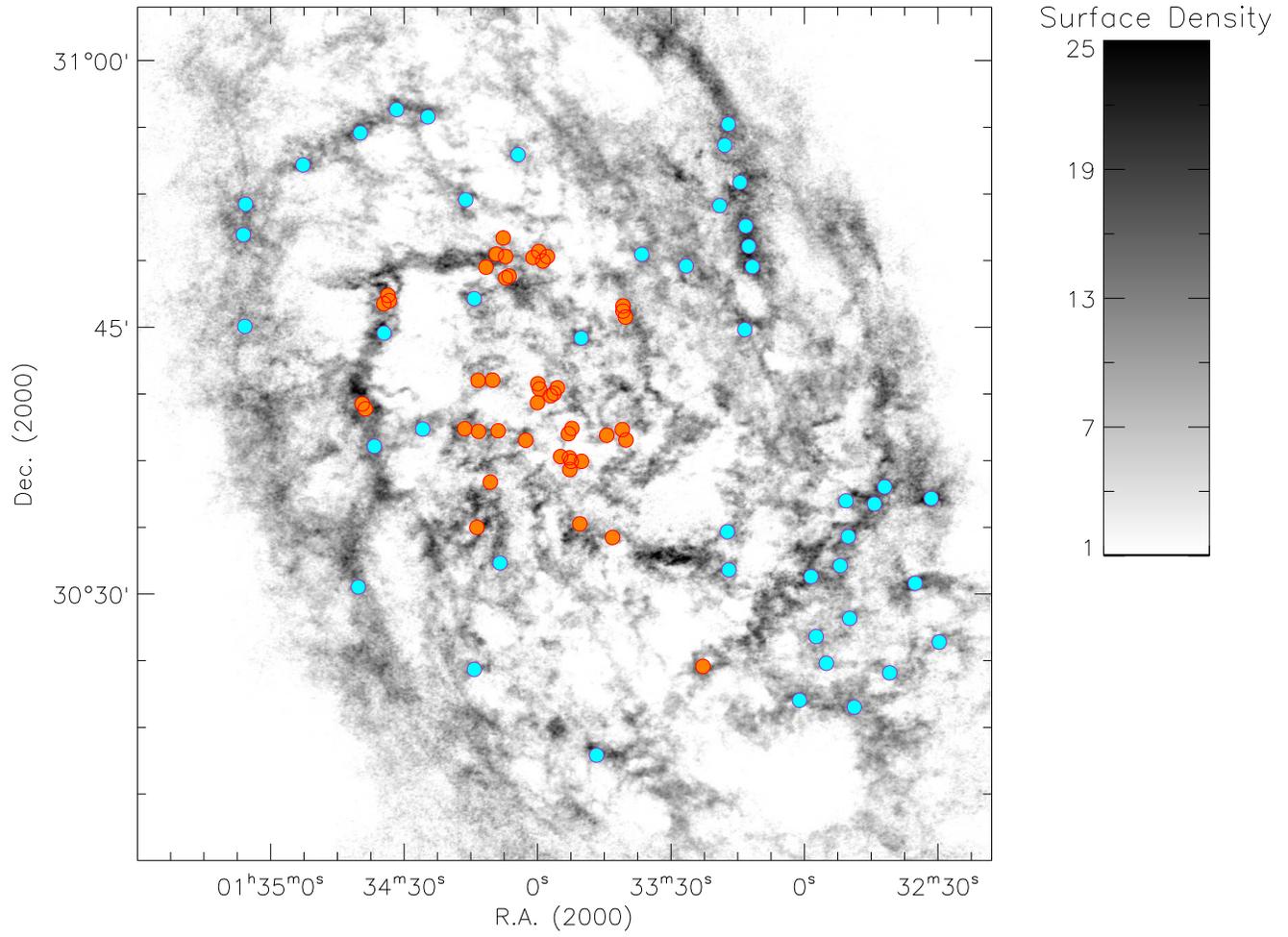}
\caption{M33: Grayscale image in units of \sunits~of the 21-cm emission of the central $45^\prime\times 45^\prime$ field. The locations of \HI~regions containing molecular clouds are overlaid in orange, and the locations of \HI~regions without observed molecular clouds are overlaid in blue. }\label{fig:m33_fake}
\end{figure*}

\begin{figure*}[htbp]
\includegraphics[scale=.8]{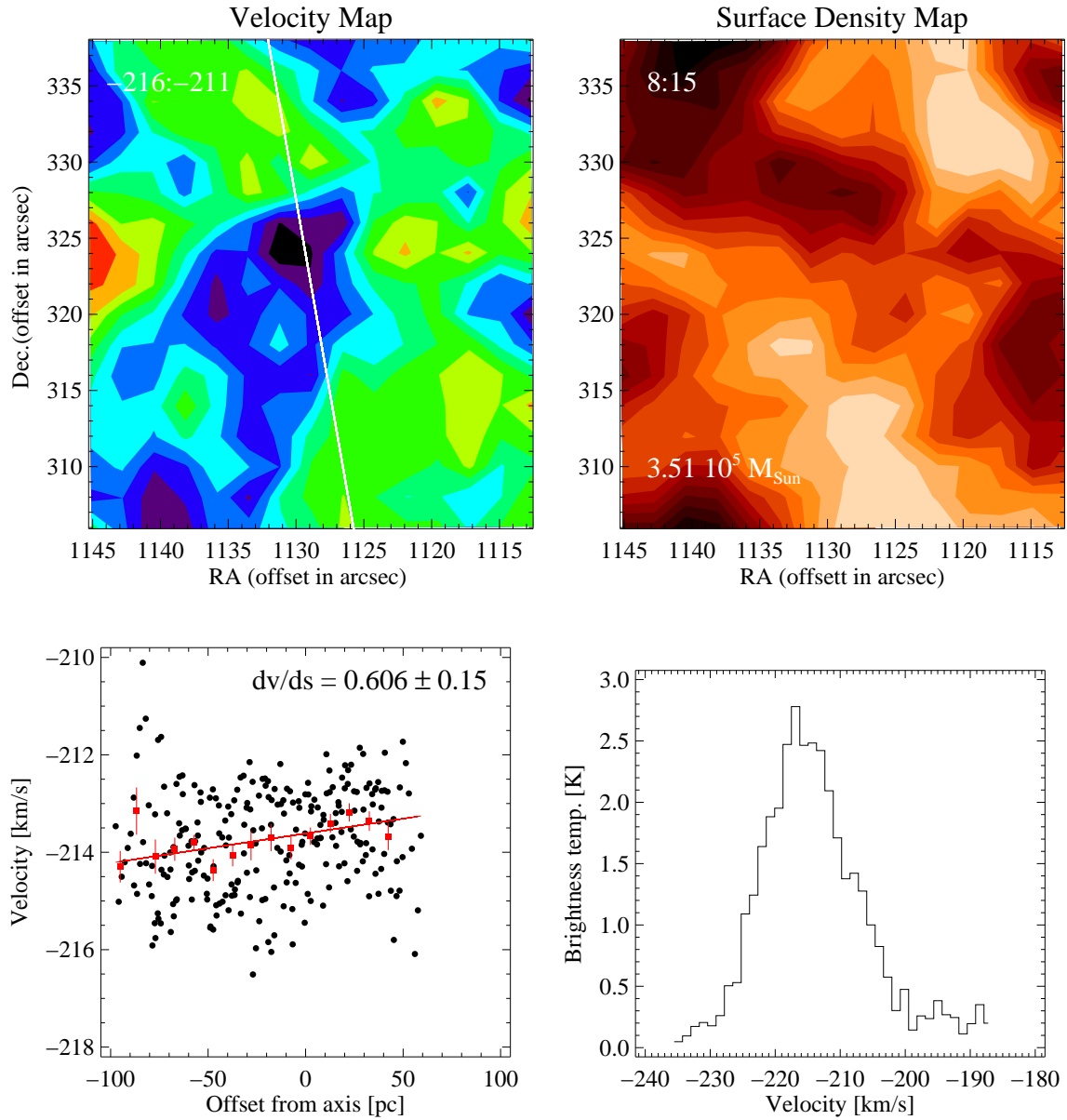}
\caption{Cloud 1A: Same as Figure \ref{fig:grad1}, accept for a region in which GMCs have not been observed.  }\label{fig:fake1}
\end{figure*}

\begin{figure*}[htbp]
\includegraphics[scale=.8]{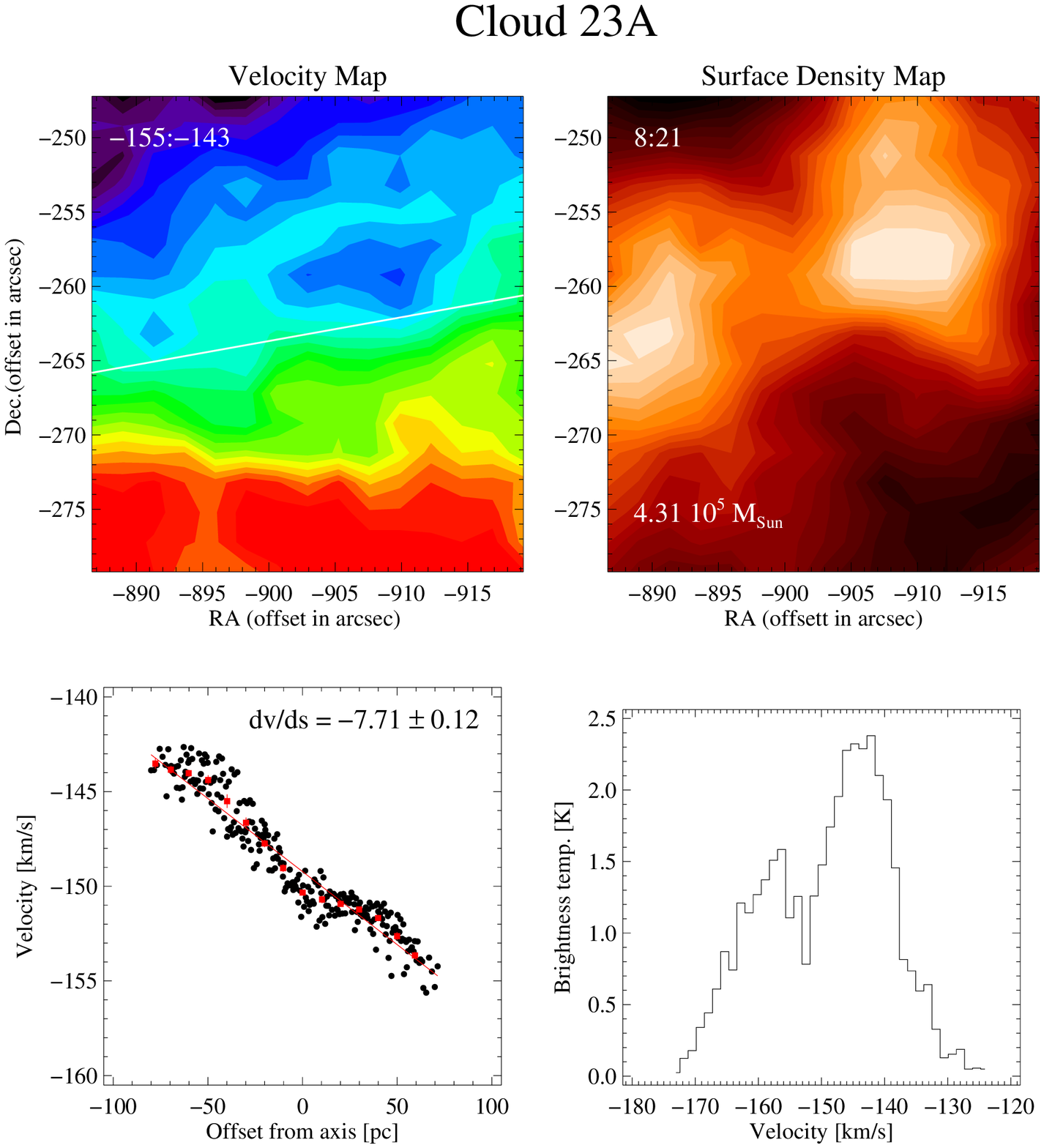}
\caption{Cloud 23A: Same as Figure \ref{fig:fake1} }\label{fig:fake2}
\end{figure*}

\begin{figure*}[htbp]
\includegraphics[scale=.8]{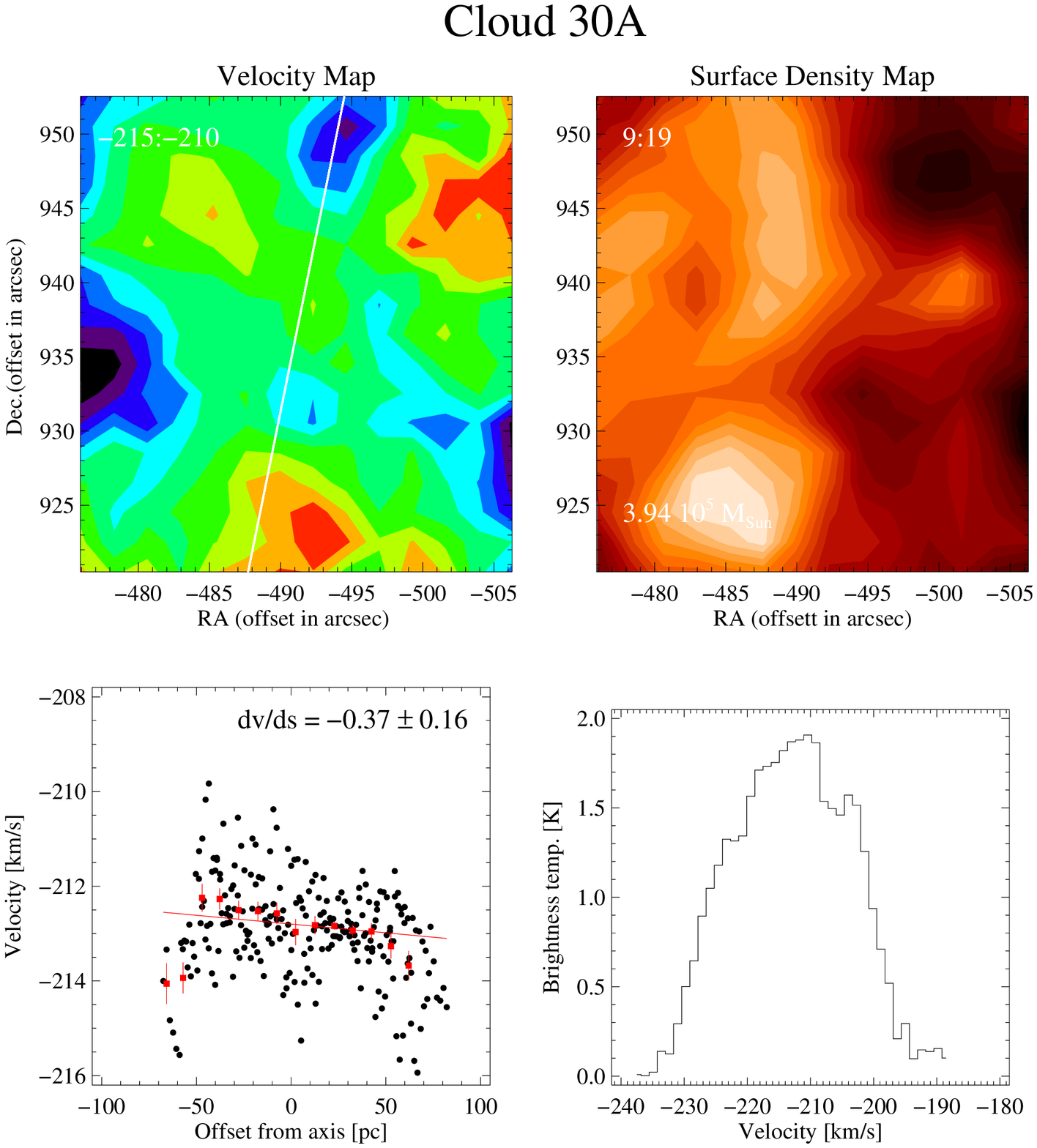}
\caption{Cloud 30A: Same as Figure \ref{fig:fake1}}\label{fig:fake3}
\end{figure*}

We select a population of 45 over-dense \HI~regions---corresponding to the number of GMCs in the Rosolowsky et al. (2003) catalog---in the M33 map along \HI~filaments in which GMCs have not been observed.  The regions, displayed in Figure \ref {fig:m33_fake},  must have a mean surface density of at least 10 \sunits~within 140 pc $\times$ 140 pc, the same size as the regions of \HI~associated with GMCs.  Examples of the non-GMC surface density maps and \HI~spectra are shown in the right-hand panels of Figures \ref{fig:fake1} -- \ref{fig:fake3}.

To  choose a relevant range of velocities at each location, we begin by estimating the mean velocity at which the emission peaks.   The mean \HI~velocity, $v_0$, at a given location is derived from the intensity-weighted first moment of the entire velocity distribution.  For the \HI~associated with GMCs, we select velocities falling within $\pm 25$ \kms~of the CO line center.  For the non-GMC \HI, the surface density and velocity maps are created by integrating over velocities within $\pm 25$ \kms~of $v_0$.  Table 2 lists the coordinates, central velocities, and other properties of the non-GMC \HI~regions.

\subsection{Velocity Gradients}\label{subsec:gradients}
We create intensity-weighted first moment maps both of the 140 pc $\times$ 140 pc $\times$ 50 \kms~\HI~regions surrounding the molecular clouds and of the atomic gas not observed to contain molecular clouds.  Following Imara \& Blitz (2010), we fit planes to the resulting velocity maps and solve for the coefficients of the fit, which define the velocity gradient, $\Omega_\t{HI}$, and the gradient direction, $\theta_\t{HI}$.  The maps of regions containing GMCs are shown in the upper-left panels of Figures \ref{fig:grad1} --  \ref{fig:grad6}  and, of non-GMC regions, in Figures \ref{fig:fake1} -- \ref{fig:fake3}.  The results of the fits are given in Tables 1 and 2.

\begin{figure*}[htbp]\centering
\includegraphics[scale=.5]{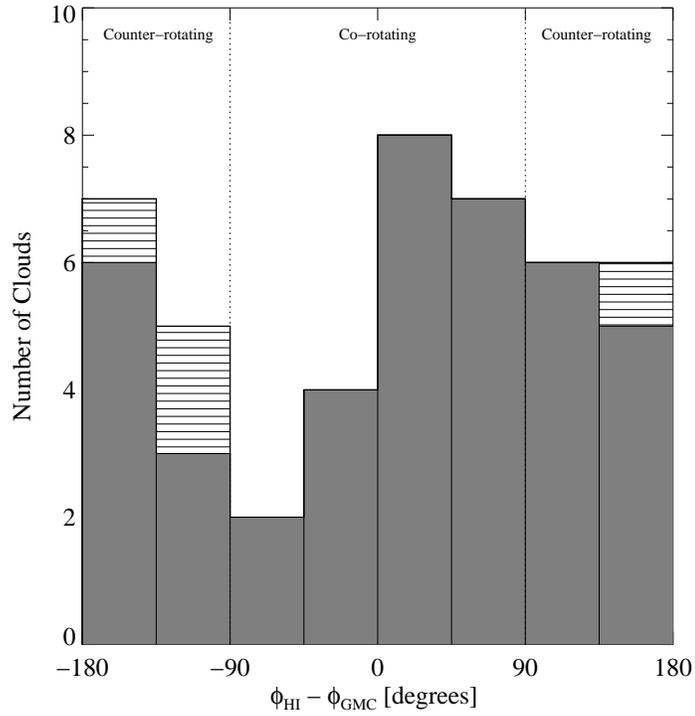}
\caption{Comparison of \HI~and GMC position angles.  \emph{If} the velocity gradients indicate rotation, then most GMCs ($\sim 53\%$) are not rotating in the same sense as the associated \HI. The hatched portions of the histograms in (b) and (c) represent regions having non-linear gradients. }\label{fig:m33_hist2}
\end{figure*}

We find that (1) approximately 41 out of 45 of the \HI~patches harboring GMCs have significant linear velocity gradients; the mean for this group is $0.050\pm 0.004$ \vunits~(Figure \ref{fig:m33_hist1}); (2) the mean of the gradients in the non-GMC \HI, $0.033\pm 0.003$ \vunits, are generally smaller than those measured in atomic gas observed to contain GMCs; (3) the position angles of the GMC rotation axes are generally \emph{not} aligned with the axes of the surrounding atomic gas, as seen in  Figure \ref{fig:m33_hist2}; and (4) about $62\%$ of the \HI~regions associated GMCs have position angles that differ from the axis of the galaxy by more than $90^\circ$ (Figure \ref{fig:m33_hist3}).

\subsubsection{Gradient Magnitudes}\label{subsubsec:magnitudes}

\begin{figure*}[htbp]
\includegraphics[scale=.8]{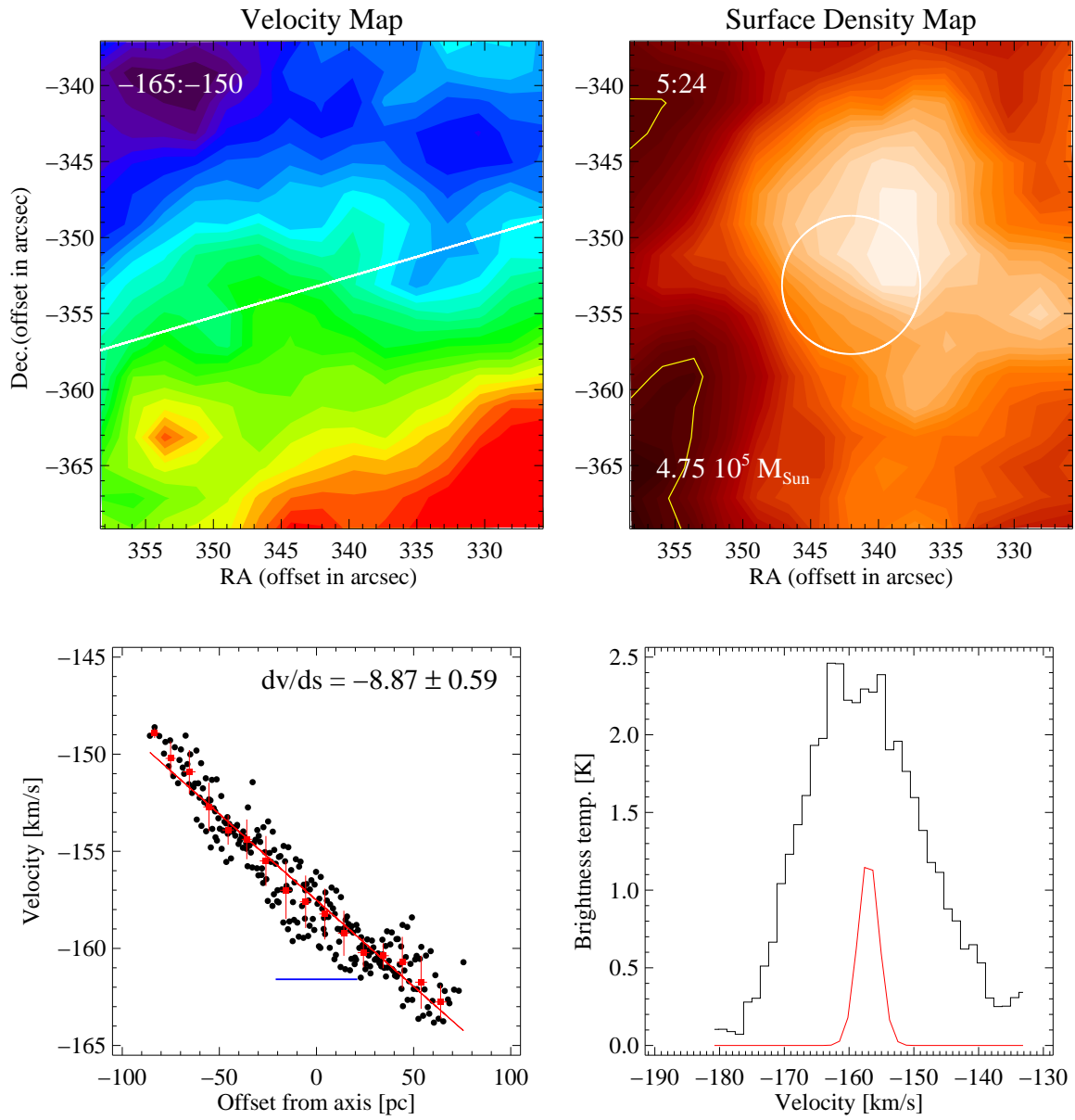}
\caption{Cloud 39: Same as Figure \ref{fig:grad1}.}\label{fig:grad5}
\end{figure*}

The upper-left panels of Figures \ref{fig:grad1} -- \ref{fig:grad6} show the first moment maps of the atomic gas within 140 pc $\times$ 140 pc of six of the GMCs in this study.\footnote{See the online appendix for figures of the remaining \HI~regions.}   
The position-velocity plots on the bottom-left indicate  whether planes are good fits to the first moment maps.  These plots show the mean velocity at a given location in the velocity map as a function of perpendicular offset from the rotation axis.  When a position-velocity plot can be fitted with a non-zero slope, $m$, this indicates that a plane is good fit to the first-moment map and that the velocity gradient can be assumed to be linear.   The slope of the position-velocity plot and the uncertainty in the least-squares fit of the slope, $\sigma_m$, are calculated for each gradient individually and are indicated in the figures; the ratio of these parameters is listed in Tables 1 and 2.

Based on examination of the first moment maps and the position-velocity plots, we set the following criteria for degree of linearity in the gradients.  Regions of \HI~having position-velocity plots $< 2$ $\sigma_m$ have non-linear, random velocity fields.  Regions with $m\approx 2$--$3$ $\sigma_m$ have ``marginally'' linear velocity gradients;  Figure \ref{fig:grad5} is an example of such a borderline case.  Regions with $m\ge 3$ $\sigma_m$ have unambiguously linear gradients.  We find that 39 out of 45 \HI~regions associated with GMCs fall into this last category.  Applying the same criteria to the non-GMC atomic gas, 43 regions have $m\ge 3$ $\sigma_m$.    Examples of the velocity fields in these regions are displayed in Figures \ref{fig:fake1} -- \ref{fig:fake3}.  We do not observe a significant correlation between the gradient magnitudes of molecular and atomic gas (Figure \ref{fig:vplot_m33}).  Tables 1 and 2 list the minimum linear extent of the gradients, as well as their magnitudes and directions.

Figure \ref{fig:m33_hist1} displays the distribution of gradient magnitudes in the molecular and atomic gas.  The mean of the magnitudes in the atomic gas surrounding GMCs, $0.050\pm 0.004$ \vunits, are comparable to those measured in clouds in the Milky Way (Imara \& Blitz 2010).  On average, the magnitudes in the \HI~around GMCs are less than the typical $0.07$ \vunits~observed in GMCs.    The average gradient observed in non-GMC \HI~is $0.033\pm 0.003$ \vunits.

\subsubsection{Gradient Directions}\label{subsubsec:directions}

 Many authors have argued that the linear velocity gradients observed in molecular clouds are caused by rotation (e.g., Kutner et al. 1977; Blitz 1993; Phillips 1999).  If this is the case, the velocity gradient magnitude of a cloud is a measure of its angular speed and the gradient direction, $\theta$, is perpendicular to its spin axis.  Figure \ref{fig:m33_hist2} demonstrates that 53\% of GMCs have position angles that differ from the those measured in the surrounding atomic gas by more than $90^\circ$. Thus, if the molecular clouds are rotating,  slightly more than half of the clouds are counter-rotating with respect to the atomic gas with which they are surrounded. 

\begin{figure*}[htbp]\centering
\includegraphics[scale=.5]{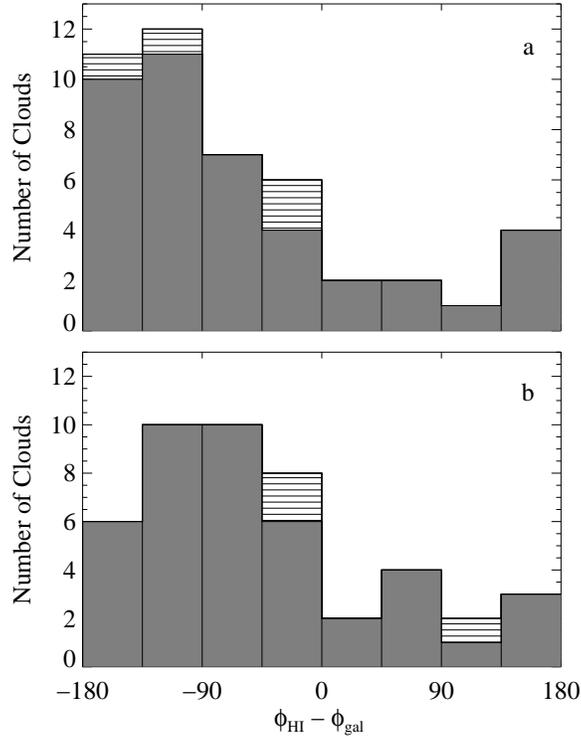}
\caption{Position angles of (a) GMC-harboring \HI~and (b) non-GMC \HI, with respect to that of M33. The hatched portions of the histograms in (b) and (c) represent regions having non-linear gradients. }\label{fig:m33_hist3}
\end{figure*}

Rosolowsky et al. (2003) show that for a galactic kinematic position angle of $\phi_\t{gal}=21^\circ$, GMCs are preferentially aligned with the galaxy and approximately $40^\circ$ are retrograde rotators.  We find that localized regions of \HI---both with GMCs and not observed to contain GMCs---have gradient directions that are generally \emph{unaligned} with respect to the galactic axis.  Figure \ref{fig:m33_hist3} shows that  62\% of the \HI~regions with GMCs have gradient position angles that differ by more than $90^\circ$ from the sense of galactic rotation.  The median difference between \HI~position angles and $\phi_\t{gal}$ is $-90^\circ$ for \HI~containing GMCs and $-60^\circ$ for non-GMC \HI.  

Rosolowsky et al. (2003) demonstrate that GMCs lying close together tend to have their gradients aligned.  Similarly, we find that neighboring regions of \HI~containing GMCs have comparable gradient directions, within $1^\circ$ to $15^\circ$ of each other. This is not surprising, since for neigboring GMCs we are taking measurements in overlapping regions of atomic gas.  In fact, sixteen of the molecular clouds in this sample have accumulation radii which overlap by at least $(1/2) R_A$, raising the question of whether the distribution of $\phi_\t{HI}-\phi_\t{gal}$, shown in Figure \ref{fig:m33_hist3}, is biased due to over-counting. We consider this by taking the average $\phi_\t{HI}$ for regions with accumulation radii overlapping by at least $1/2$.  We count such regions once and find that the distribution of $\phi_\t{HI}-\phi_\t{gal}$ is not significantly altered.  The median difference between between  $\phi_\t{HI}$ and $\phi_\t{gal}$, in this case, is still $-90^\circ$.

\begin{figure*}[htbp]\centering
\includegraphics[scale=.6]{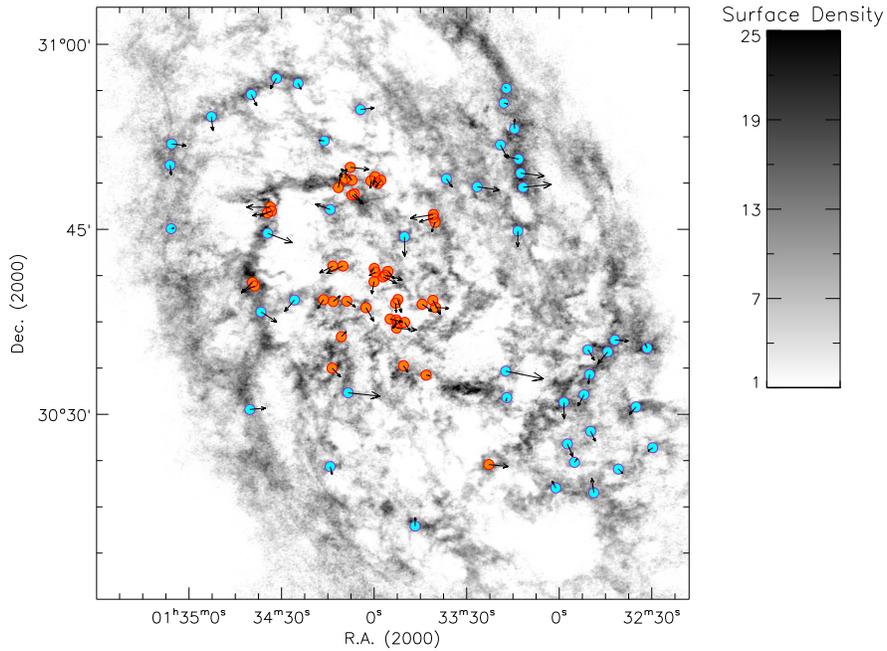}
\caption{M33: The directions of the gradients in the atomic gas are plotted for \HI~regions containing molecular clouds (orange) and for \HI~regions without observed molecular clouds (blue).  The arrows point in the direction of increasing velocity and have lengths proportional to the gradient magnitude.   The gradient directions of the individual velocity fields where GMCs are observed (or, where GMCs may potentially be in the process of forming) do not appear to make up a large-scale, systematic pattern.    }\label{fig:m33_directions}
\end{figure*}

Figure \ref{fig:m33_directions} displays the velocity vector field of both groups of atomic gas.  The arrows point in the direction of increasing velocity and have sizes proportional to the gradient magnitudes.  Along the \HI~filament in the the North-East of M33, the non-GMC \HI~regions tend to have gradient directions pointing toward the South and South-West.  And a string of non-GMC \HI~regions along the prominent filament in the South-West---Clouds 23A, 21A, 19A, 13A, 12A, 17A, and 15A---all have gradients pointing within $40^\circ$ of due South.  Gradient alignment is not always observed along filaments, however.  The non-GMC \HI~regions in the nearly vertical filament at a right ascension of of $\sim 01^\t{h}33^\t{m}30^\t{s}$ have a seemingly random distribution of gradient position angles.  We also observe adjacent regions of \HI~associated with GMCs having large differences in gradient directions---for instance, Clouds 20 and 23 located near $\sim 01^\t{h}34^\t{m}$, $30^\circ 39$ have gradients pointing in nearly opposite directions.  Overall, neither population of \HI~appears to have a global, systematic pattern.  

\section{Implications for GMC Evolution}\label{sec:discuss}
Assuming that molecular clouds are rotating, measurements of their angular momentum  may reveal information about their past.  The angular momentum per unit mass, $j$, is a useful quantity for comparing the angular momenta in different regions having comparable mass.  For a rotating body with a power-law density  profile, the specific angular momentum is
\begin{equation}
j=\beta\Omega R^2, \label{eq:j}
\end{equation}
where $\beta$ is a constant determined by the mass distribution, and $R$ is the cloud radius.  The constant $\beta$ ranges from 0.33 for oblong bodies to 0.5 for disks.  We adopt the intermediate value, $\beta=0.4$ (used for spherical structures having constant surface density), for GMCs.  Note that because we do not know the inclination, $i$, of a given cloud along our line of sight, our measurement of $\Omega$ underestimates the true value, $\Omega/\sin i$ and, hence, our measurement of $j$ is also underestimated. The position angle of the gradient is the direction of the total angular momentum vector.    

Equation \ref{eq:j} may also be used to estimate the specific angular momentum initially imparted to a GMC by the ISM from which it forms.  This depends on the process of GMC formation, and here, following Rosolowsky et al. (2003) and Paper I, we assume that GMCs collapse or condense from the surrounding atomic gas via a ``top-down'' mechanism, such as a large-scale gravitational instability. In this case, $\Omega$ is the velocity gradient in measured in the galactic disk at the location of the GMC, the accumulation radius $R_A$ (Equation \ref{eq:accum}) is the size of the region that collapsed to form the GMC, and $\beta=0.5$. 

\begin{figure*}[htbp]\centering
\includegraphics[scale=0.5]{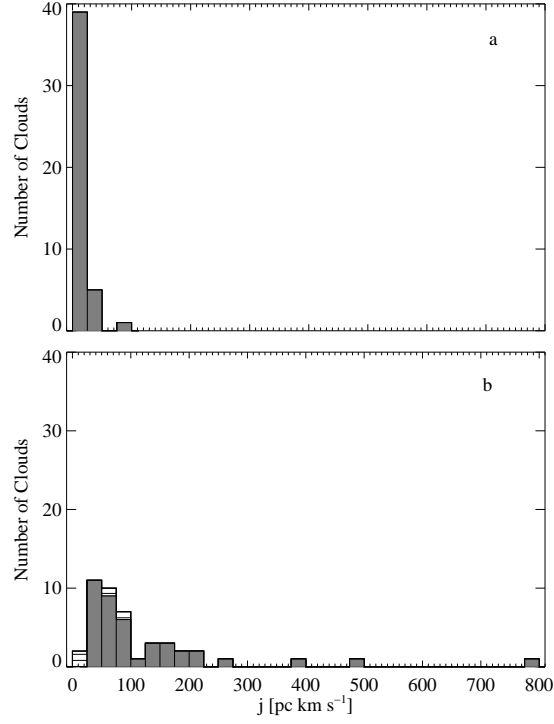}
\caption{Distribution of specific angular momentum for (a) GMCs and for (b) \HI~clouds containing GMCs.  }\label{fig:m33_hist4}
\end{figure*}

\begin{figure*}[htbp]\centering
\includegraphics[scale=.5]{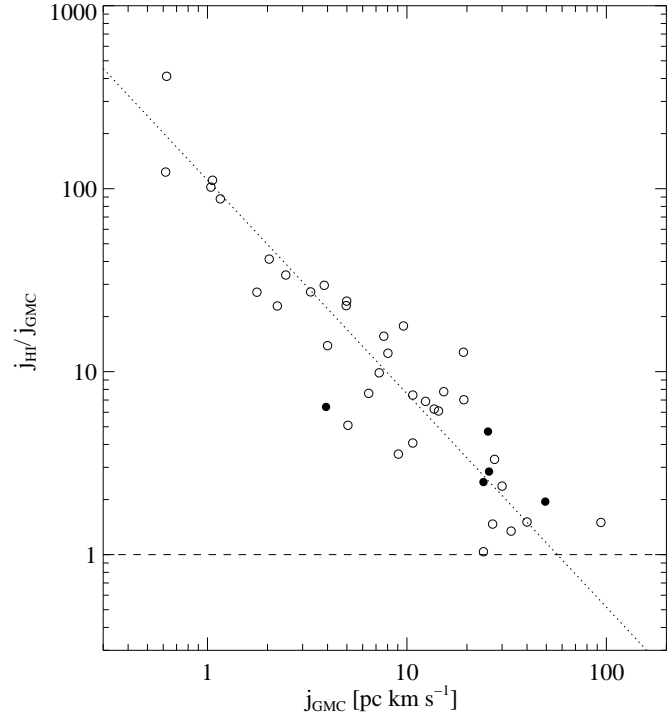}
\caption{Ratio of specific angular momenta in atomic gas and GMCs, $j_\t{HI}/j_\t{GMC}$, versus specific angular momentum in the 36 resolved GMCs.  This plot shows that $j_\t{HI}>j_\t{GMC}$ is always the case.  The dotted line shows the least-squares fit to the data: $(j_\t{HI}/j_\t{GMC})\propto j_\t{GMC}^{-1.17\pm 0.05}$.  The median  $j_\t{HI}/j_\t{GMC}$ is 13 and the average is 27.   Data points for Milky Way GMCs are overplotted in filled circles (but are not included in the fit).}\label{fig:jplot_m33}
\end{figure*}

Figure \ref{fig:m33_hist4} displays the distribution of specific angular momenta for GMCs and the predicted $j$ for the two populations of \HI~regions.  The histograms show that, if the gradients observed in the molecular clouds are due to rotation, and assuming simple top-down formation, \emph{the specific angular momentum of the molecular clouds is less than that of the atomic material from which they presumably formed}.  Rosolowsky et al. (2003) found that the angular momentum expected from simple formation theories, including the Toomre and Parker instabilities, was higher than the observed value by an average factor of 5 and by as much as an order of magnitude.  We measure an average factor of 27, a median of 13, with 11 clouds having a ratio of more than an order of magnitude.    Figure \ref{fig:jplot_m33} shows that $j_\t{HI}>j_\t{GMC}$ is always the case for the 36 resolved GMCs in the Rosolowsky et al. (2003) catalog.  The reason that we are measuring larger ratios between the expected and observed $j$ is because we have taken direct measurements of the \HI~velocity gradients, $\Omega_\t{HI}$, whereas Rosolowsky et al. used a galactic rotation curve (Corbelli \& Schneider 1997) to estimate $\Omega_\t{HI}$.  In other words, we have measured higher velocity gradients in the atomic gas surrounding GMCs than those determined from the galactic rotation curve.  
Overplotted in Figure \ref{fig:jplot_m33} are data points from our study of five molecular clouds in the Milky Way (Paper I).   The Galactic GMCs, including Perseus, Orion A, NGC 2264, MonR2, and the Rosette,  follow the trend of the M33 clouds in that the ratio of the specific angular momenta, $j_\t{HI}/j_\t{GMC}$,  is always greater than unity and decreases with increasing $j_\t{GMC}$ as $(j_\t{HI}/j_\t{GMC})\propto j_\t{GMC}^{-1.17\pm 0.05} $.  This proportionality is driven by the equality between $j_\t{GMC}$ and its inverse, as we do not observe a significant correlation between $j_\t{HI}$ and $j_\t{GMC}$.

\begin{figure*}[htbp]\centering
\includegraphics[scale=1.]{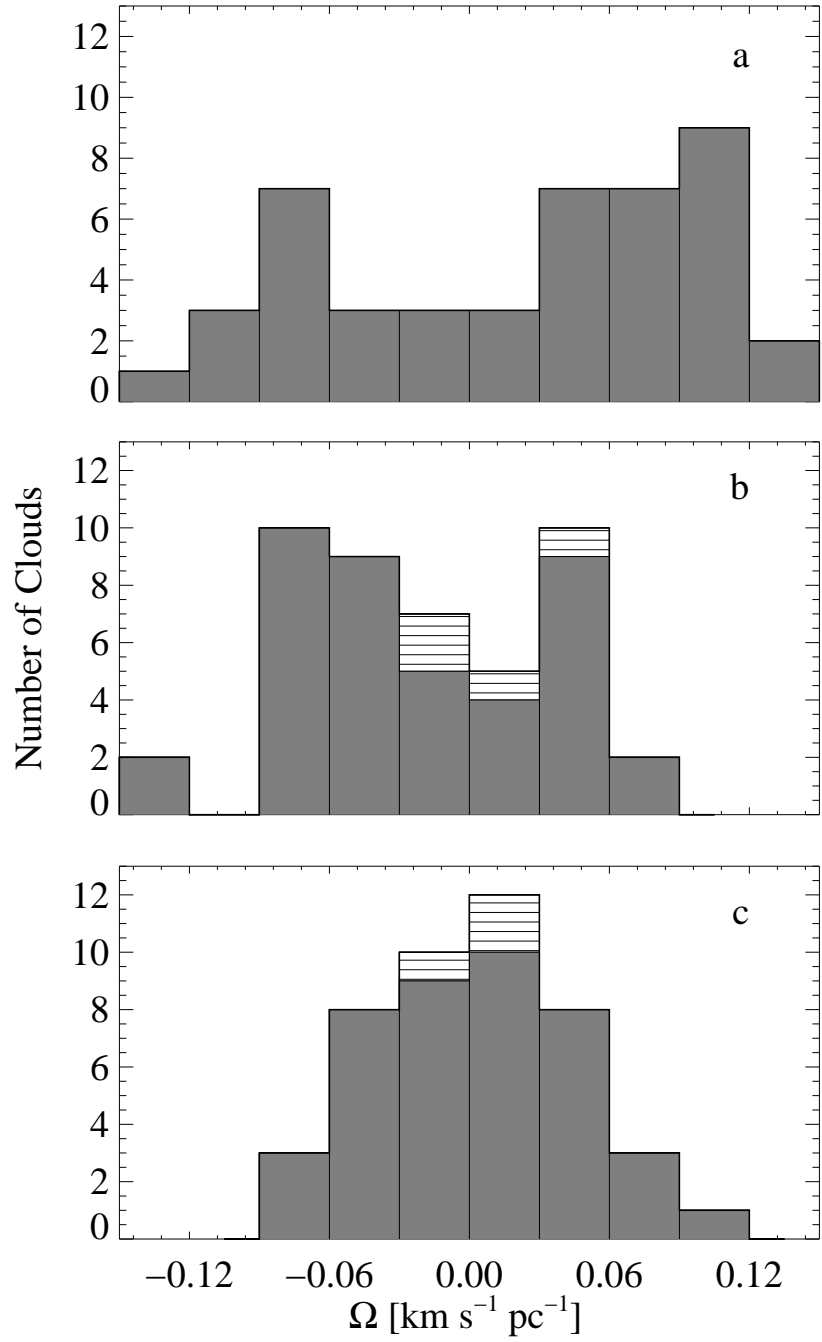}
\caption{Gradient magnitudes for (a) GMCs, (b) \HI~clouds containing GMCs, and (c) \HI~clouds without observed GMCs in M33.  Clouds having a position angle differing from the galaxy by more than $90^\circ$ are given negative values. The hatched portions of the histograms in (b) and (c) represent regions having non-linear linear gradients.}\label{fig:m33_hist1}
\end{figure*}

\begin{figure*}[htbp]\centering
\includegraphics[scale=.5]{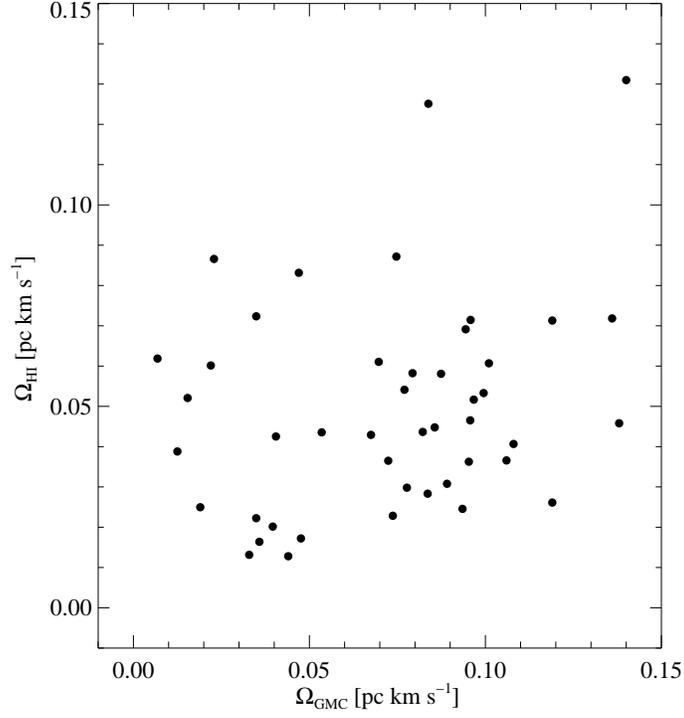}
\caption{Gradient magnitudes in the atomic gas versus GMC gradient magnitudes.  There is no significant correlation between $\Omega_\t{HI}$ and $\Omega_\t{GMC}$.  }\label{fig:vplot_m33}
\end{figure*}

If GMCs inherit their rotation from the ambient \HI~from which they condense, one might expect their gradients to be much larger than observed.  By conservation of angular momentum, in the absence of external forces, the ratio of the velocity gradient magnitudes before and after contraction is proportional to the radii of the rotating bodies: $\Omega_\t{after}/\Omega_\t{before}=(R_\t{before}/R_\t{after})^2$.  Thus, to conserve angular momentum, a cloud that contracts from an accumulation length of 70 pc to a typical GMC radius of 25 pc would have to spin up by a factor of $\sim 8$.  Figures \ref{fig:m33_hist1}(a), \ref{fig:m33_hist1}(b), and \ref{fig:vplot_m33} show, however, that the magnitudes of GMC gradients are rarely much greater than those of the \HI~with which they are associated. From Figure \ref{fig:vplot_m33} and Table 1 we see that the average value of $\Omega_\t{GMC}/\Omega_\t{HI}$ is 1.8 and reaches a maximum of 4.6.  From conservation of angular momentum, we might also expect to find molecular clouds rotating in the same sense as the surrounding ISM from which they formed. As previously shown in Figure \ref{fig:m33_hist2}, however, the  position angles in the GMCs and associated \HI~differ by at least $90^\circ$  in over half of the regions.  

These key results may imply one of the following: (1) Since GMCs formed, external torques, such as magnetic fields, may have caused the redistribution of angular momentum in the molecular clouds; (2) the \HI~with which GMCs are presently associated is unrelated to and unrepresentative of the atomic gas which originally formed the molecular clouds;  or (3) the \HI~and GMCs are to some extent associated, but the origin of the GMC velocity gradients is not rotation.  We consider each of these in turn below.

1. First, given that the interstellar media of galaxies are magnetized---the mean field strength of M33 is $\sim 6 \pm 2$ $\mu$G (Beck 2000)---magnetic fields potentially play dominant roles in the formation and evolution of GMCs.  For instance, many authors have advanced the Parker instability as a mechanism for GMC formation (e.g., Mouschovias et al. 1974; Blitz \& Shu 1980; Shibata \& Matsumoto 1991).  Parker (1966) showed that in a gravitational field, a vertically layered gas coupled to a magnetic field is unstable to long-wavelength perturbations.  The attractiveness of this magnetohydrodynamic (MHD) instability is that it has time and length scales that are comparable to the estimates of GMC lifetimes and to the distance observed between GMCs.  Thus, if MHD effects play a significant role in the formation of GMCs, presumably they might also be responsible for redistributing angular momentum during later stages of GMC evolution.  

While Rosolowsky et al. (2003) ruled out the Parker instability as the dominant mechanism for GMC formation because it over-predicts the amount of angular momentum observed in GMCs, they suggested that magnetic braking \emph{could} slow down the rotation of GMCs, since the Alfv\'{e}n speed in the ISM of M33  ($\sim 6$ \kms) is comparable to the timescales of certain instabilities they evaluated.  Yet because only 10\% of the GMCs in their catalog have values of $j$ approaching values expected from $j(\HI)$, they propose that magnetic braking must occur early on in the lifetime of GMCs or else occurs in the atomic gas surrounding GMCs.  Mouschovias \& Paleologou (1979) predicted a set of observational consequences for a rotating cloud collapsing out of an ISM threaded by a magnetic field that is initially perpendicular to the cloud's axis of symmetry.  As the cloud collapses, one expectation is that the surrounding medium in the vicinity of the cloud will be set into corotation with cloud, a phenomenon that we do not observe in our results.  Mouschovias  \& Paleologou also predicted that magnetic braking is highly efficient at slowing down rotation.  According to their simulations,  a rotating cloud with mass $10^3$ \msun~and density $10^3$ cm$^{-3}$ will have its angular momentum reduced by at least 99\% within 1 Myr, and the efficiency of magnetic braking will increase as the cloud contracts and as the density contrast between it and the surrounding medium increases. It would be worthwhile to find out what updated models predict for the rate of angular momentum loss in clouds having the masses and densities observed in GMCs.

2. If the atomic gas presently surrounding GMCs is unrelated to the past formation history of GMCs, the comparisons we are making may be invalid. In order to resolve this issue we need some sort of control field with which to compare the \HI~currently harboring GMCs. It is for precisely this reason that we performed the analysis, outlined in \S \ref{subsec:non-gmc}, on ``non-GMC'' \HI~regions.  We found that the two populations of \HI~have distinctly different distributions of gradient magnitudes and position angles.  The atomic gas associated with GMCs has a typical $\Omega$ of $0.05$ \vunits~and the non-GMC \HI~has a typical value of $0.03$ \vunits.  Also, the median difference between the position angle of the \HI~and that of the galaxy is $-90^\circ$ for the former population and $-60^\circ$ for the latter.  These differences in $\Omega$ and $\phi$ between the two groups indicate that something unique has occurred in atomic gas associated with GMCs---presumably, something that has either \emph{caused} the formation of the GMCs or has \emph{resulted from} their formation.

We have already seen that the mean gradient magnitude in atomic gas associated with GMCs is greater than the typical gradient in non-GMC \HI~(\S \ref{subsec:gradients}).  This implies that \HI~containing GMCs has \emph{higher} angular momentum (and/or shear) than non-GMC \HI.  We can evaluate how significant the difference is between the two populations by investigating the parameter $\beta_\t{rot}$, the ratio of a cloud's rotational energy to its self-gravitational energy (Goodman et al. 1993):

\begin{equation} 
\beta_\t{rot}\equiv\frac{1}{3}\frac{\Omega^2 R^2}{GM/R}. \label{eq:beta}
\end{equation}

\begin{figure*}[htbp]\centering
\includegraphics[scale=.5]{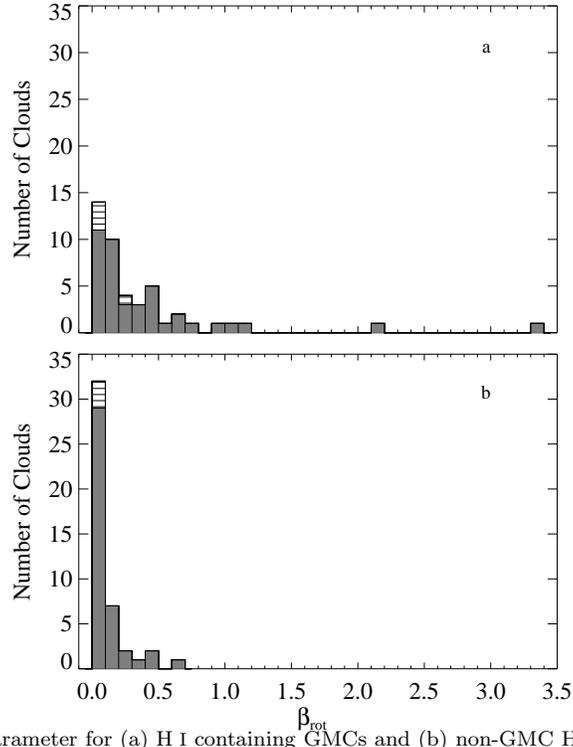}
\vspace{-1.0em}
\caption{Distribution of the $\beta_\t{rot}$ parameter for (a) \HI~containing GMCs and (b) non-GMC \HI.  }\label{fig:bplot_m33}
\end{figure*}

Figure \ref{fig:bplot_m33} shows that non-GMC \HI~has a narrower distribution of $\beta_\t{rot}$ and a lower mean, $\beta_\t{rot}= 0.10\pm 0.02$, than does atomic gas containing GMCs.   Thirty-two of 45 of the non-GMC \HI~regions have $\beta_\t{rot}<0.1$, whereas only 14 of the \HI~regions  containing GMCs  have values of $\beta_\t{rot}<0.1$.   Note that for \HI~regions containing GMCs, we calculate $\beta_\t{rot}$ excluding the gravitational potential energy due to the GMCs within them (Figure \ref{fig:bplot_m33} [a]).  This group has an average $\beta_\t{rot}$ of $0.39\pm 0.09$, with 4 clouds having $\beta_\t{rot}>1$ (Clouds 1, 2, 17, and 43), implying that these regions may have a considerable amount of rotational energy. Note that for \HI~regions containing GMCs, we calculate $\beta_\t{rot}$ excluding the gravitational potential energy due to the GMC within it (Figure \ref{fig:bplot_m33} [a]).  If we calculate $\beta_\t{rot}$ including \emph{all} of the gas within a given region, the average for \HI~containing GMCs goes down to $0.24\pm 0.04$.  The regions listed above still appear to have a considerable amount of rotational energy in comparison to their gravitational energy: $\beta_\t{rot}=0.90$, 0.75, 0.88, and 1.1 for clouds 1, 2, 17, and 43, respectively.   Clouds having large values of $\beta_\t{rot}$ can potentially become stable against gravitational instabilities. It is difficult to see how GMCs could have formed under such conditions.  These observations suggest that between the time prior to the onset of GMC formation to the time after formation, processes occur that increase $\Omega$ observed in the \HI~associated with GMCs.  An alternative is that GMCs preferentially form from gas having high $\Omega$, possibly in regions that are unstable to gravitational collapse.

It is also worth pointing out that \HI~regions falling into Classes 2 and 3 (see \S \ref{subsec:hi})---regions  where the GMCs are not quite spatially coincident with local \HI~maxima---tend to have lower gradient magnitudes ($\Omega \sim 0.04$ \vunits~with a dispersion of 40\%) than regions in Class 1 (0.06 \vunits~with a dispersion of 50\%).  We apply a t-test and find that the difference between the two classes is significant to the 95\% confidence level.    However, molecular clouds belonging to either class do not appear to have different distributions of $\Omega$; that is GMCs in both Class 1 and Class 2 have the same average gradient ($\sim 0.07$ \vunits).  This suggests that activity within the GMCs may have more of an impact on the surroundings than the surrounding environment has on the velocity fields of the GMCs.

\begin{figure*}[htbp]\centering
\includegraphics[scale=.4]{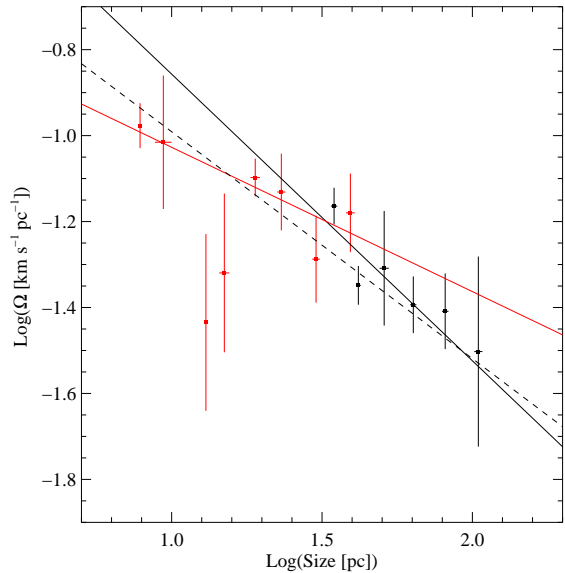}
\caption{Gradient magnitudes observed in GMCs (red) and associated atomic gas (black) as a function of size.  The data are averaged in bin sizes of $\Delta R=0.1$ dex.  The lines indicate least-squares, power-law fits to the data.  For GMCs, $\Omega_\t{GMC}\propto R^{-0.3\pm 0.2}$, and for \HI,$\Omega_\t{HI}\propto R_A^{-0.7\pm 0.2}$, where $R_A$ is the accumulation radius.  For GMCs and \HI~combined is $ \Omega\propto R^{-0.5\pm 0.1}$, which is the relationship found by Burkert \& Bodenheimer (2000) for turbulent molecular cores.   }\label{fig:r-vgrad}
\end{figure*}

3. Given that the gradient directions of the GMCs and associated atomic gas are generally not aligned, this raises the possibility that the linear velocity gradients observed in molecular clouds may not be caused by rotation.    Burkert \& Bodenheimer (2000) demonstrated that turbulent velocity fields can also cause linear gradients.  They found that the gradient magnitude of turbulent cores scales with size as $\Omega\propto R^{-0.5}$.  As shown in Figure \ref{fig:r-vgrad}, fitting a power-law relationship to the gradients observed in the \HI~surrounding GMCs as a function of the accumulation radius, we find $\Omega_\t{HI}\propto R_A^{-0.7\pm 0.2}$.   The relationship between $\Omega_\t{GMC}$ and size for the 36 resolved GMCs is  $\Omega_\t{GMC}\propto R^{-0.3\pm 0.2}$.   Figure \ref{fig:r-vgrad} also displays a combined fit for atomic gas and GMCs, which is in good agreement with the Burkert \& Bodenheimer (2000) result. 

In our previous study (Paper I), we found that star formation activity provides a reasonable explanation of the gradients observed in a small sample of Milky Way molecular clouds.  In at least three out of five cases, including Orion A, NGC 2264, and the Rosette, the velocity fields may have been produced by expansion driven by highly energetic stellar winds.  The location of \HII~regions corresponds to the highest velocities of the molecular material in these clouds.  Though the sample was too small to make a firm conclusion, it appears that the \HII~regions within these molecular clouds may also be influencing the morphology of the associated atomic gas.   In NGC 2264 and the Rosette, for instance, there are local peaks of \HI~that appear to have been swept up by expanding molecular gas.  Following up this study of M33 with an analysis comparing the locations of \HII~regions to the kinematic properties of GMCs could provide useful insight to how the velocity fields of the latter evolve.

\section{Summary}\label{sec:summary}
We have presented a detailed comparison of the velocity fields in 45 GMCs detected in M33 by Rosolowsky et al. (2003) using BIMA observations and the atomic hydrogen with which they are associated.  Using high-resolution VLA 21-cm observations to create surface density and intensity-weighted first-moment maps of the \HI, we also compared the properties of atomic gas containing molecular clouds with atomic gas in which molecular clouds have not been detected.  Based on our measurements, including the velocity gradient magnitudes and directions of these regions, we make the following conclusions:

1. The average surface density of atomic hydrogen associated with the GMCs is $\sim 10$ \sunits, similar to the saturation level of \HI~above which gas becomes primarily molecular in other galaxies, including the Milky Way (Martin \& Kennicutt 2001; Wong \& Blitz 2002; Bigiel et al. 2008).  A power-law relationship exists between the \HI~surface density and GMC mass: $\Sigma_\t{HI} \propto M_\t{GMC}^{0.27 \pm 0.06}$.

2. We observe three categories of GMCs, based on their proximity to local peaks in the atomic gas.  The majority of GMCs (64\%) coincide spatially and kinematically with local \HI~peaks.  Twenty-nine percent of GMCs are located near the edge of an \HI~peak, or sit between two peaks.  Clouds in this category tend to have their mean LSR velocities offset from the mean \HI~velocity by a few \kms.  The remaining three clouds (7\%) are not associated with \HI~maxima and the mean velocities of the molecular cloud and atomic gas are offset by $\sim 10$ \kms.

3. Thirty-nine out of 45 of the \HI~regions in the vicinity of GMCs have linear velocity gradients of around $0.050$ \vunits~and spanning $0.013$ to $0.13$ \vunits.  If GMCs are rotating and initially inherited their angular velocity from the surrounding \HI, conservation of angular momentum would require that the angular speed of a typical GMC ($R\approx 25$ pc) speed up by a factor of 8.   The average value of $\Omega_\t{GMC}/\Omega_\t{HI}$, however, is only 1.8.  Magnetic braking has been used by some authors to explain how the slowing of GMC rotation may have occurred.  But if magnetic braking is as efficient at slowing rotation as predicted (e.g., Mouschovias \& Paleologou 1979), we would expect most GMCs to have much \emph{lower} values of $\Omega_\t{GMC}$---as much as an order of magnitude less---than observed.  Alternatively, magnetic braking may be less efficient, or it may operate over longer timescales.

4. Fifty-three percent of the molecular clouds have gradients whose directions differ from the gradient direction in the local \HI~by more than $90^\circ$.  If the gradients in the GMCs were caused by rotation, this implies that over half of them are counter-rotating with respect to the atomic gas with which they are surrounded.  This measurement is difficult to reconcile with the notion that gradients are due to rotation and that GMCs form in a simple top-down matter.

5. Gradients in the atomic gas associated with GMCs generally have larger magnitudes than expected from galactic differential rotation alone.  Also, 62\% of these regions have gradient position angles that differ from the sense of galactic rotation by more than $90^\circ$.

6. We examined the properties of high-density \HI~regions in which molecular clouds have not been detected and found that they have a lower range of velocity gradients, $\sim 0.03$ \vunits, than regions where GMCs are observed.  This suggests that something occurs during the course of GMC evolution that may increase the shear of the atomic gas.  Neither population of atomic gas has gradient directions that are preferentially aligned with the kinematic position angle of the galaxy, nor did we find a correlation between gradient magnitude and direction in either population.

7. A power-law relationship exists between gradient magnitude and size in both the molecular clouds and the \HI~surrounding them.  For GMCs, $\Omega_\t{GMC}\propto R^{-0.3\pm 0.2}$; for \HI, $\Omega_\t{HI}\propto R_A^{-0.7\pm 0.2}$, where $R_A$ is the accumulation length.  The combined relationship is $\Omega \propto R^{-0.5\pm 0.1}$, consistent with what Burkert \& Bodenheimer (2000) found for the velocity fields of turbulent molecular cores.

8. Our analysis raises considerable doubt to the hypothesis that the origin of GMC velocity gradients is rotation.  Alternative explanations worth exploring include turbulence, shear, and star formation activity.

\clearpage

\vspace{-7.0em}
\begin{table*}[htbp]\scriptsize\centering
\scalebox{1.}{
\begin{tabular}{lccccccccccc}
\multicolumn{11}{c}{\textbf{TABLE 1}} \\
\multicolumn{11}{c}{\textbf{Cloud Properties in M33}} \\
\hline \hline 
Cloud &  R.A.  & Dec.   &   $\Omega_\t{GMC}$ & $\Omega_\t{HI}$  & $\theta_\t{HI}$ & $\theta_\t{HI}-\theta_\t{GMC}$ & Extent & $M_\t{HI}$ &  $M_\t{GMC}$ &  $|m|/\sigma_m$ \\
\cline{2-3} 
  & \multicolumn{2}{c}{[J200]} & \multicolumn{2}{c}{[$0.01~\vunits$]} & [deg]     &  [deg]      & [pc] & [$10^5 ~\msun$] &  [$10^4 ~\msun$] & \\\hline
 
   1 & 1 33 52 & 30 39.3 &  8.38 & 12.51 & -175 & 6 & 140 &      1.8 & 26  &  11.20  \\ 
   2 & 1 33 53 & 30 39.0 &  2.29 & 8.66 & -178 & -114 & 140 &      1.9 & 6.7  &   5.08  \\
   3 & 1 33 44 & 30 38.9 &  9.58 & 7.14 & -159 & -77 & 160 &      2.8 & 3.6  & 10.51  \\
   4 & 1 33 56 & 30 41.3 & 10.10 & 6.07 & -134 & -90 & 120 &      2.1 & 4.2  &  6.61  \\
   5 & 1 33 55 & 30 41.6 & 13.60 & 7.18 & -166 & -106 & 160 &      2.3 & 14  &   9.33  \\
   6 & 1 33 57 & 30 41.1 & 11.90 & 7.13 & -151 & -170 & 180 &      2.0 & 2.9 &  10.80  \\
   7 & 1 33 50 & 30 37.5 &  8.36 & 2.83 & -133 & 83 & 170 &      1.1 &  9.9  & 3.45  \\
   8 & 1 33 52 & 30 37.7 &  3.49 & 7.24 & 177 & -149 & 140 &      1.4 & 2.9  &  5.63  \\
   9 & 1 33 52 & 30 37.5 &  2.20 & 6.02 & 168 & 177 & 130 &      1.5 &  5.3  & 5.38  \\
  10 & 1 33 59 & 30 41.5 &  1.90 & 2.50 & 108 & 124 & 160 &      1.7 &  12  & 3.06  \\
  11 & 1 34 00 & 30 40.8 &  9.53 & 3.63 & 170 & 146 & 120 &      3.0 &  41  & 3.55  \\
  12 & 1 33 54 & 30 37.7 & 10.60 & [3.66] & [-103] & [-135] & [\ldots] &   1.4 & 6.7  &  1.50  \\
  13 & 1 33 59 & 30 41.8 &  3.29 & [1.31] & [121] & [167] & [\ldots] &      1.0 & 11 &   0.92  \\
  14 & 1 33 52 & 30 37.0 &  7.70 & 5.41 & -108 & -17 & 170 &      1.7 &3.7   &  5.91  \\
  15 & 1 33 40 & 30 39.2 &  9.57 & 4.65 & -151 & -84 & 120 &      2.4 & 45  &  5.60  \\
  16 & 1 33 40 & 30 38.7 &  1.25 & [3.88] & [-93] & [147] & [\ldots] &  1.5 & 7.0  &     2.57  \\
  17 & 1 34 02 & 30 38.6 & 14.00 & 13.10 & -171 & -167 & 160 &      3.8 &12   & 19.09  \\
  18 & 1 34 10 & 30 42.0 &  4.70 & 8.31 & 149 & -34 & 120 &      2.3 & 3.4  & 11.37  \\
  19 & 1 33 50 & 30 33.9 &  0.69 & 6.19 & -169 & 138 & 160 &      2.9 &  23 & 16.25  \\
  20 & 1 34 08 & 30 39.2 &  8.74 & 5.81 & -157 & -140 & 170 &      2.5 & 15 &  17.93  \\
  21 & 1 33 43 & 30 33.2 &  4.40 & [1.28] & [-96] & [126] & [\ldots] &  3.1 & 12  &  0.46  \\
  22 & 1 34 13 & 30 42.0 &  6.75 & 4.29 & 116 & 23 & 140 &      2.3 & 5.5 &   4.59  \\
  23 & 1 34 13 & 30 39.1 &  7.77 & 2.98 & -63 & 59 & 180 &      3.1 & 6.2 &   9.78  \\
  24 & 1 34 07 & 30 47.8 &  4.05 & 4.25 & -128 & -123 & 180 &      3.7 &  6.7&  11.49  \\
  25 & 1 34 06 & 30 47.9 &  7.24 & 3.65 & -137 & -28 & 110 &      3.6 &  15 &  5.84  \\
  26 & 1 33 40 & 30 45.6 & 11.90 & 2.61 & 150 & -69 & 60 &      3.2 &  8.5 &  3.52  \\
  27 & 1 33 40 & 30 45.9 &  1.54 & 5.21 & 135 & 23 & 130 &      3.8 & 5.5  &  6.09  \\
  28 & 1 34 10 & 30 36.3 &  3.49 & 2.22 & -57 & -33 & 150 &      3.1 & 27 &   4.54  \\
  29 & 1 34 16 & 30 39.3 &  8.91 & 3.08 & 146 & -43 & 180 &      4.2 & 21 &   8.28  \\
  30 & 1 33 40 & 30 46.2 &  9.44 & 6.91 & 121 & -81 & 190 &      3.7 & 16 &  21.49  \\
  31 & 1 33 58 & 30 48.7 &  3.58 & [1.64] & [67] & [153] & [\ldots] &      2.7 & 32  &  2.17  \\
  32 & 1 34 01 & 30 48.9 &  4.76 & [1.72] & [174] & [111] & [\ldots] &      2.1 & 5.8  &  0.94  \\
  33 & 1 34 11 & 30 48.4 &  5.35 & 4.35 & -10 & 44 & 100 &      4.2 & 8.0   &  7.42  \\
  34 & 1 34 07 & 30 49.0 & 13.80 & 4.58 & 43 & -25 & 110 &      2.9 &  7.5 &  3.80  \\
  35 & 1 33 57 & 30 49.0 &  9.35 & 2.45 & 128 & -96 & 170 &      2.7 & 6.2  &  8.03  \\
  36 & 1 33 59 & 30 49.3 &  3.96 & 2.02 & 156 & -149 & 110 &      2.7 & 24 &   6.89  \\
  37 & 1 34 09 & 30 49.1 &  7.37 & 2.28 & -6 & -30 & 130 &      4.9 &  78 &  3.75  \\
  38 & 1 34 07 & 30 50.0 &  9.67 & 5.17 & -111 & 132 & 160 &      1.7 & 7.1 &   5.02  \\
  39 & 1 34 13 & 30 33.7 &  7.47 & 8.72 & -165 & -55 & 150 &      4.8 & 32 &   14.86  \\
  40 & 1 34 33 & 30 46.5 &  7.93 & 5.82 & 150 & -65 & 140 &      4.8 & 9.6 &   6.94  \\
  41 & 1 34 33 & 30 46.8 &  6.97 & 6.11 & 90 & -84 & 120 &      4.8 & 28  &  5.87  \\
  42 & 1 34 34 & 30 46.3 &  8.22 & 4.37 & 100 & -129 & 90 &      4.1 & 30 &   4.67  \\
  43 & 1 33 22 & 30 25.9 &  9.95 & 5.33 & -109 & -5 & 150 &      0.2 & 4.6 &   5.87  \\
  44 & 1 34 38 & 30 40.4 & 10.80 & 4.07 & 115 & 145 & 170 &      4.2 & 6.7 &   5.91  \\
  45 & 1 34 39 & 30 40.7 &  8.56 & 4.48 & 133 & 101 & 140 &      5.0 & 25  &  7.09  \\

\hline
\end{tabular}
}
\caption*{The cloud numbers correspond to those in the Rosolowsky et al. (2003) catalog. The gradient magnitudes and directions are $\Omega$ and $\theta$, respectively.  The difference between the \HI~and GMC gradient directions is $\theta_\t{HI}-\theta_\t{GMC}$. The extent of the gradient is listed for \HI~regions having linear gradients.  The gradient measurements of clouds in which linear gradients have not been detected ($|m|/\sigma_m<3$) are listed in brackets. }
\end{table*}

\begin{table*}[htbp]\footnotesize \centering
\scalebox{1.}{
\begin{tabular}{lcccccccc}
\multicolumn{8}{c}{\textbf{TABLE 2}} \\
\multicolumn{8}{c}{\textbf{Properties of Non-GMC \HI~in M33}} \\
\hline \hline 
Cloud &  R.A.  & Dec.   & $\Omega_\t{HI}$ & $\theta_\t{HI}$ &  $M_\t{HI}$ & $v_0$ & $m/\sigma_m$ \\
\cline{2-3} 
      & \multicolumn{2}{c}{[J2000]} &[$0.01~\vunits$] & [deg]          & [$10^5 ~\msun$] & [\kms]&   \\\hline

 1A & 1 35 05 & 30 45.1 & [0.62] & [-78] &      3.5 &     -211 &  3.98  \\
   2A & 1 35 06 & 30 50.2 & 1.76 & -172 &      4.6 &     -231 & 13.82  \\
   3A & 1 35 05 & 30 51.9 & 3.30 & -96 &      4.0 &     -233 & 16.50  \\
   4A & 1 34 52 & 30 54.1 & 3.21 & -173 &      4.1 &     -246 & 27.62  \\
   5A & 1 34 39 & 30 55.9 & 2.41 & -148 &      4.0 &     -260 & 26.35  \\
   6A & 1 34 31 & 30 57.2 & 2.32 & 144 &      3.7 &     -267 & 17.39  \\
   7A & 1 34 24 & 30 56.8 & 1.25 & -147 &      3.8 &     -267 & 20.41  \\
   8A & 1 34 16 & 30 52.2 & 1.11 & 84 &      3.1 &     -263 &  8.01  \\
   9A & 1 34 04 & 30 54.7 & 3.14 & -82 &      3.3 &     -258 & 22.09  \\
  10A & 1 33 16 & 30 31.4 & 0.67 & 168 &      4.0 &     -119 &  8.54  \\
  11A & 1 33 17 & 30 33.5 & 9.54 & -127 &      3.4 &     -133 & 41.02  \\
  12A & 1 32 58 & 30 31.0 & 3.67 & -178 &      5.2 &     -128 & 22.84  \\
  13A & 1 32 51 & 30 31.6 & 2.46 & 147 &      4.2 &     -135 & 21.41  \\
  14A & 1 32 35 & 30 30.6 & 2.01 & 133 &      5.0 &     -126 & 25.62  \\
  15A & 1 32 57 & 30 27.6 & 2.61 & -151 &      3.1 &     -116 & 31.86  \\
  16A & 1 32 55 & 30 26.1 & 0.91 & -43 &      2.6 &     -109 & 14.65  \\
  17A & 1 32 49 & 30 28.6 & 2.08 & -148 &      3.2 &     -118 & 19.45  \\
  18A & 1 32 29 & 30 27.3 & 1.21 & 120 &      3.8 &     -115 &  6.07  \\
  19A & 1 32 50 & 30 33.2 & 1.70 & 171 &      4.8 &     -140 & 13.11  \\
  20A & 1 32 31 & 30 35.4 & 0.83 & 31 &      6.1 &     -140 &  8.85  \\
  21A & 1 32 44 & 30 35.1 & 3.49 & 140 &      5.3 &     -147 & 29.62  \\
  22A & 1 32 41 & 30 36.0 & 3.06 & -93 &      4.9 &     -150 & 12.16  \\
  23A & 1 32 50 & 30 35.2 & 7.96 & -170 &      4.3 &     -147 & 61.35  \\
  24A & 1 33 50 & 30 44.4 & 4.44 & -178 &      4.0 &     -219 & 24.53  \\
  25A & 1 33 11 & 30 48.4 & 6.11 & -68 &      5.9 &     -201 & 49.19  \\
  26A & 1 33 12 & 30 49.6 & 5.22 & -113 &      6.2 &     -202 & 39.74  \\
  27A & 1 33 13 & 30 50.7 & 3.19 & 78 &      5.4 &     -206 & 22.55  \\
  28A & 1 33 19 & 30 51.8 & 2.82 & -146 &      3.9 &     -216 & 18.78  \\
  29A & 1 33 14 & 30 53.2 & 1.79 & 0 &      5.4 &     -213 &  6.82  \\
  30A & 1 33 17 & 30 55.2 & [0.82] & [-101] &      3.9 &     -212 &  2.24  \\
  31A & 1 33 17 & 30 56.4 & [0.35] & [55] &      5.1 &     -214 &  1.65  \\
  32A & 1 33 26 & 30 48.4 & 5.07 & -117 &      4.0 &     -206 & 45.15  \\
  33A & 1 33 36 & 30 49.1 & 6.26 & -169 &      3.5 &     -224 & 67.19  \\
  34A & 1 34 40 & 30 30.4 & 3.58 & -84 &      3.6 &     -162 & 23.48  \\
  35A & 1 34 14 & 30 25.8 & 6.03 & -176 &      3.8 &     -126 & 34.44  \\
  36A & 1 33 46 & 30 20.9 & 1.58 & -1 &      3.5 &     -109 & 15.13  \\
  37A & 1 33 01 & 30 24.0 & 5.15 & 8 &      3.6 &     -106 & 30.46  \\
  38A & 1 32 48 & 30 23.6 & 2.67 & 9 &      3.4 &     -104 & 14.41  \\
  39A & 1 32 40 & 30 25.6 & 1.26 & -127 &      3.0 &     -115 & 15.00  \\
  40A & 1 34 34 & 30 44.7 & 8.61 & -144 &      4.1 &     -220 & 81.84  \\
  41A & 1 34 36 & 30 38.3 & 4.25 & -122 &      4.3 &     -189 & 51.84  \\
  42A & 1 34 25 & 30 39.3 & 3.42 & 139 &      2.3 &     -202 & 25.70  \\
  43A & 1 34 08 & 30 31.7 & 6.90 & -111 &      2.5 &     -148 & 43.44  \\
  44A & 1 34 14 & 30 46.6 & 3.56 & 70 &      4.0 &     -243 & 46.75  \\
  45A & 1 33 13 & 30 44.9 & 3.55 & 179 &      4.1 &     -187 & 19.76  \\
\hline
\end{tabular}
}
\caption*{Properties of \HI~regions in which GMCs have not been detected.}
\end{table*}

\clearpage

\appendix
The following figures display the velocity maps, position-velocity plots, surface density maps, and spectra of the \HI~regions associated with the remaining molecular clouds not displayed in our paper on giant molecular cloud (GMC) angular momentum in M33 (Imara, Bigiel \& Blitz 2010).  If for a given region the GMC was resolved in the Rosolosky et al. (2003) catalog, a circle matching to the size of the GMC is overlaid on the surface density map.

\vspace{2cm}
\begin{figure*}[htbp]
\includegraphics[scale=0.75]{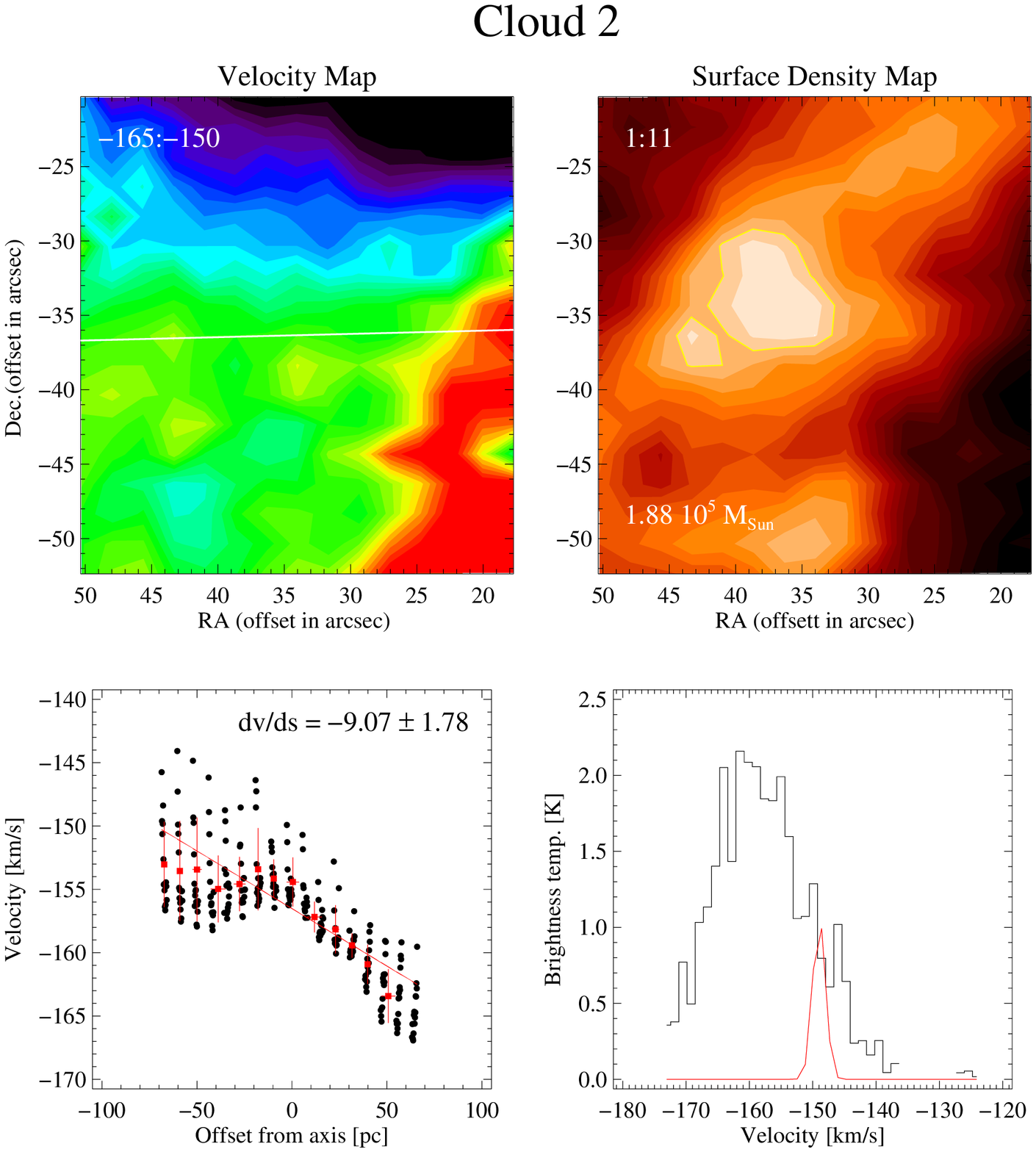}
\caption{The top left figure shows the intensity-weighted first moment map of the \HI~with the gradient axis overlaid.  The velocity range of the map is indicated in the top left corner in units of \kms; red represents the maximum speed.  Below this figure is a plot of the central velocity at a given location in the first-moment map versus perpendicular offset from the gradient axis; the linearity of the plot indicates that a plane is a good fit to the first-moment map; the radial extent of the GMC is demarcated by the horizontal blue line. The top right figure is a surface density map of the \HI~overlaid with the 10 $\sunits$ contour in yellow.  The range of \HI~surface densities displayed in the map are in the top left corner in units of \sunits, and the total \HI~mass in the region is written in the bottom left corner.  Below is a plot of the average spectra of \H I~emission (black) and CO emission (red) toward the region.}
\label{fig:f22}
\end{figure*}

\newpage
\vspace*{2cm}
\begin{figure*}[htbp]
\includegraphics[scale=0.8]{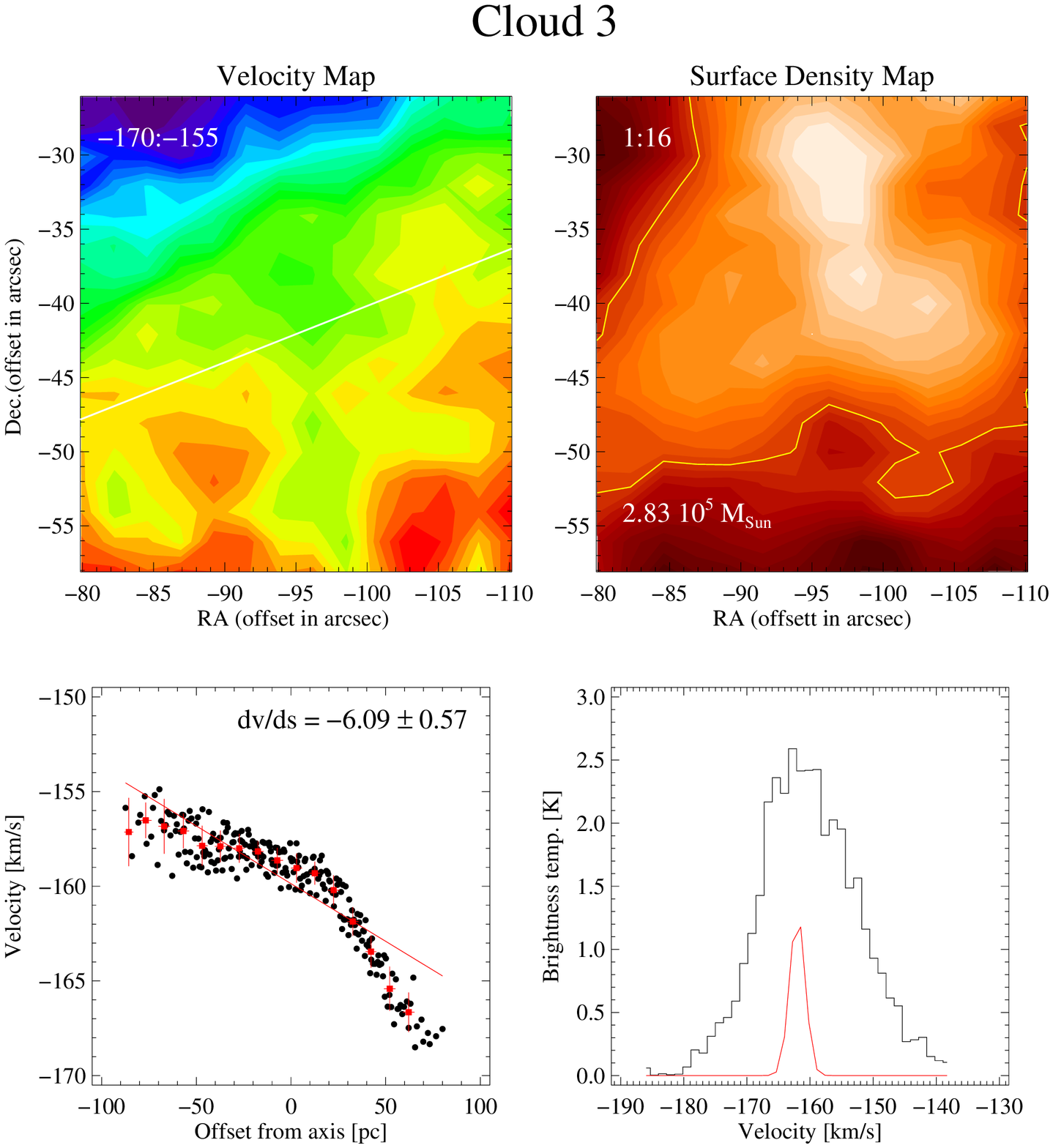}
\caption{See Figure \ref{fig:f22}.}
\end{figure*}

\newpage
\vspace*{2cm}
\begin{figure*}[htbp]
\includegraphics[scale=0.8]{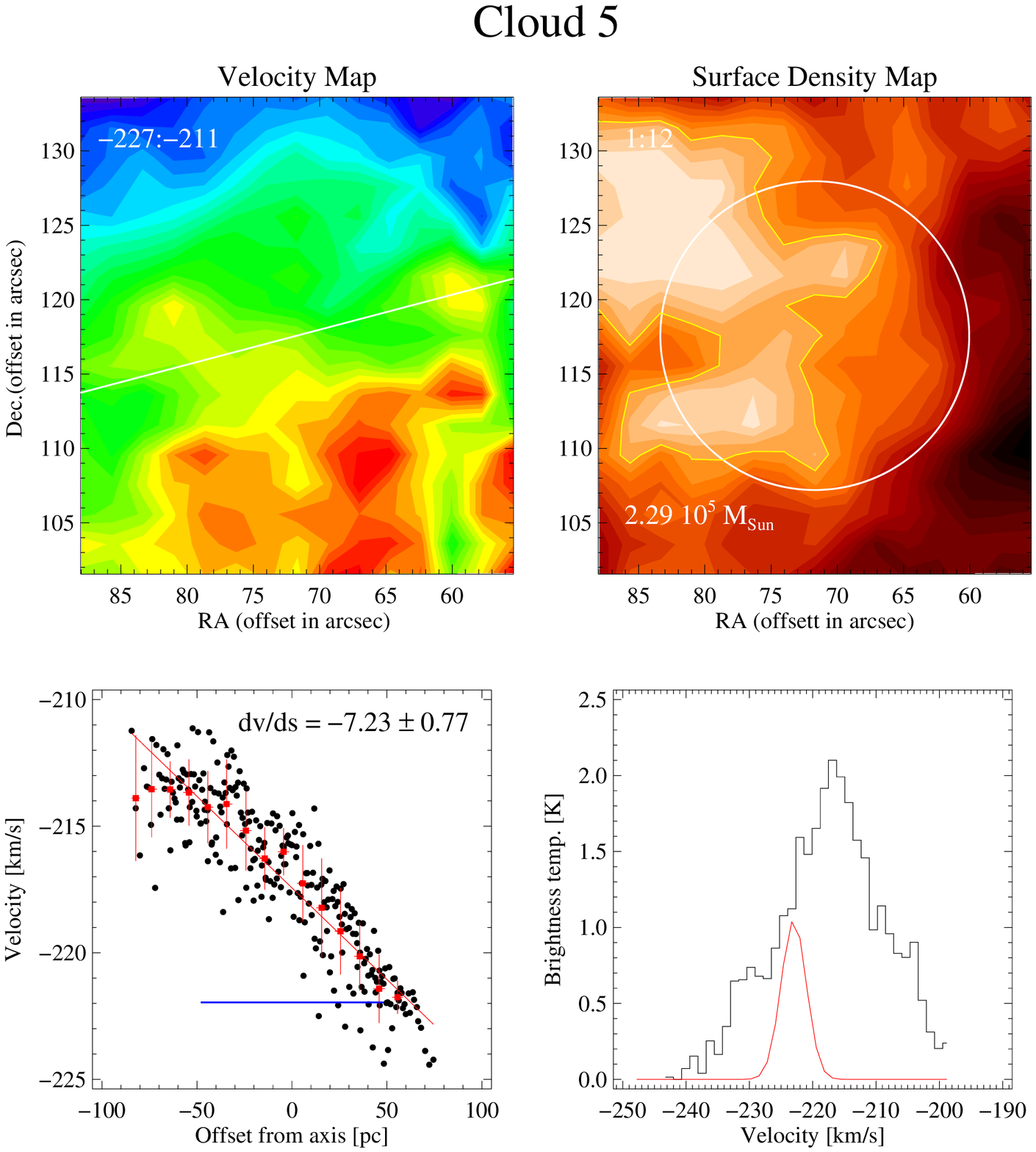}
\caption{See Figure \ref{fig:f22}.}
\end{figure*}

\newpage
\vspace*{2cm}
\begin{figure*}[htbp]
\includegraphics[scale=0.8]{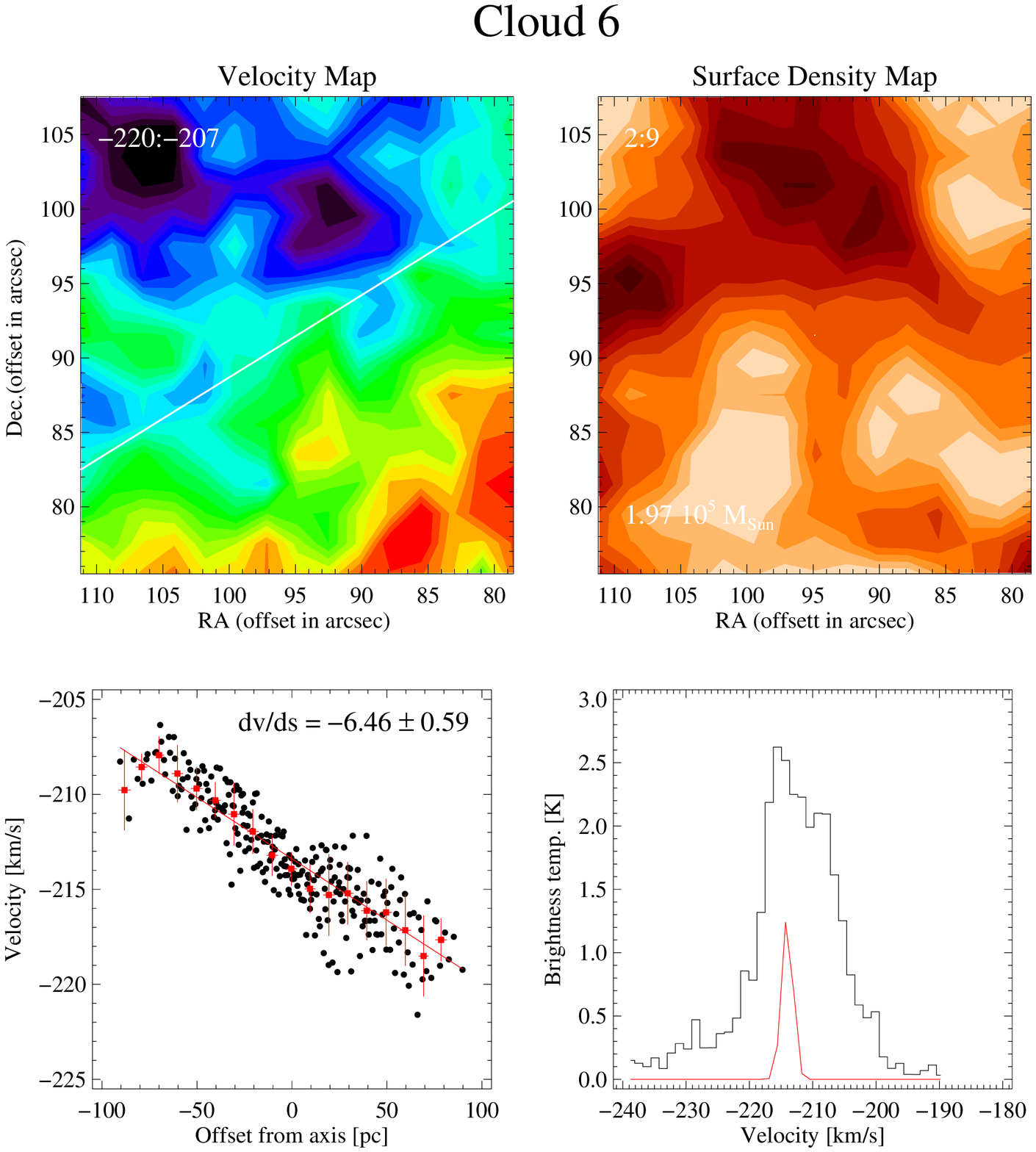}
\caption{See Figure \ref{fig:f22}.}
\end{figure*}

\newpage
\vspace*{2cm}
\begin{figure*}[htbp]
\includegraphics[scale=0.8]{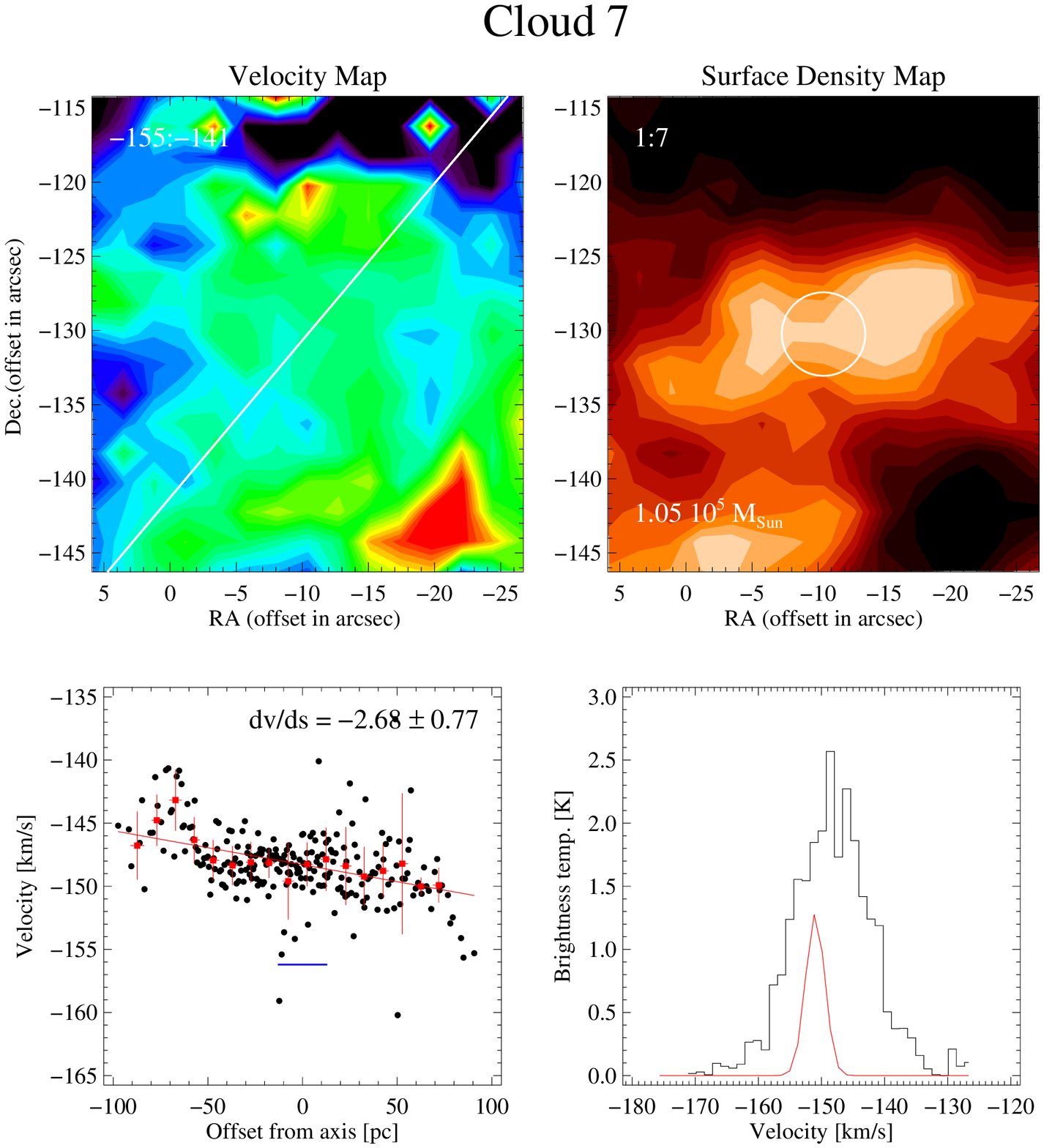}
\caption{See Figure \ref{fig:f22}.}
\end{figure*}

\newpage
\vspace*{2cm}
\begin{figure*}[htbp]
\includegraphics[scale=0.8]{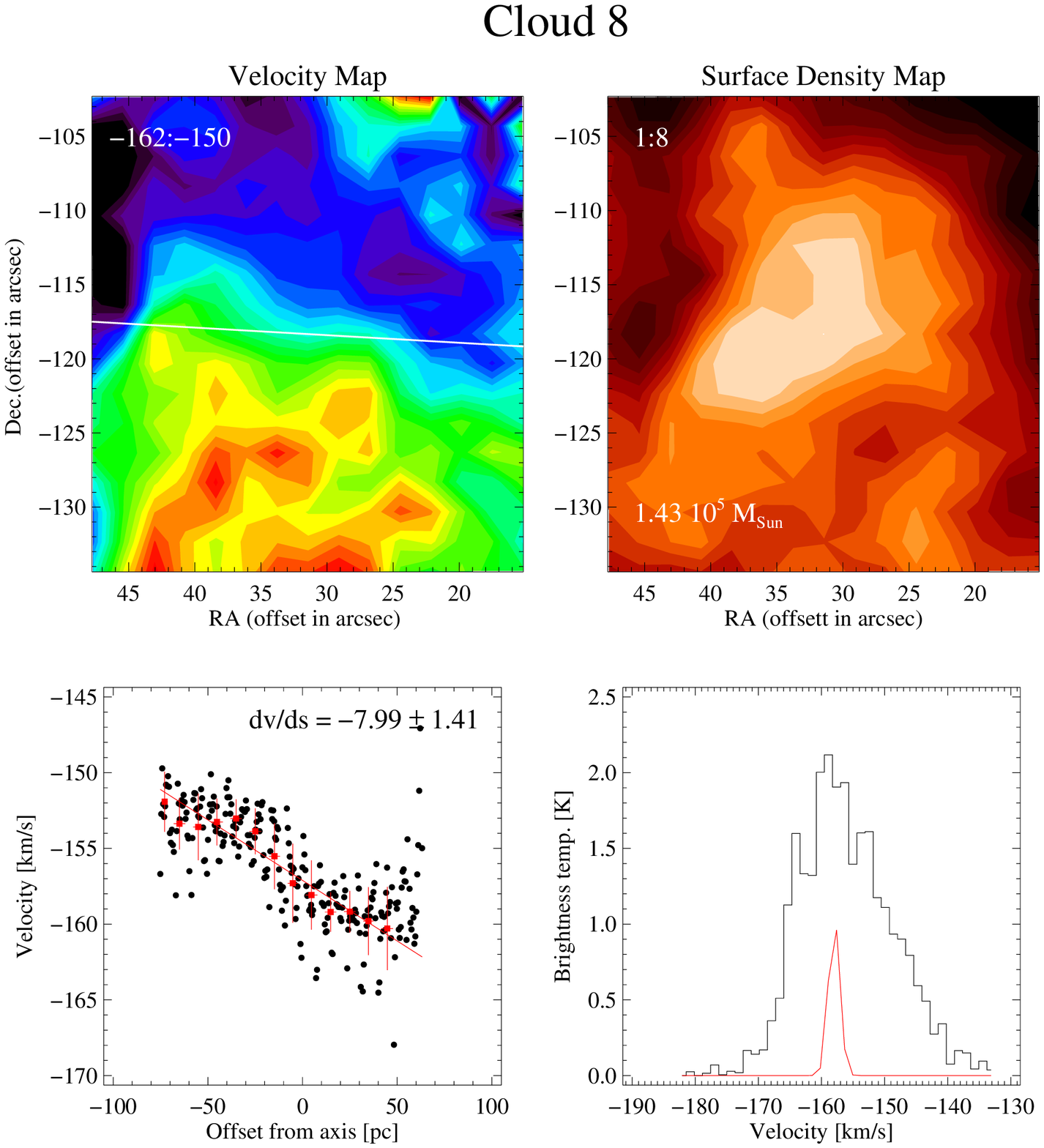}
\caption{See Figure \ref{fig:f22}.}
\end{figure*}

\newpage
\vspace*{2cm}
\begin{figure*}[htbp]
\includegraphics[scale=0.8]{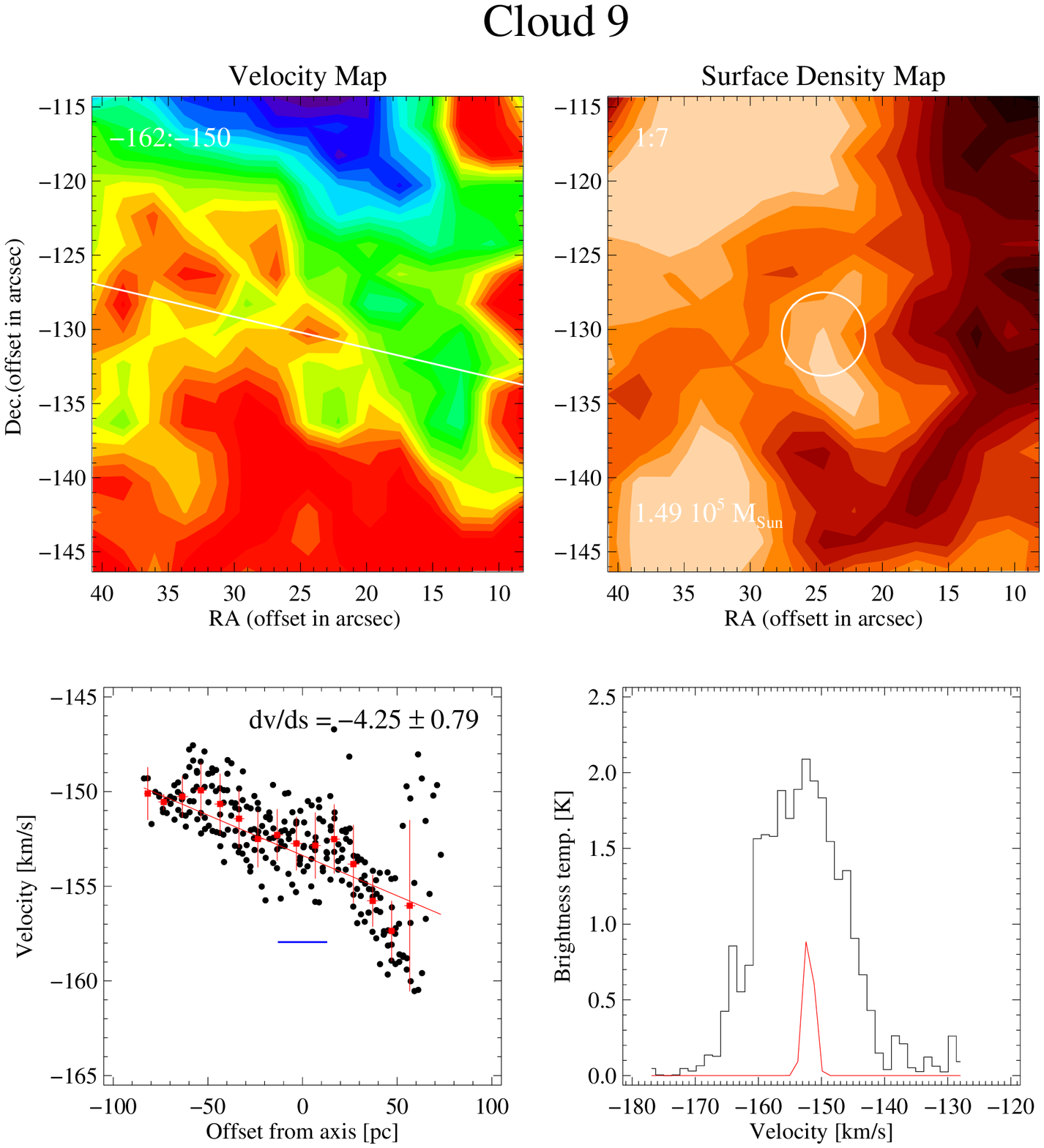}
\caption{See Figure \ref{fig:f22}.}
\end{figure*}

\newpage
\vspace*{2cm}
\begin{figure*}[htbp]
\includegraphics[scale=0.8]{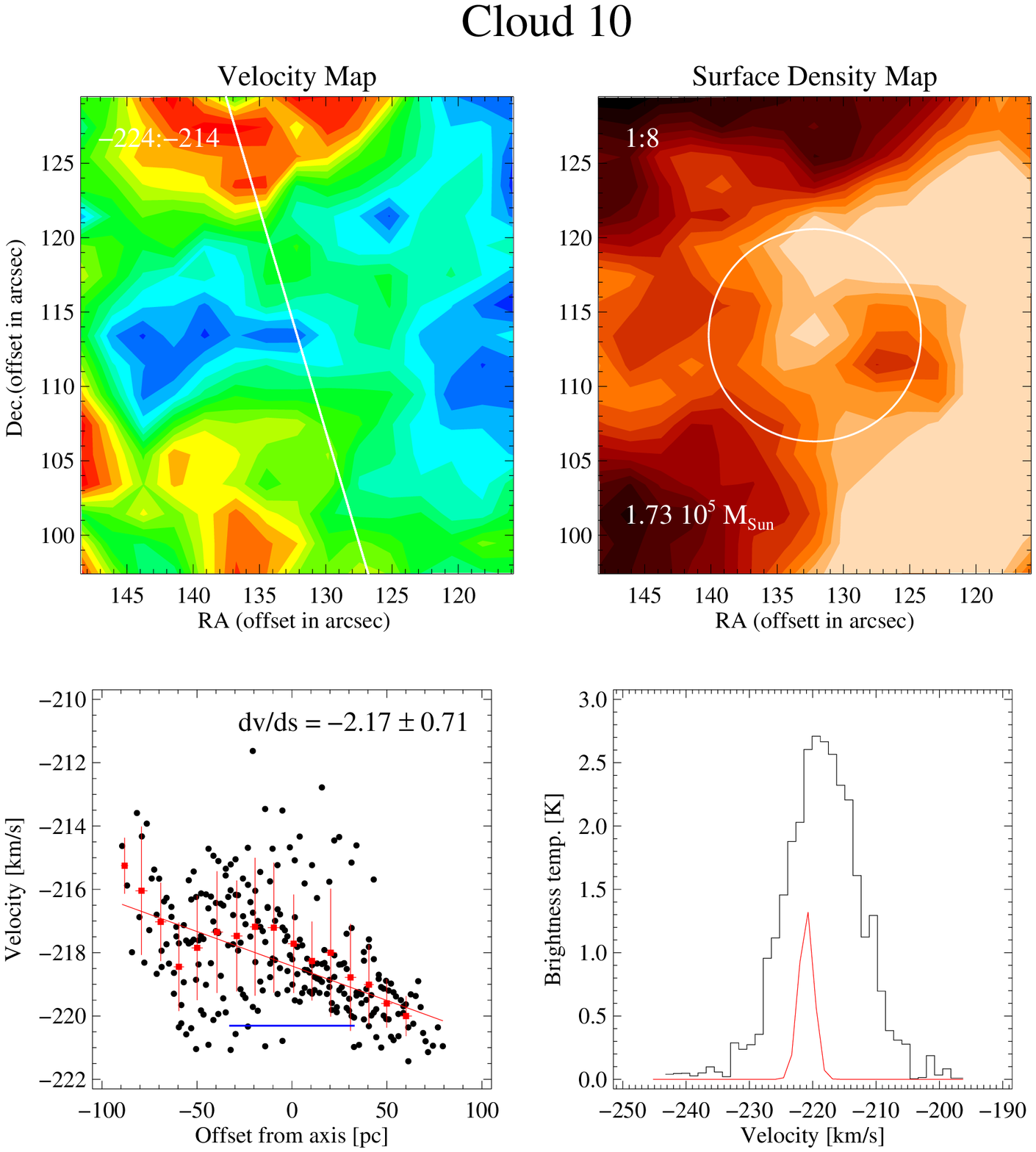}
\caption{See Figure \ref{fig:f22}.}
\end{figure*}

\newpage
\vspace*{2cm}
\begin{figure*}[htbp]
\includegraphics[scale=0.8]{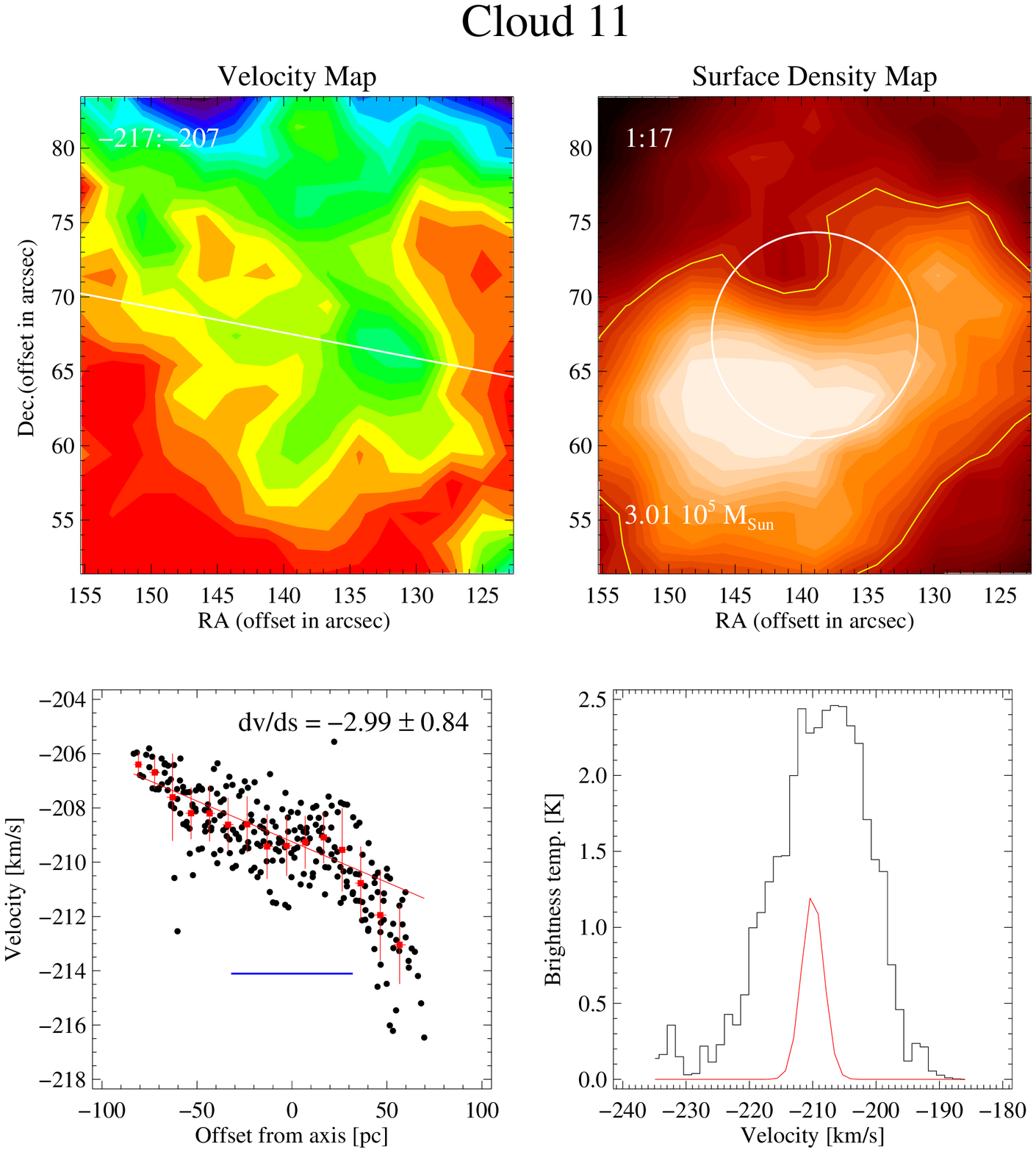}
\caption{See Figure \ref{fig:f22}.}
\end{figure*}

\newpage
\vspace*{2cm}
\begin{figure*}[htbp]
\includegraphics[scale=0.8]{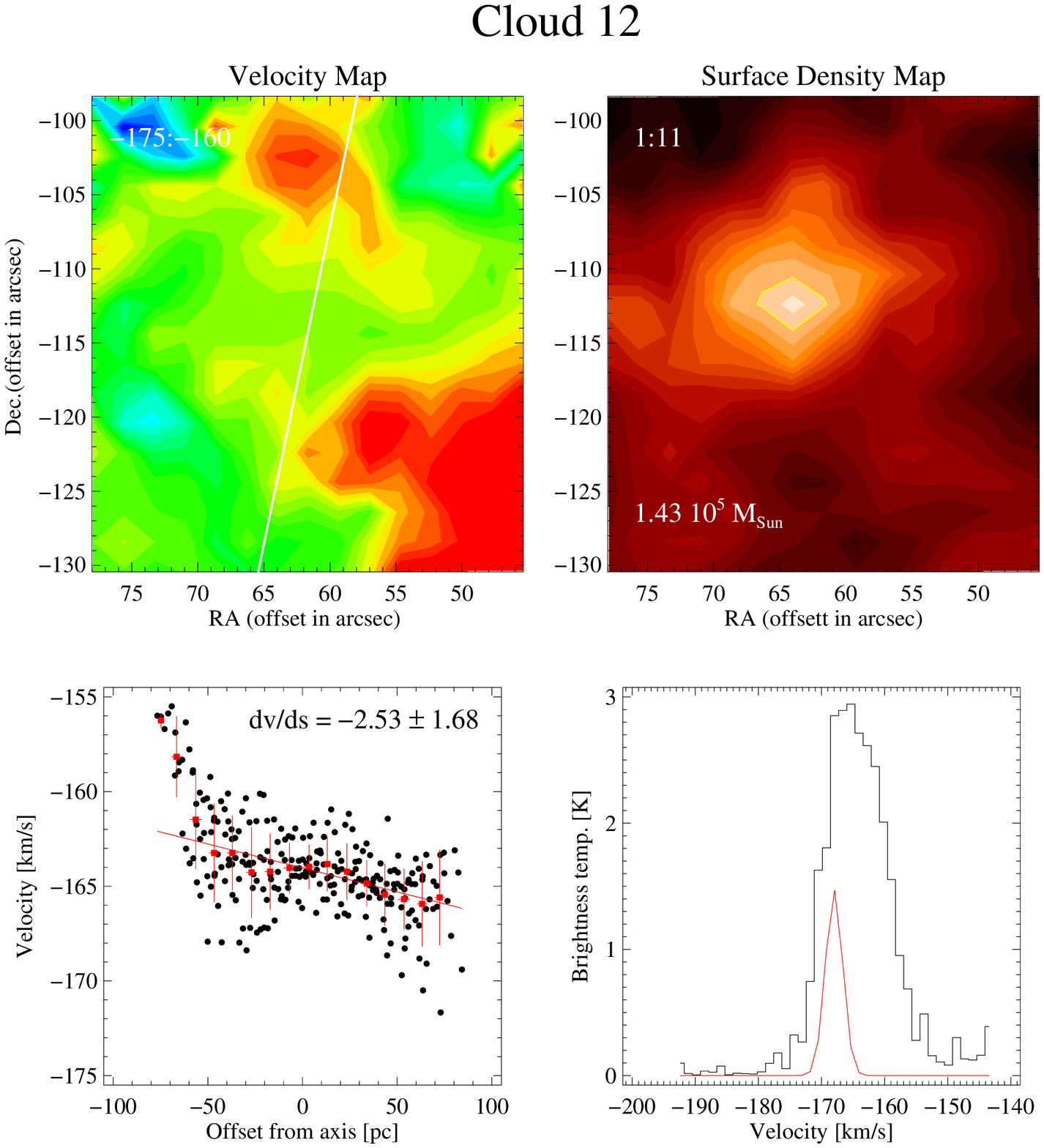}
\caption{See Figure \ref{fig:f22}.}
\end{figure*}

\newpage
\vspace*{2cm}
\begin{figure*}[htbp]
\includegraphics[scale=0.8]{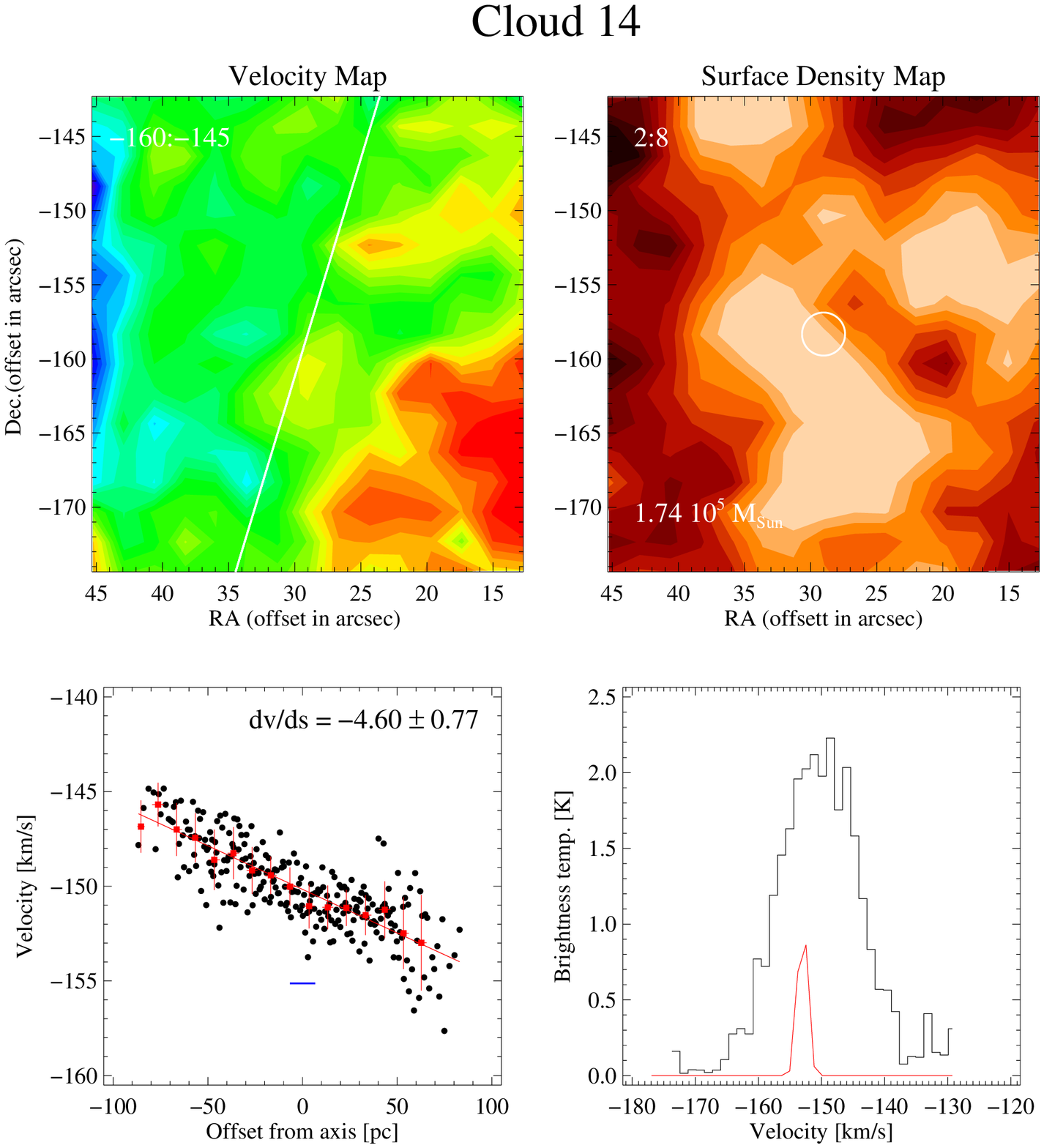}
\caption{See Figure \ref{fig:f22}.}
\end{figure*}

\newpage
\vspace*{2cm}
\begin{figure*}[htbp]
\includegraphics[scale=0.8]{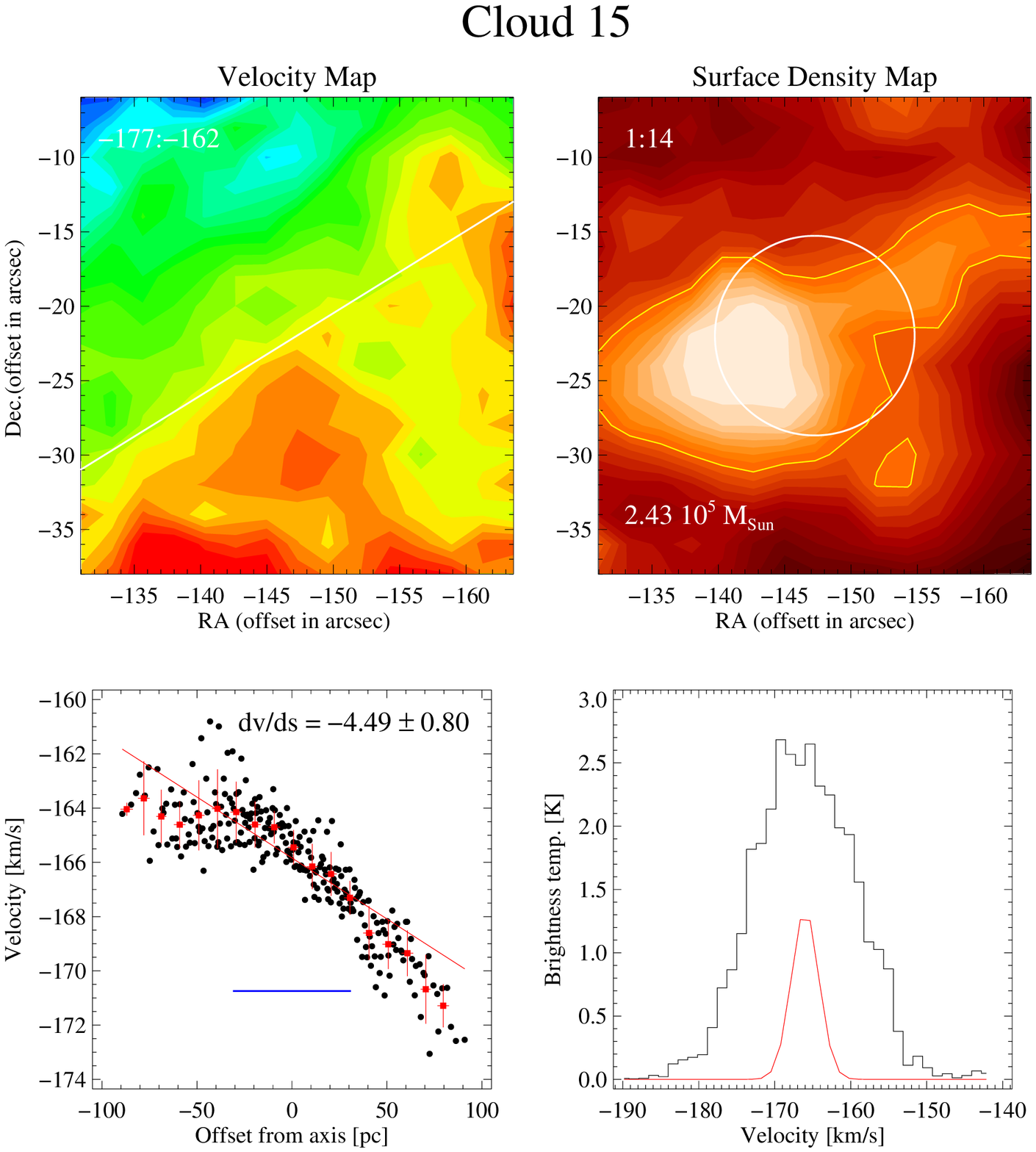}
\caption{See Figure \ref{fig:f22}.}
\end{figure*}

\newpage
\vspace*{2cm}
\begin{figure*}[htbp]
\includegraphics[scale=0.8]{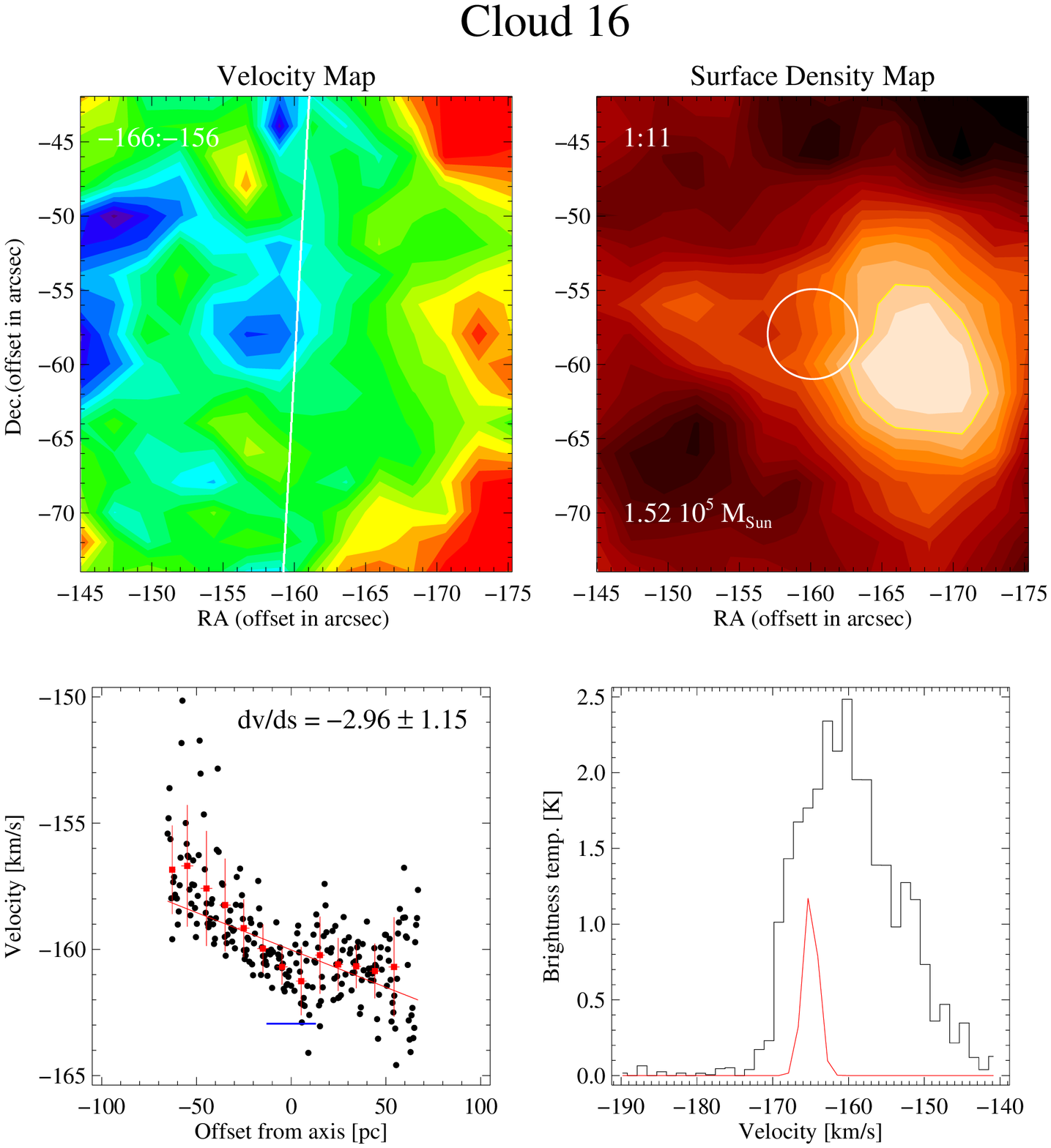}
\caption{See Figure \ref{fig:f22}.}
\end{figure*}

\newpage
\vspace*{2cm}
\begin{figure*}[htbp]
\includegraphics[scale=0.8]{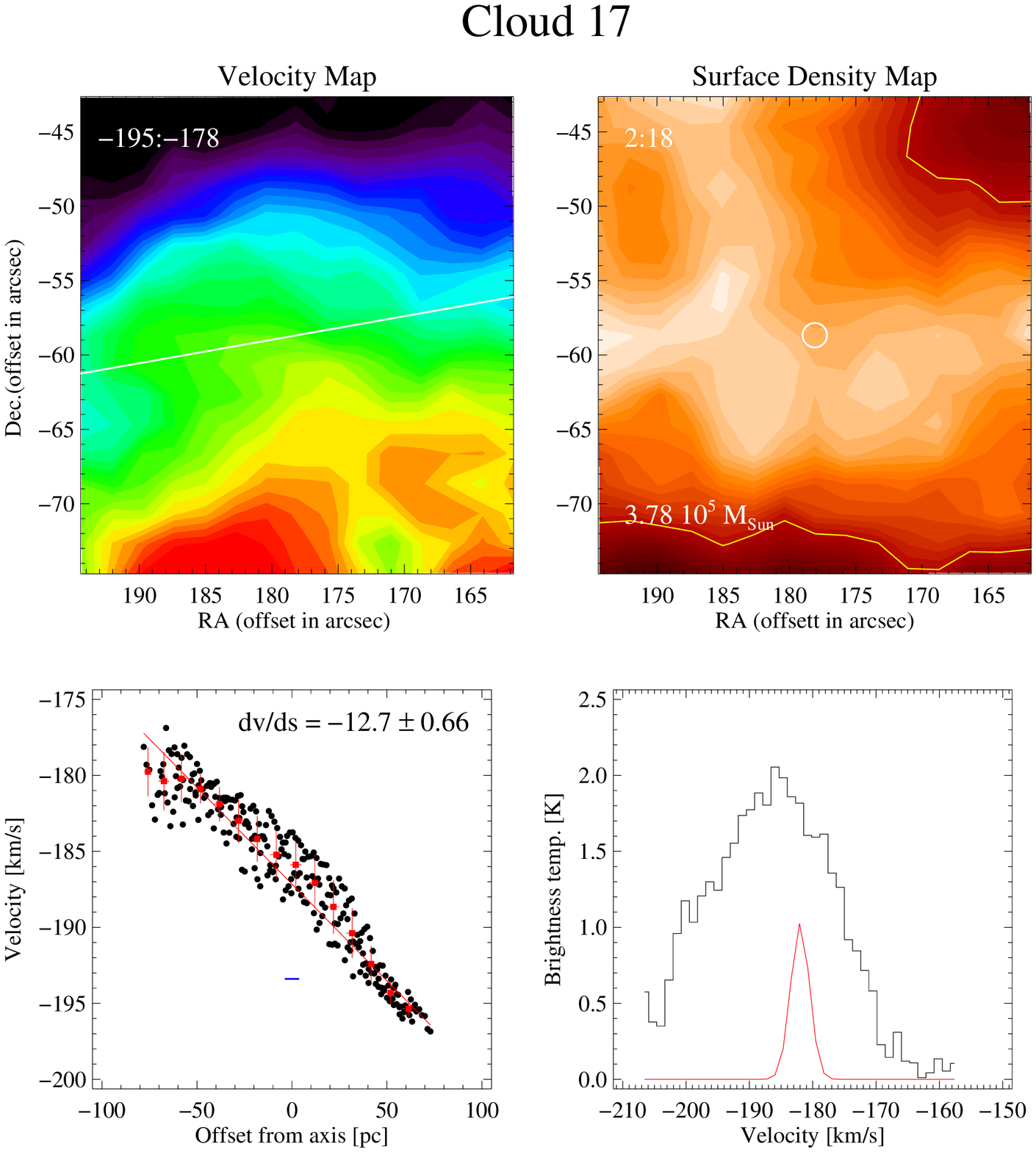}
\caption{See Figure \ref{fig:f22}.}
\end{figure*}

\newpage
\vspace*{2cm}
\begin{figure*}[htbp]
\includegraphics[scale=0.8]{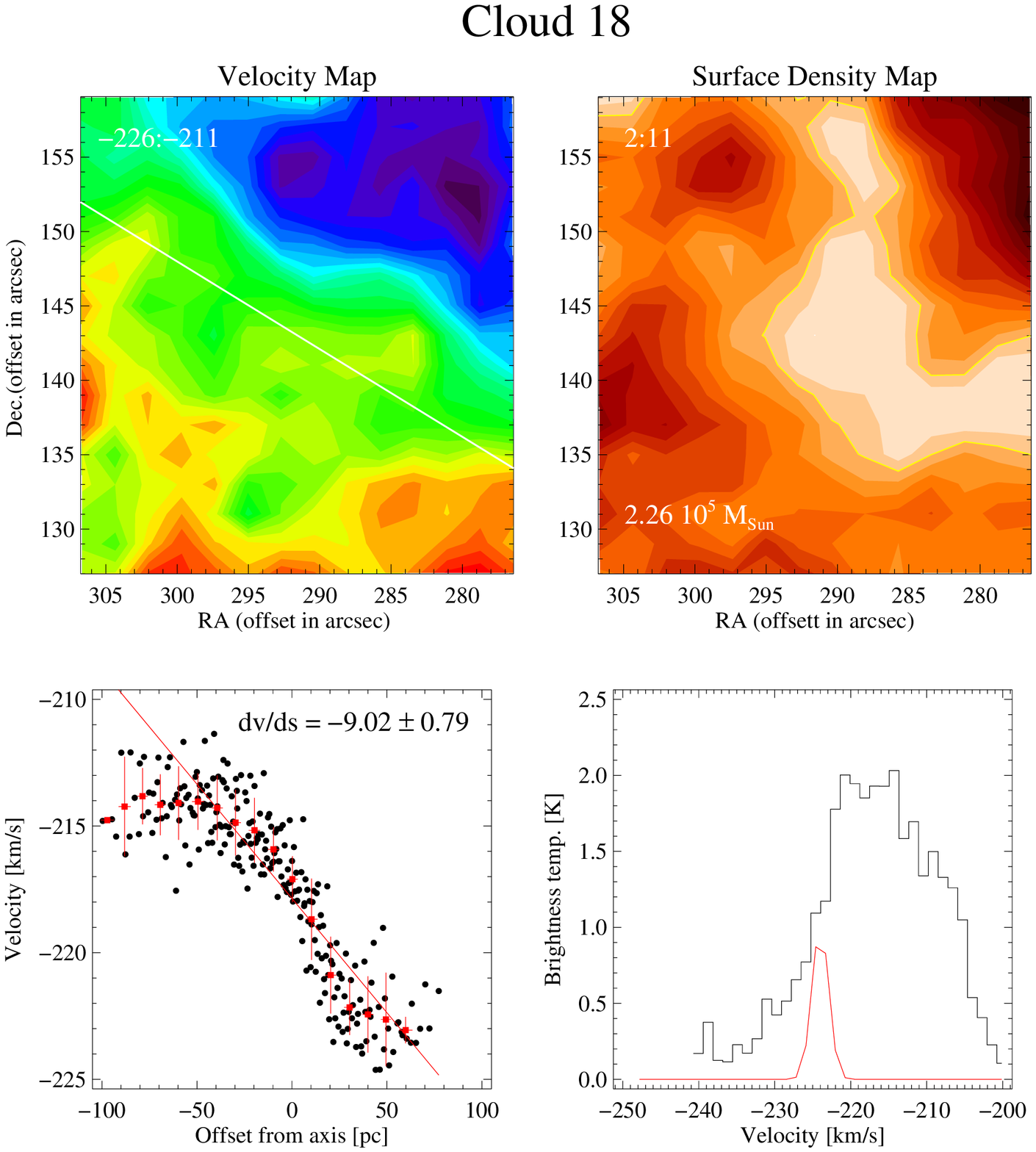}
\caption{See Figure \ref{fig:f22}.}
\end{figure*}

\newpage
\vspace*{2cm}
\begin{figure*}[htbp]
\includegraphics[scale=0.8]{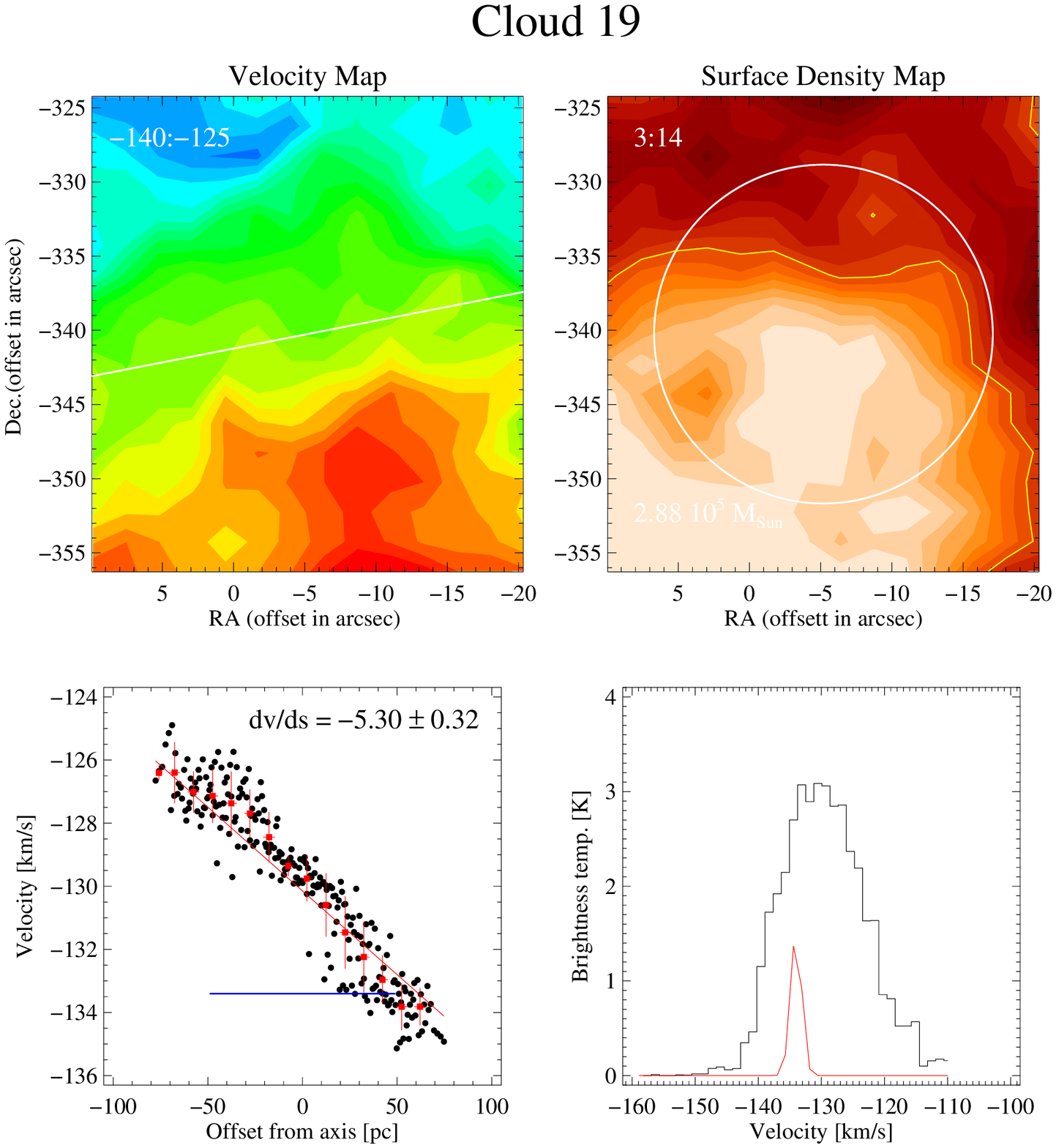}
\caption{See Figure \ref{fig:f22}.}
\end{figure*}

\newpage
\vspace*{2cm}
\begin{figure*}[htbp]
\includegraphics[scale=0.8]{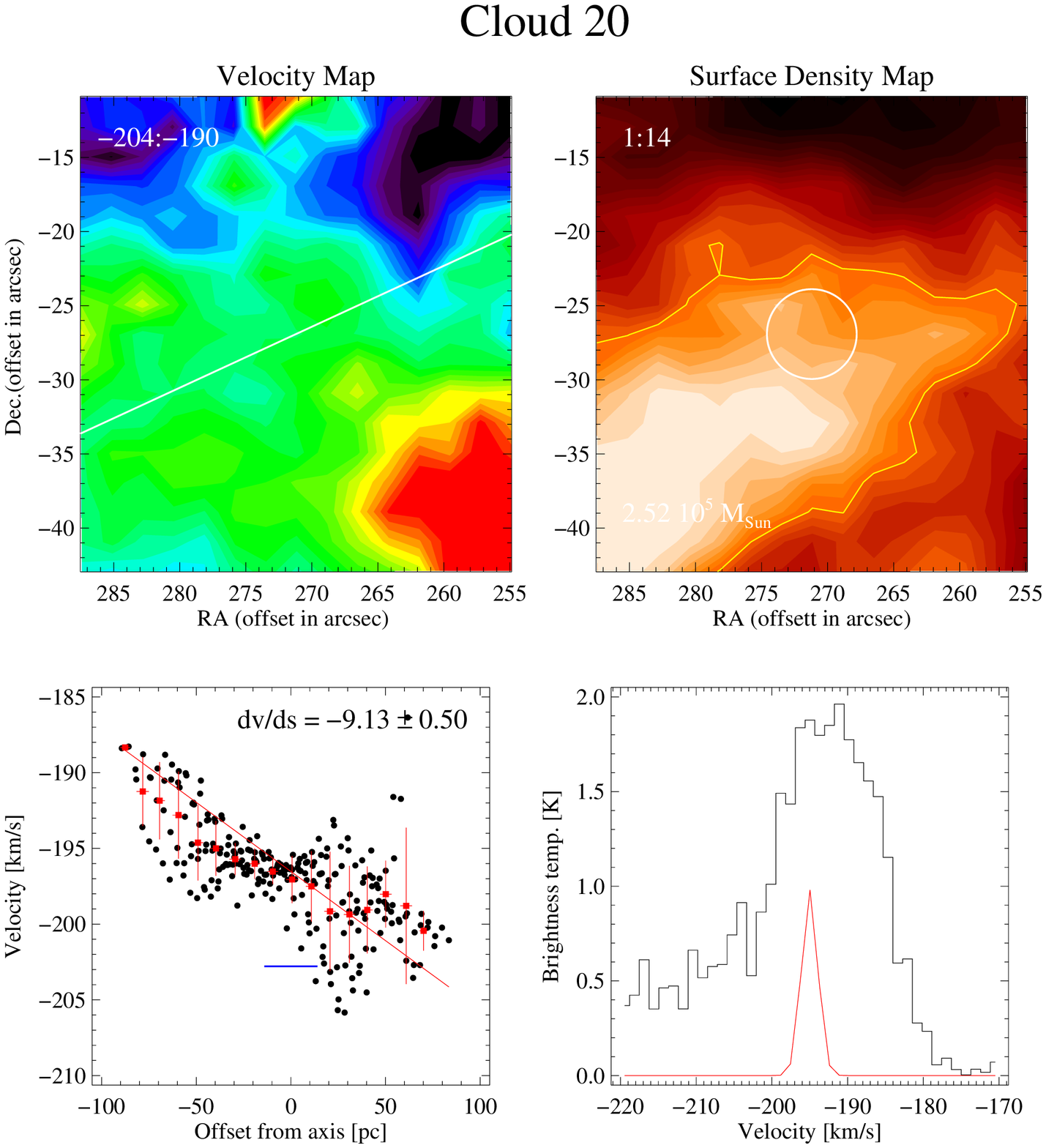}
\caption{See Figure \ref{fig:f22}.}
\end{figure*}

\newpage
\vspace*{2cm}
\begin{figure*}[htbp]
\includegraphics[scale=0.8]{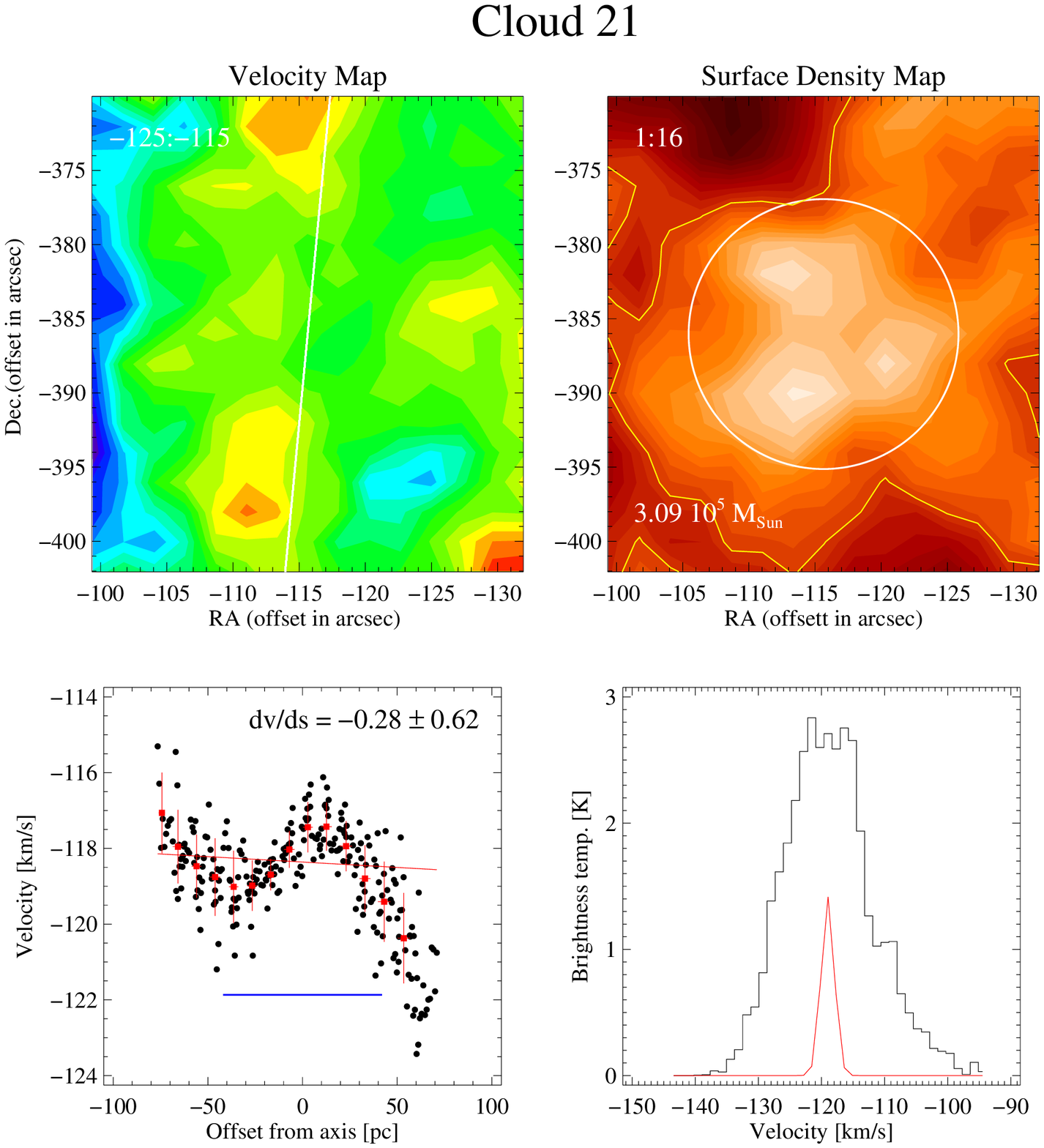}
\caption{See Figure \ref{fig:f22}.}
\end{figure*}

\newpage
\vspace*{2cm}
\begin{figure*}[htbp]
\includegraphics[scale=0.8]{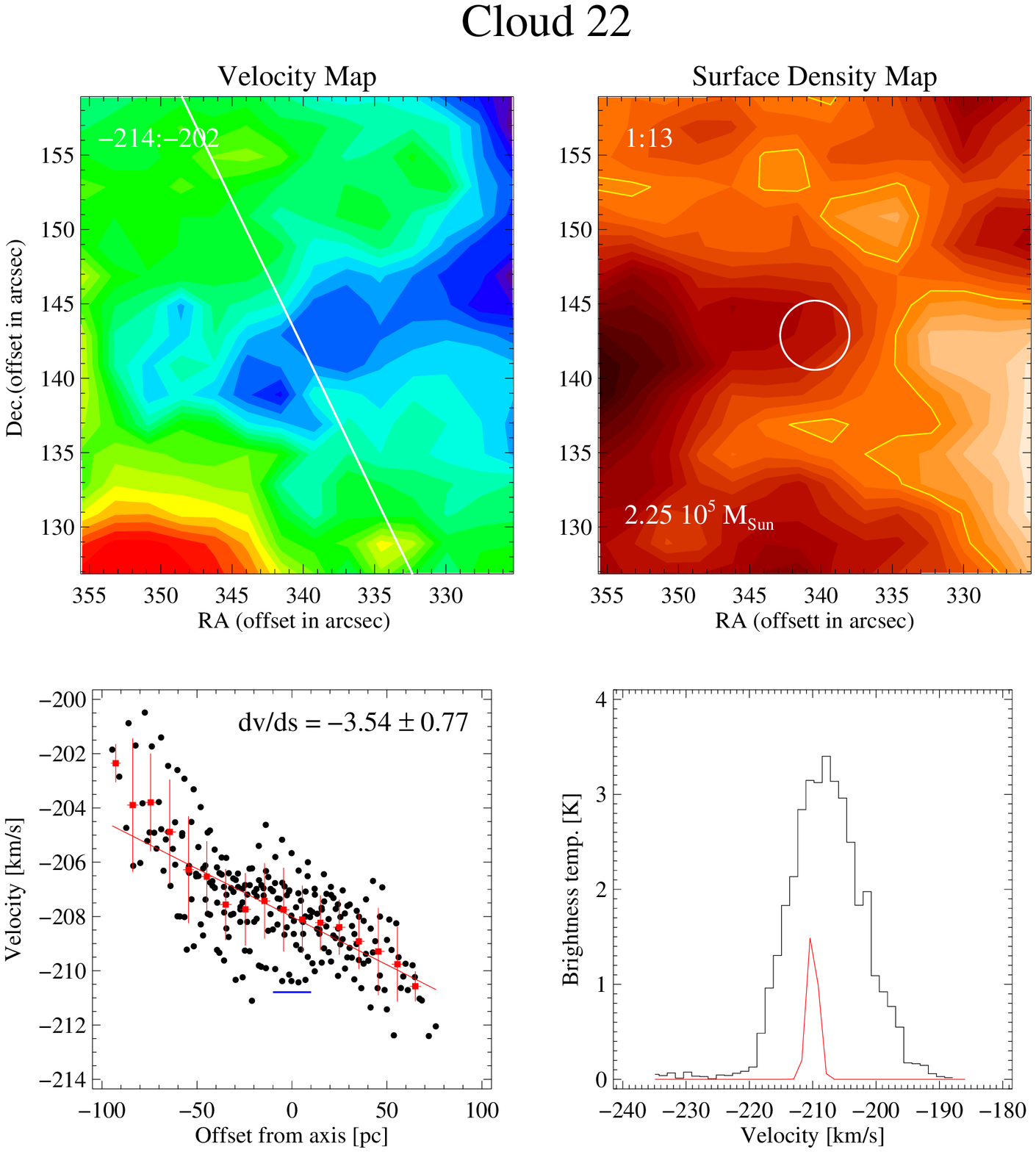}
\caption{See Figure \ref{fig:f22}.}
\end{figure*}

\newpage
\vspace*{2cm}
\begin{figure*}[htbp]
\includegraphics[scale=0.8]{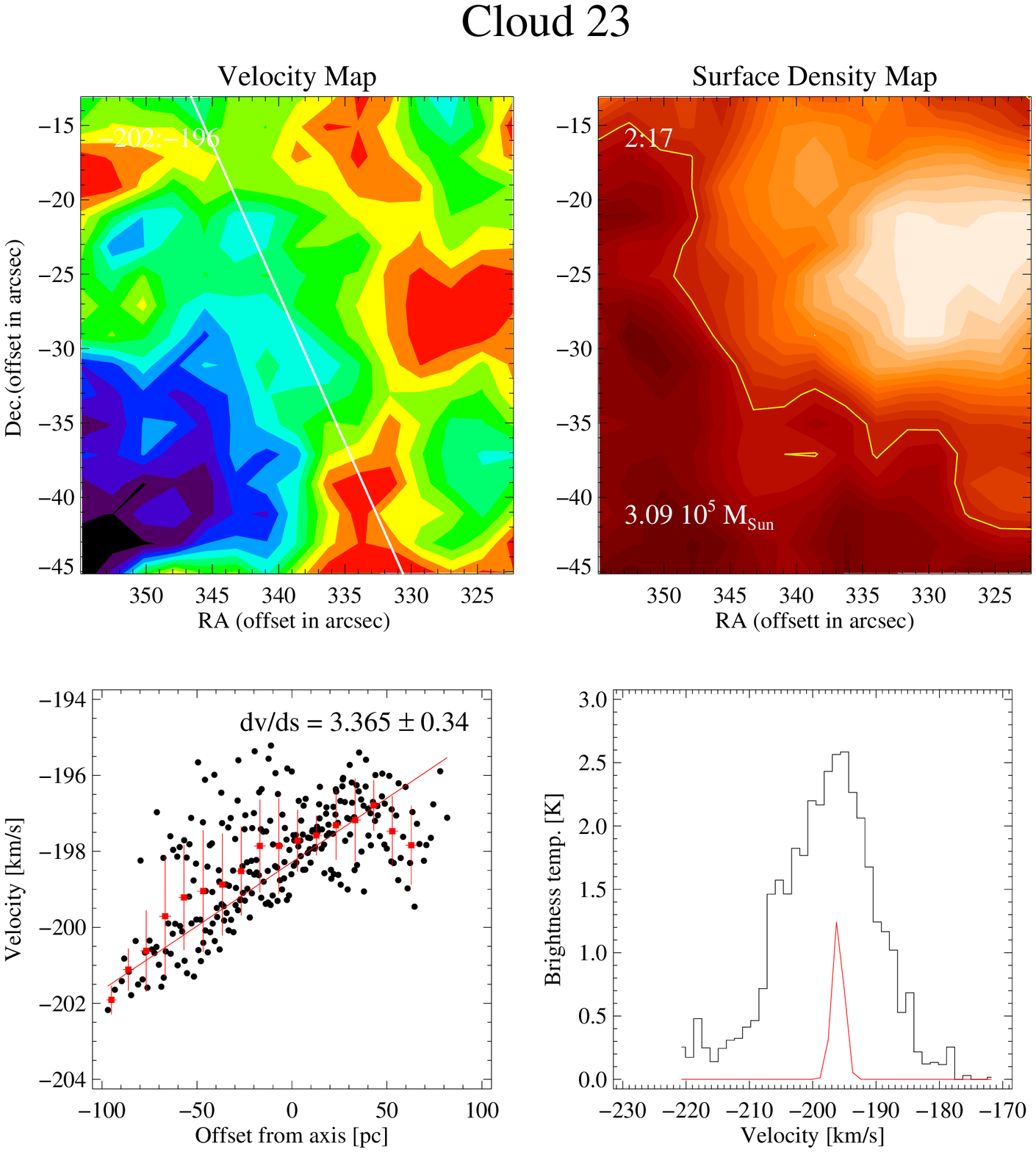}
\caption{See Figure \ref{fig:f22}.}
\end{figure*}

\newpage
\vspace*{2cm}
\begin{figure*}[htbp]
\includegraphics[scale=0.8]{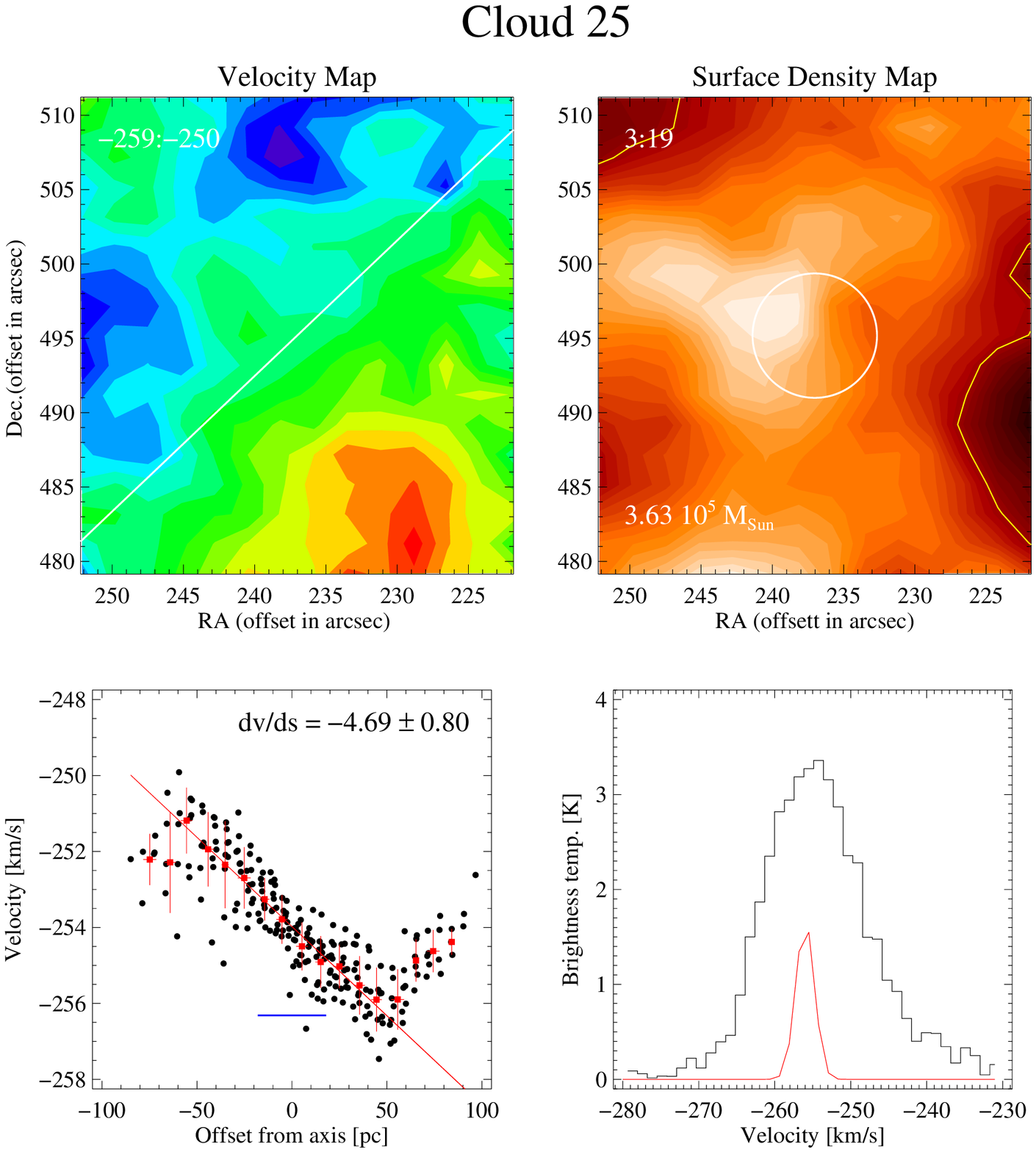}
\caption{See Figure \ref{fig:f22}.}
\end{figure*}

\newpage
\vspace*{2cm}
\begin{figure*}[htbp]
\includegraphics[scale=0.8]{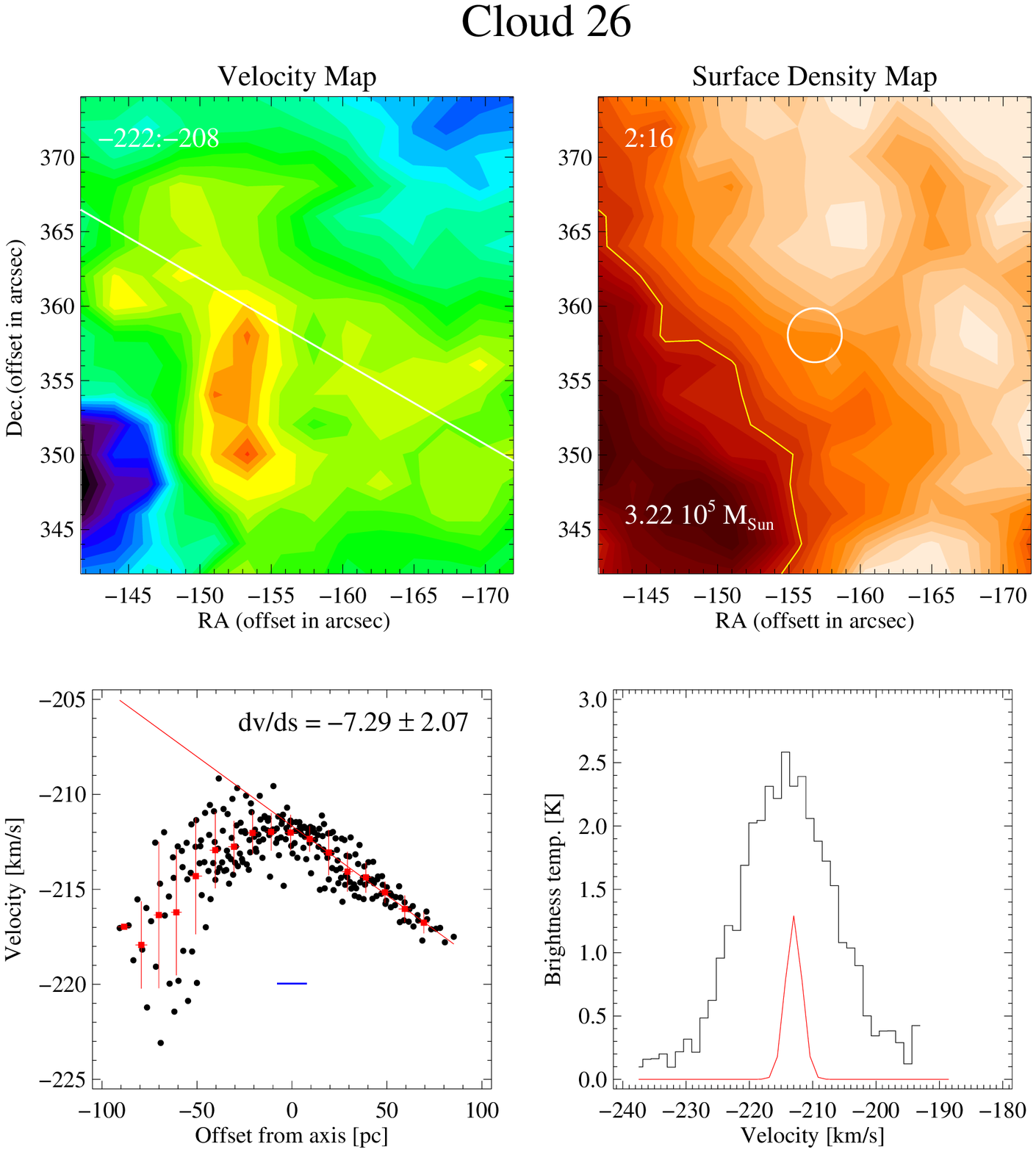}
\caption{See Figure \ref{fig:f22}.}
\end{figure*}

\newpage
\vspace*{2cm}
\begin{figure*}[htbp]
\includegraphics[scale=0.8]{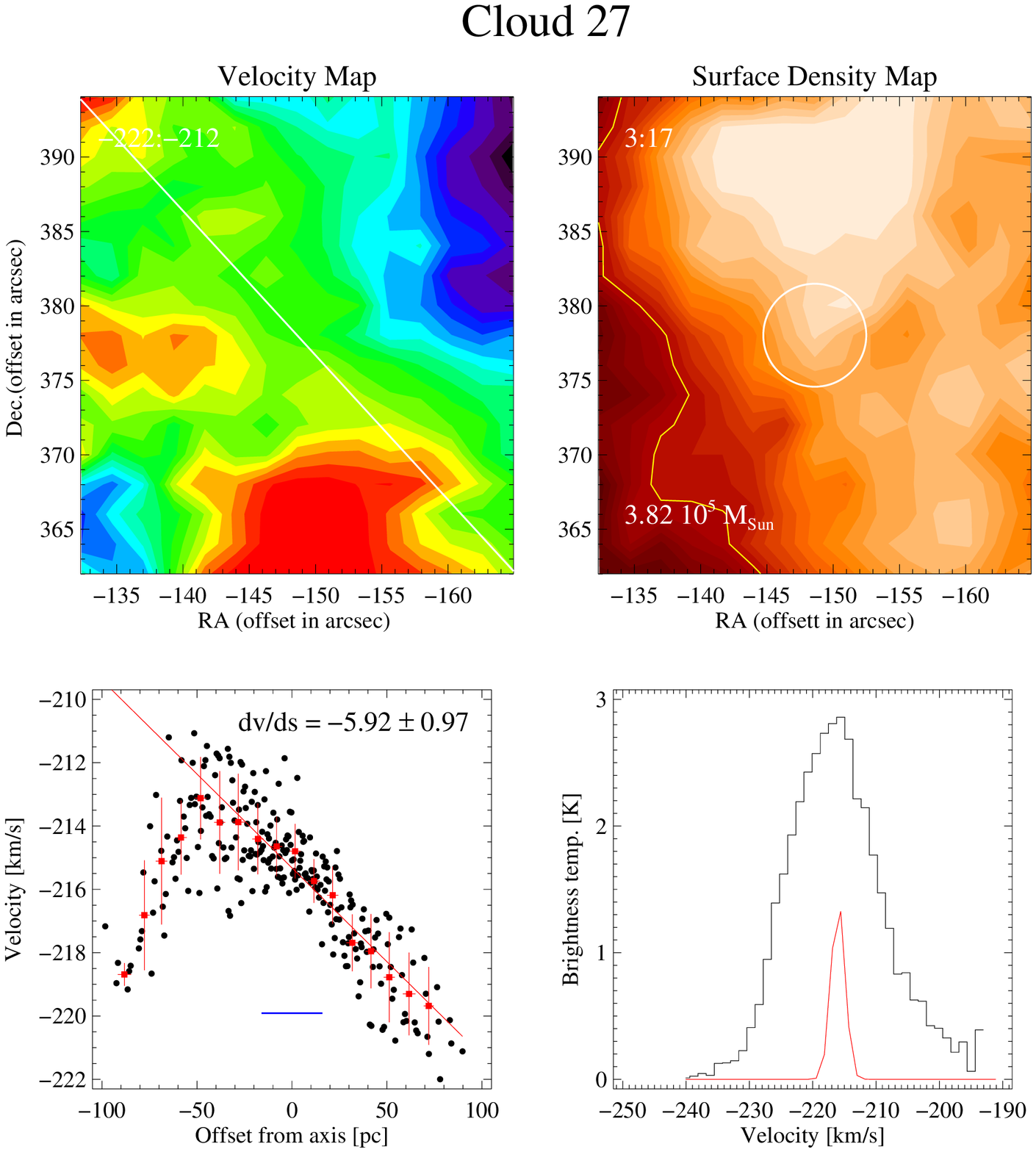}
\caption{See Figure \ref{fig:f22}.}
\end{figure*}

\newpage
\vspace*{2cm}
\begin{figure*}[htbp]
\includegraphics[scale=0.8]{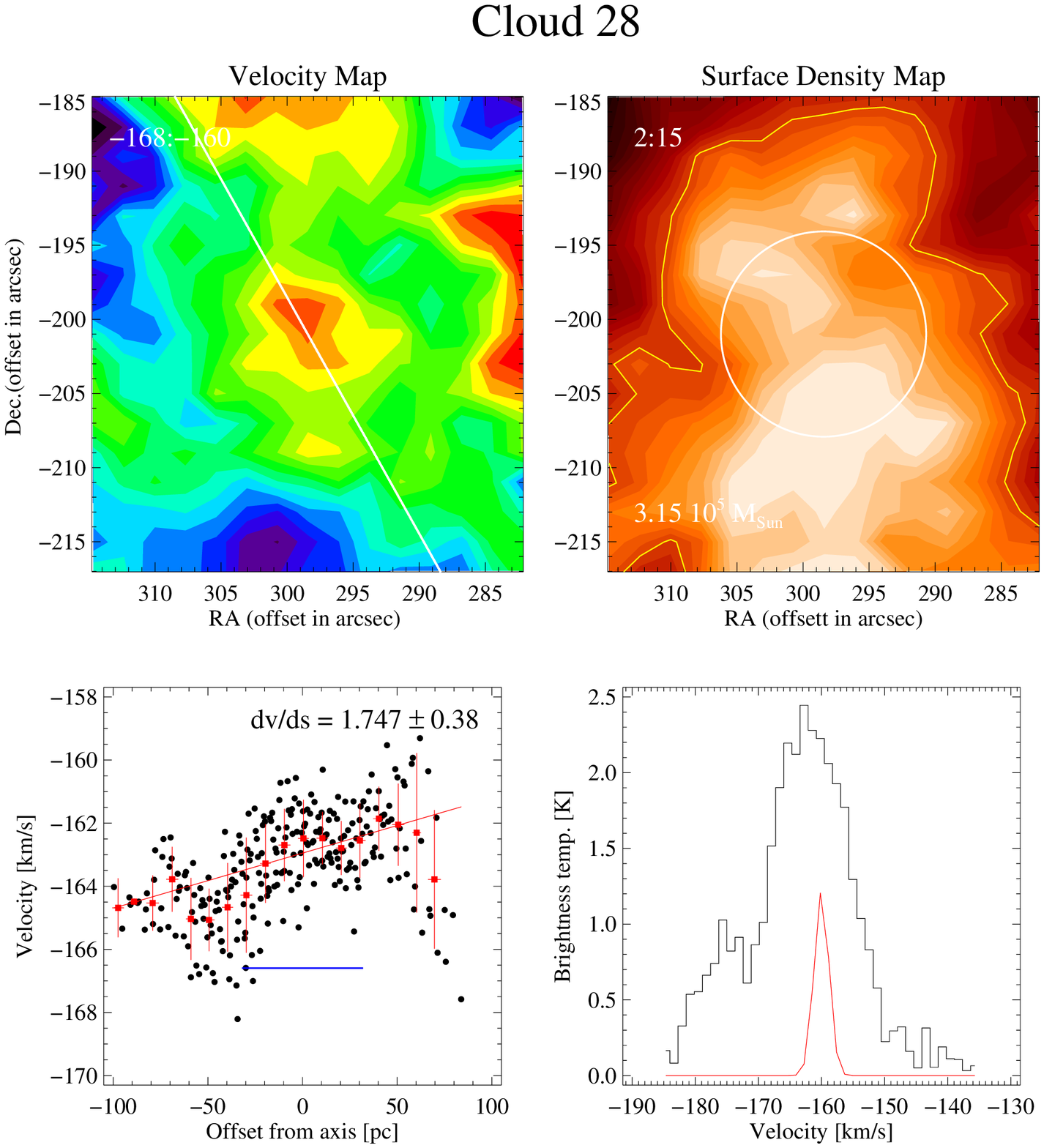}
\caption{See Figure \ref{fig:f22}.}
\end{figure*}

\newpage
\vspace*{2cm}
\begin{figure*}[htbp]
\includegraphics[scale=0.8]{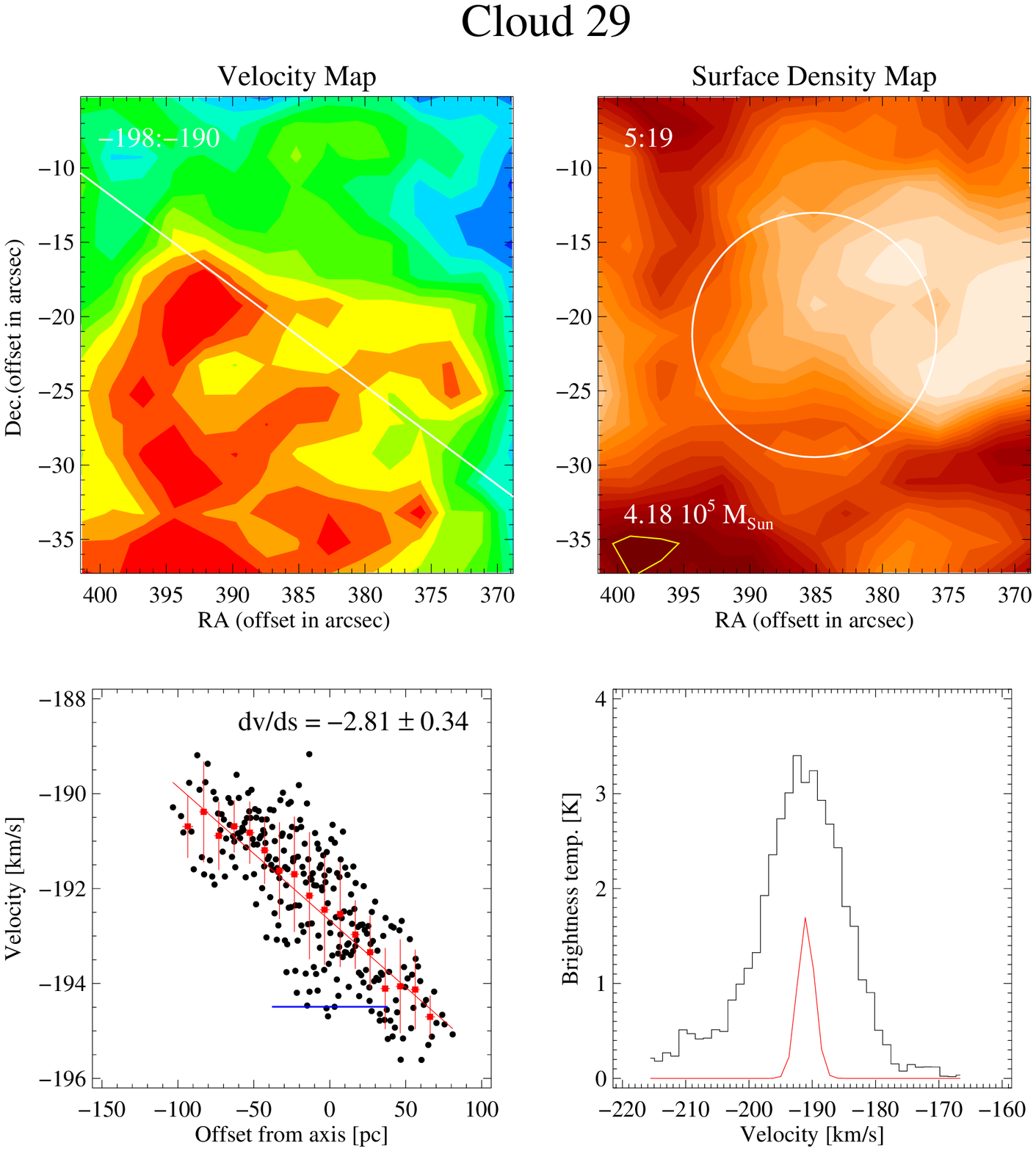}
\caption{See Figure \ref{fig:f22}.}
\end{figure*}

\newpage
\vspace*{2cm}
\begin{figure*}[htbp]
\includegraphics[scale=0.8]{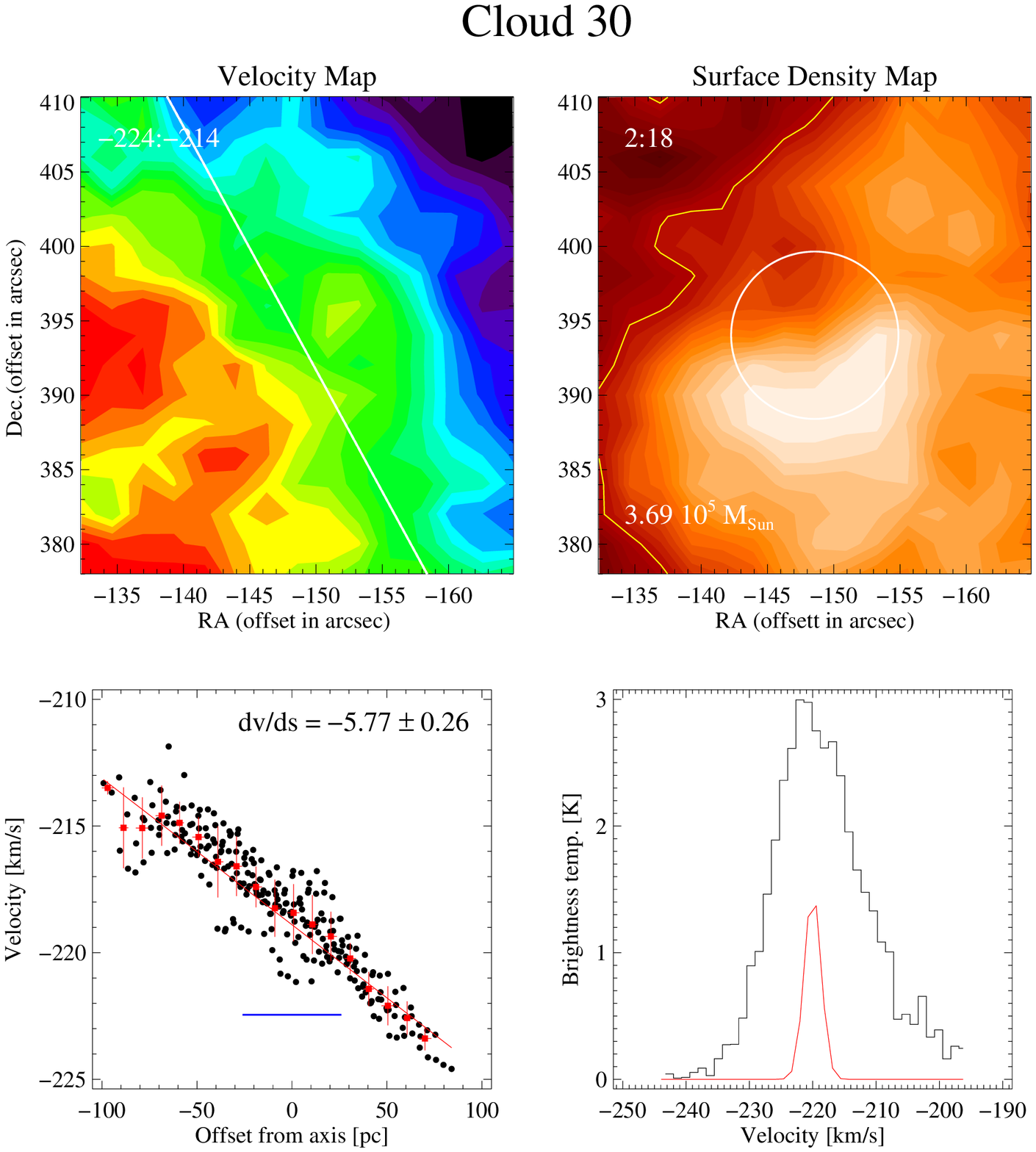}
\caption{See Figure \ref{fig:f22}.}
\end{figure*}

\newpage
\vspace*{2cm}
\begin{figure*}[htbp]
\includegraphics[scale=0.8]{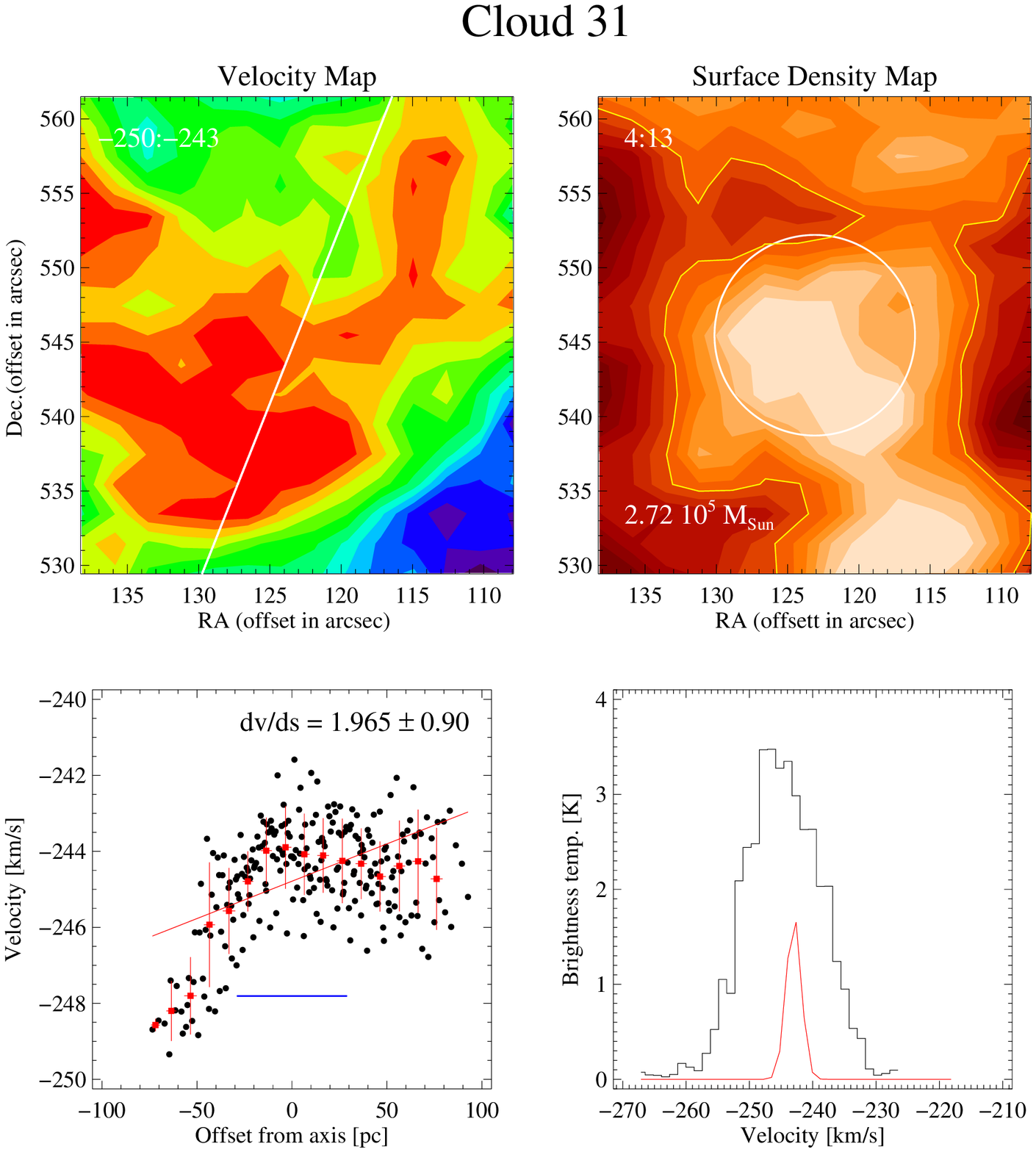}
\caption{See Figure \ref{fig:f22}.}
\end{figure*}

\newpage
\vspace*{2cm}
\begin{figure*}[htbp]
\includegraphics[scale=0.8]{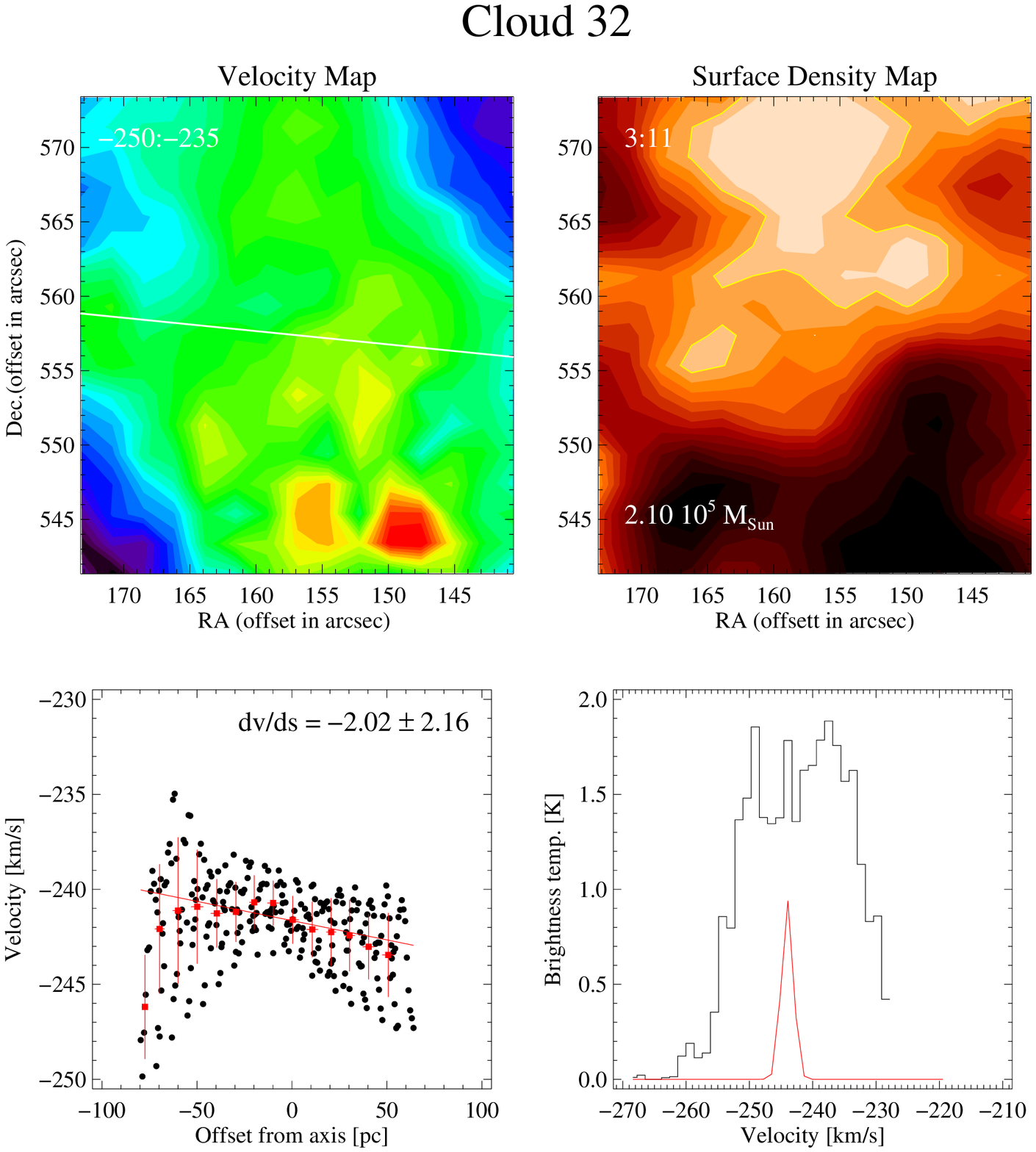}
\caption{See Figure \ref{fig:f22}.}
\end{figure*}

\newpage
\vspace*{2cm}
\begin{figure*}[htbp]
\includegraphics[scale=0.8]{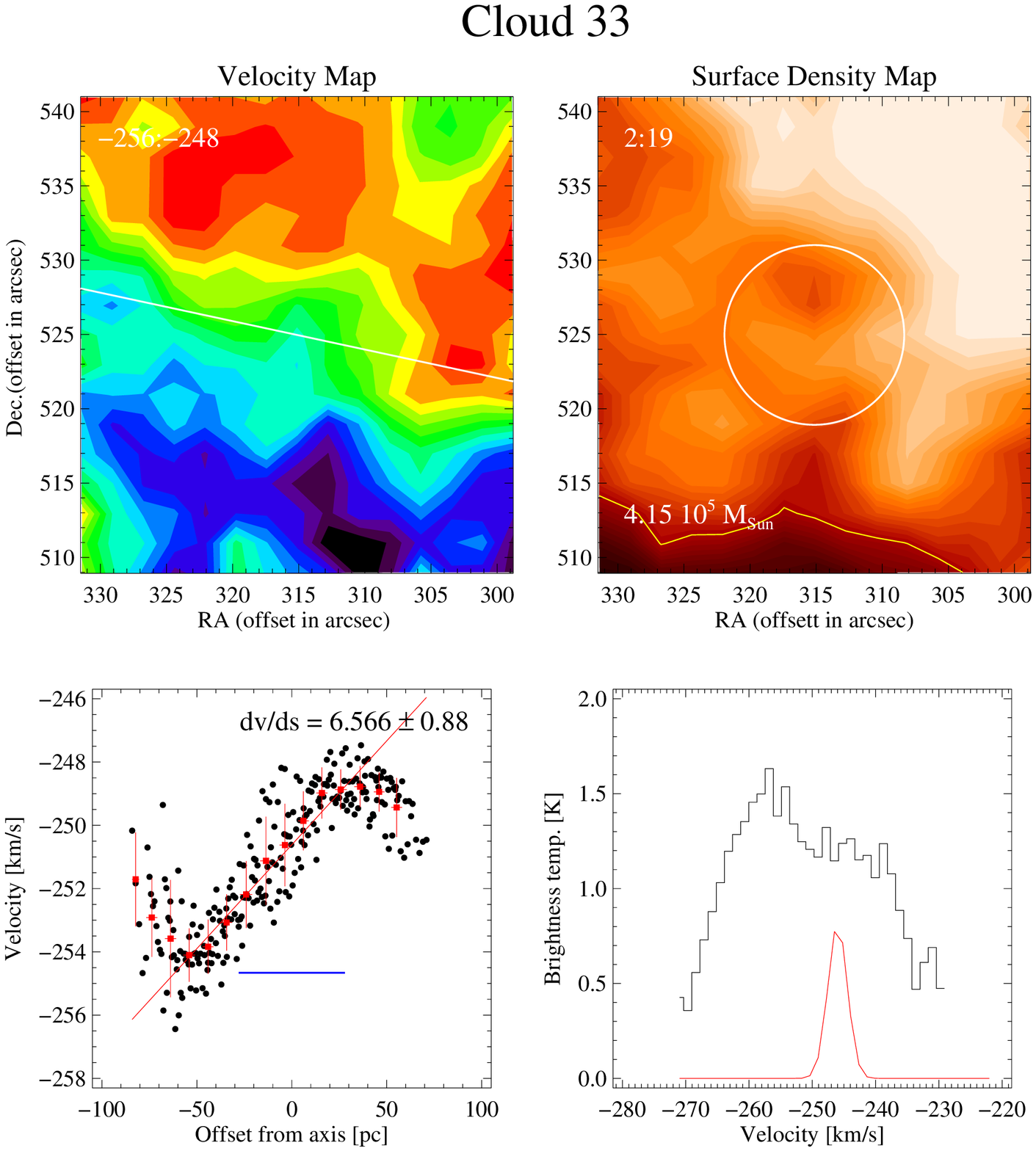}
\caption{See Figure \ref{fig:f22}.}
\end{figure*}

\newpage
\vspace*{2cm}
\begin{figure*}[htbp]
\includegraphics[scale=0.8]{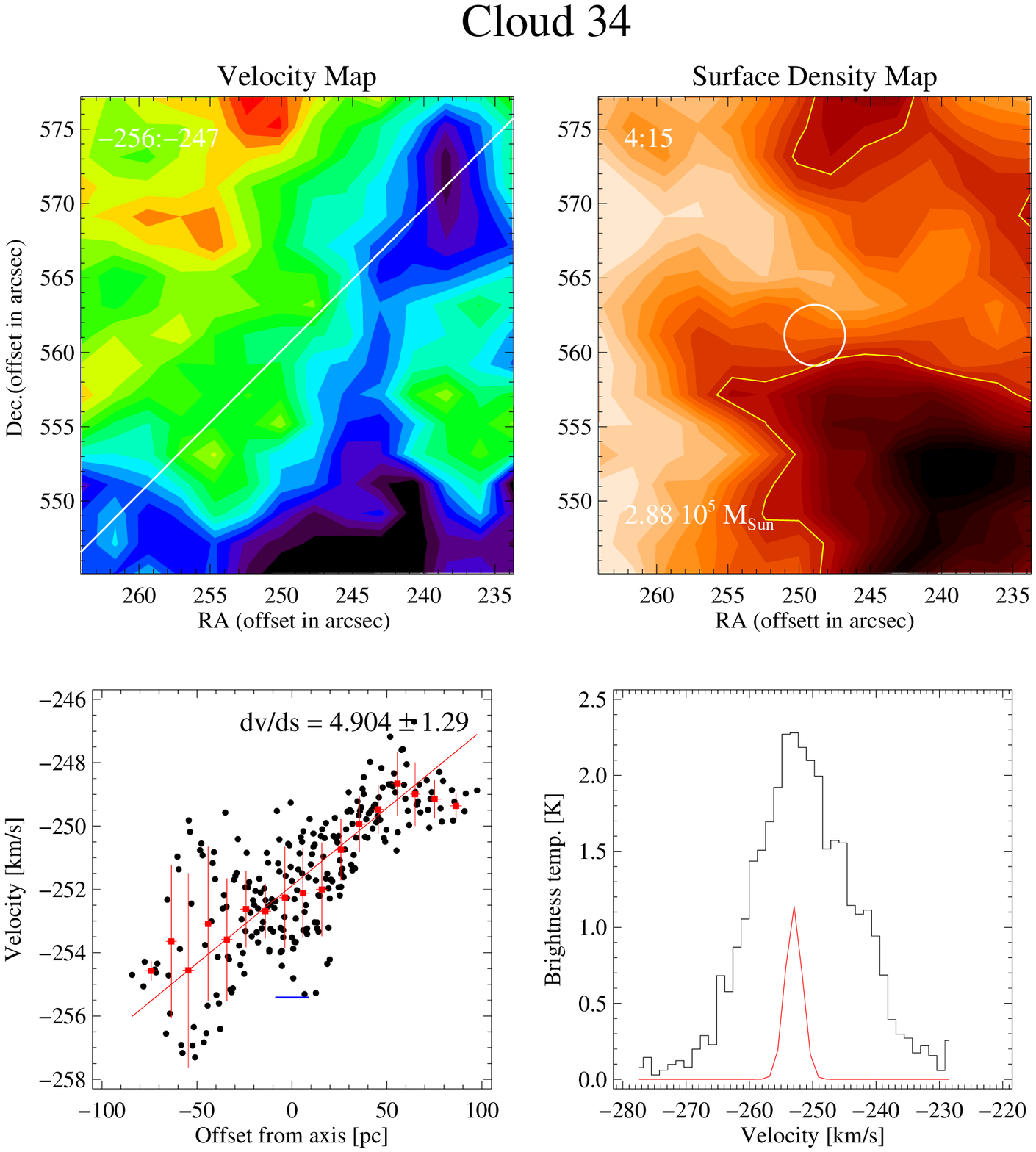}
\caption{See Figure \ref{fig:f22}.}
\end{figure*}

\newpage
\vspace*{2cm}
\begin{figure*}[htbp]
\includegraphics[scale=0.8]{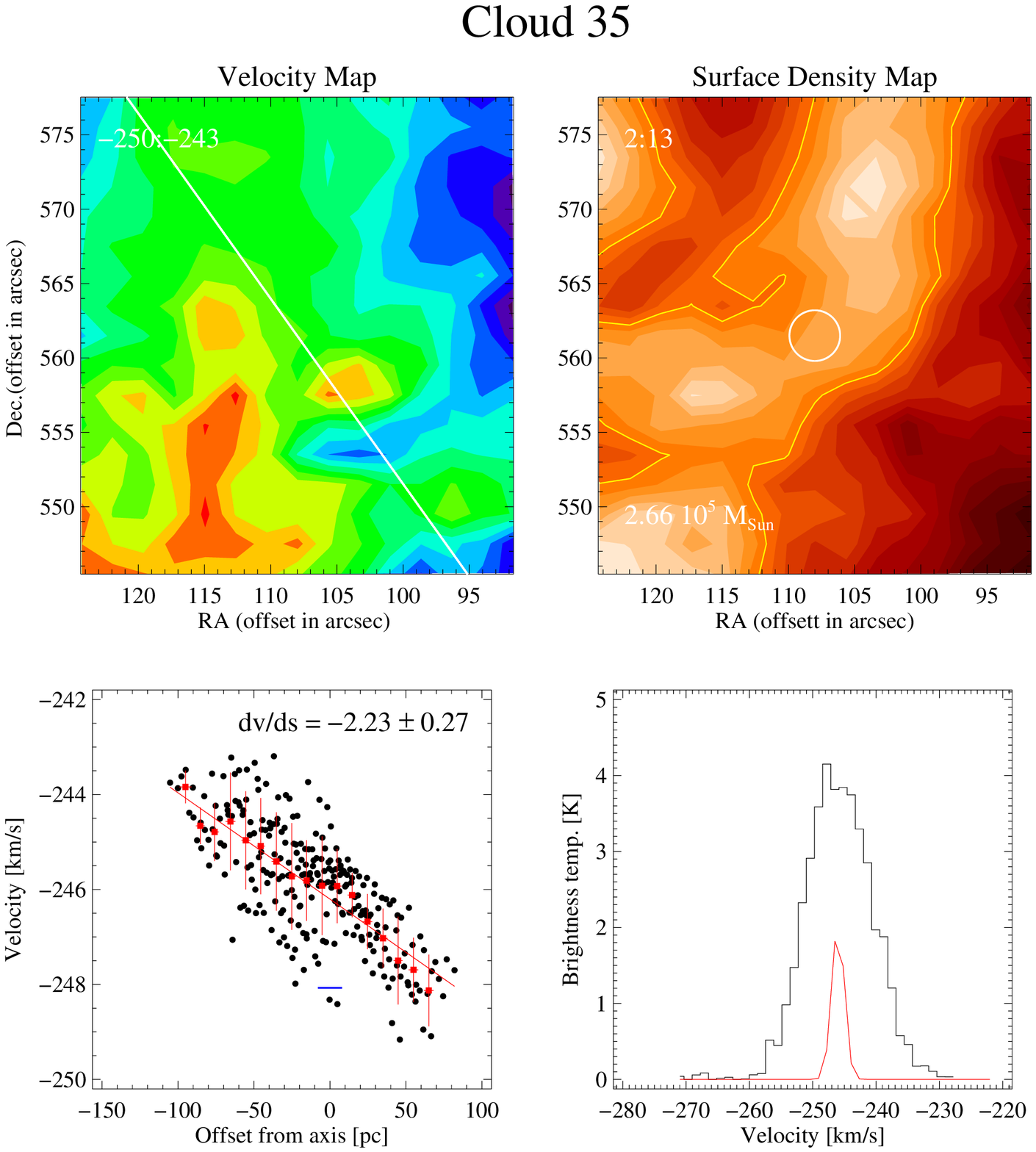}
\caption{See Figure \ref{fig:f22}.}
\end{figure*}

\newpage
\vspace*{2cm}
\begin{figure*}[htbp]
\includegraphics[scale=0.8]{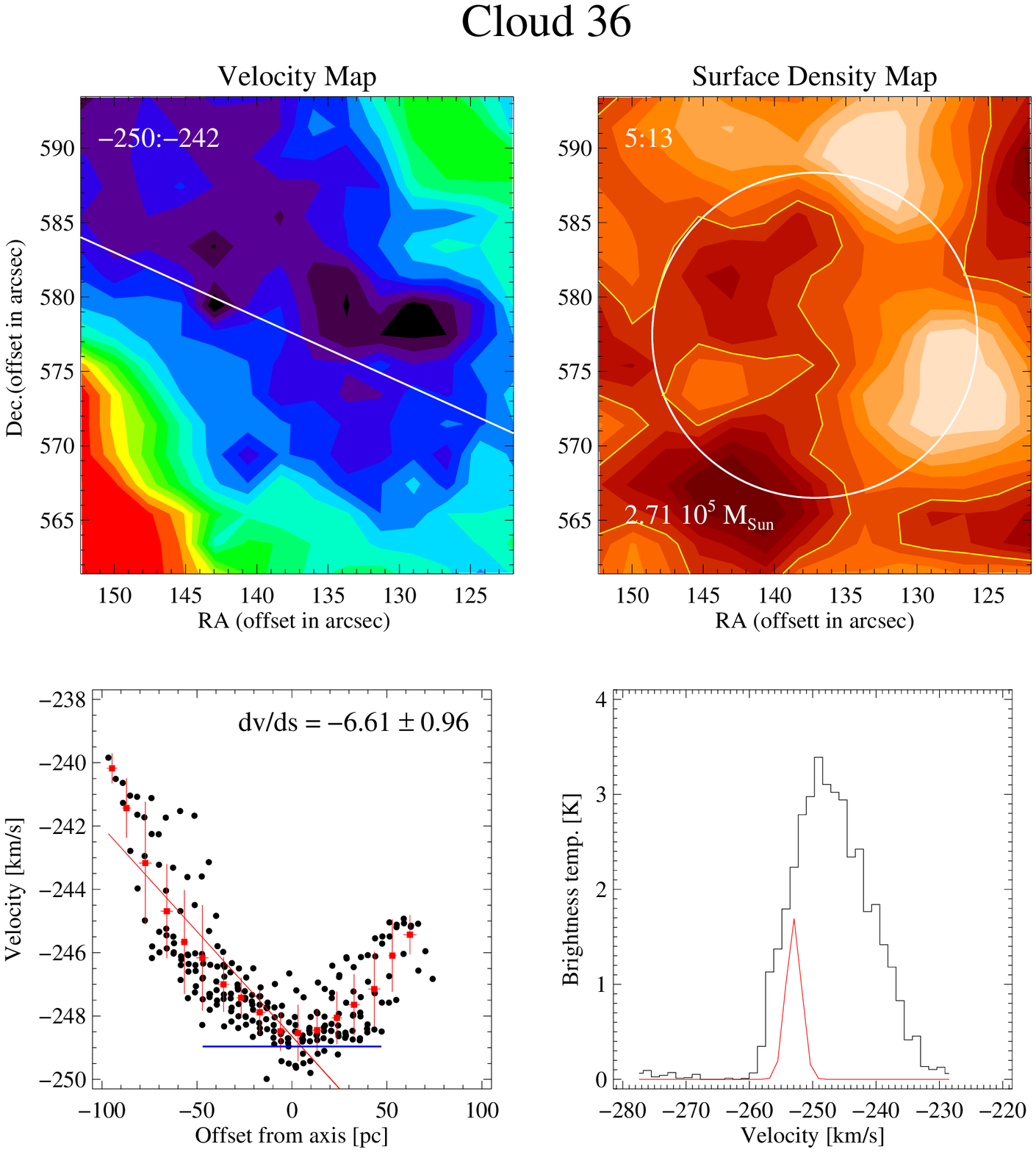}
\caption{See Figure \ref{fig:f22}.}
\end{figure*}

\newpage
\vspace*{2cm}
\begin{figure*}[htbp]
\includegraphics[scale=0.8]{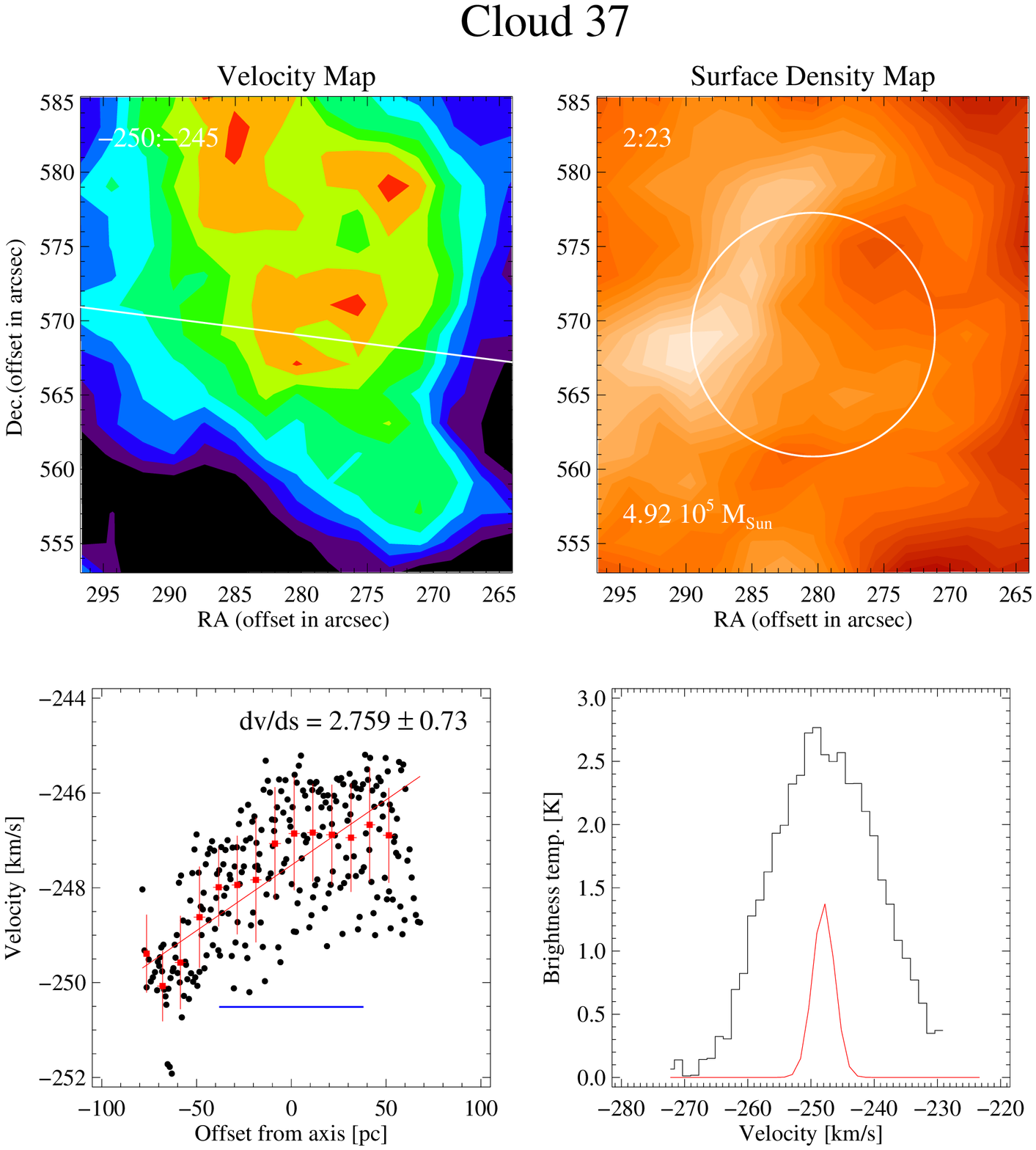}
\caption{See Figure \ref{fig:f22}.}
\end{figure*}

\newpage
\vspace*{2cm}
\begin{figure*}[htbp]
\includegraphics[scale=0.8]{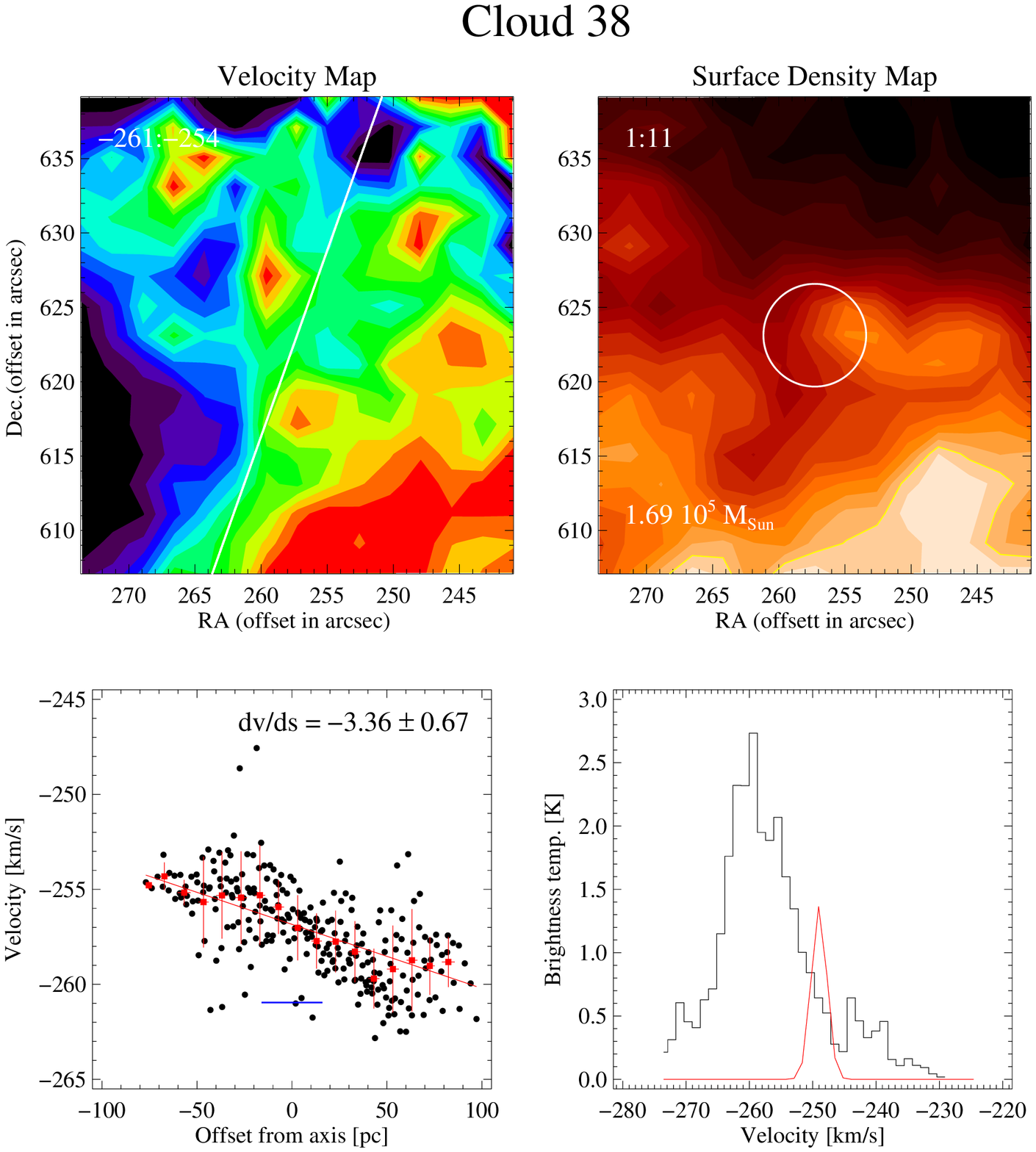}
\caption{See Figure \ref{fig:f22}.}
\end{figure*}

\newpage
\vspace*{2cm}
\begin{figure*}[htbp]
\includegraphics[scale=0.8]{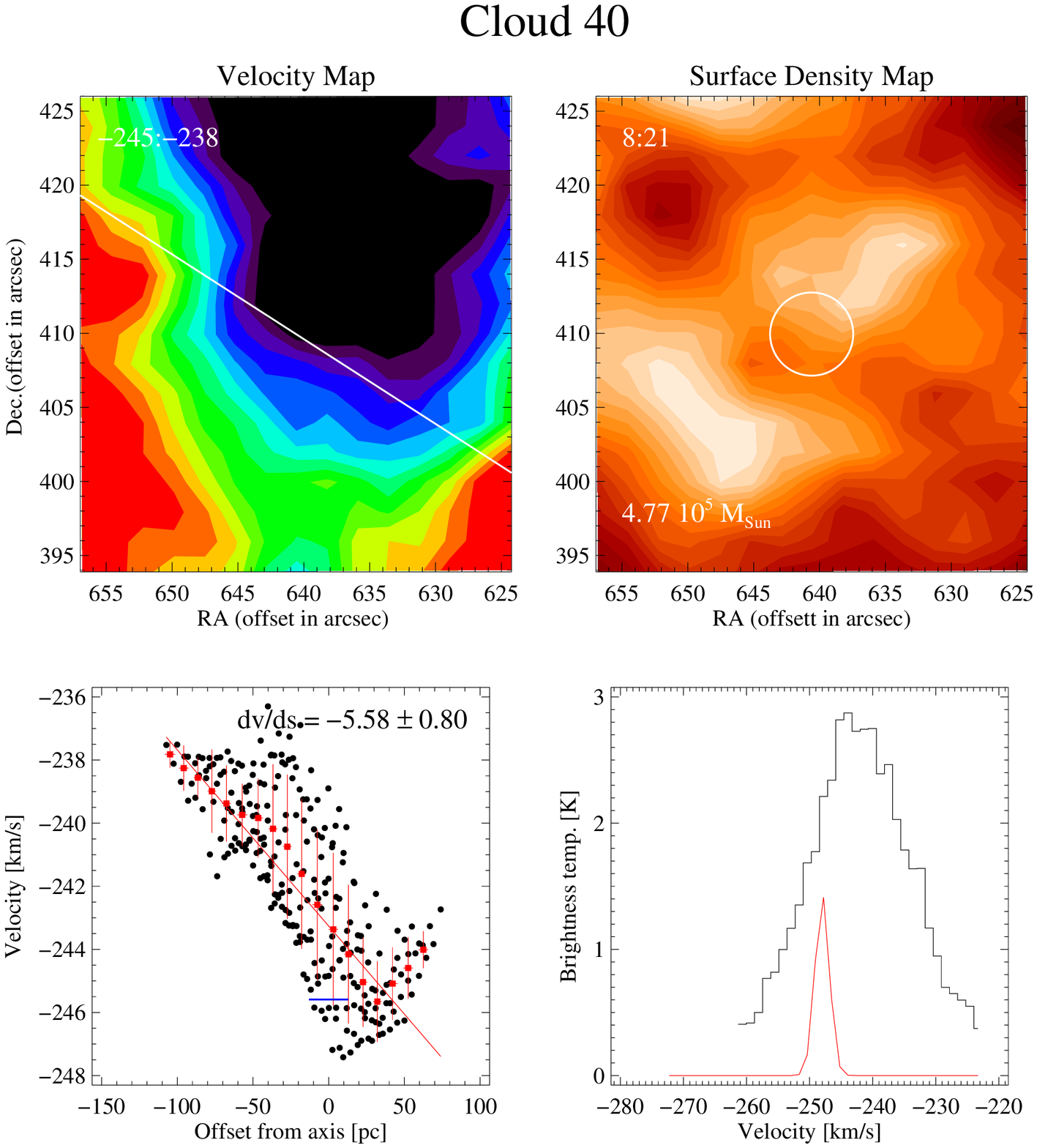}
\caption{See Figure \ref{fig:f22}.}
\end{figure*}

\newpage
\vspace*{2cm}
\begin{figure*}[htbp]
\includegraphics[scale=0.8]{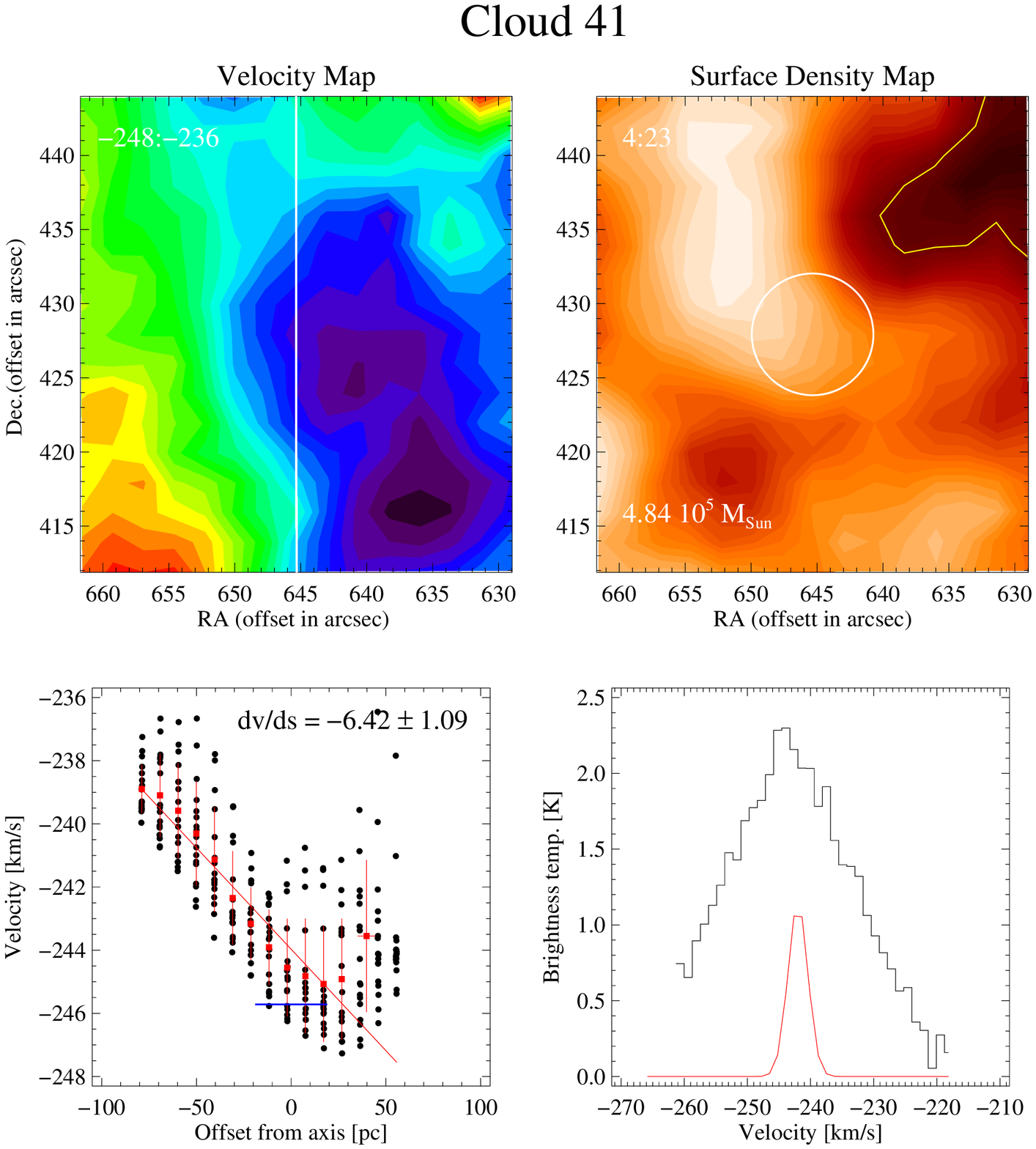}
\caption{See Figure \ref{fig:f22}.}
\end{figure*}

\newpage
\vspace*{2cm}
\begin{figure*}[htbp]
\includegraphics[scale=0.8]{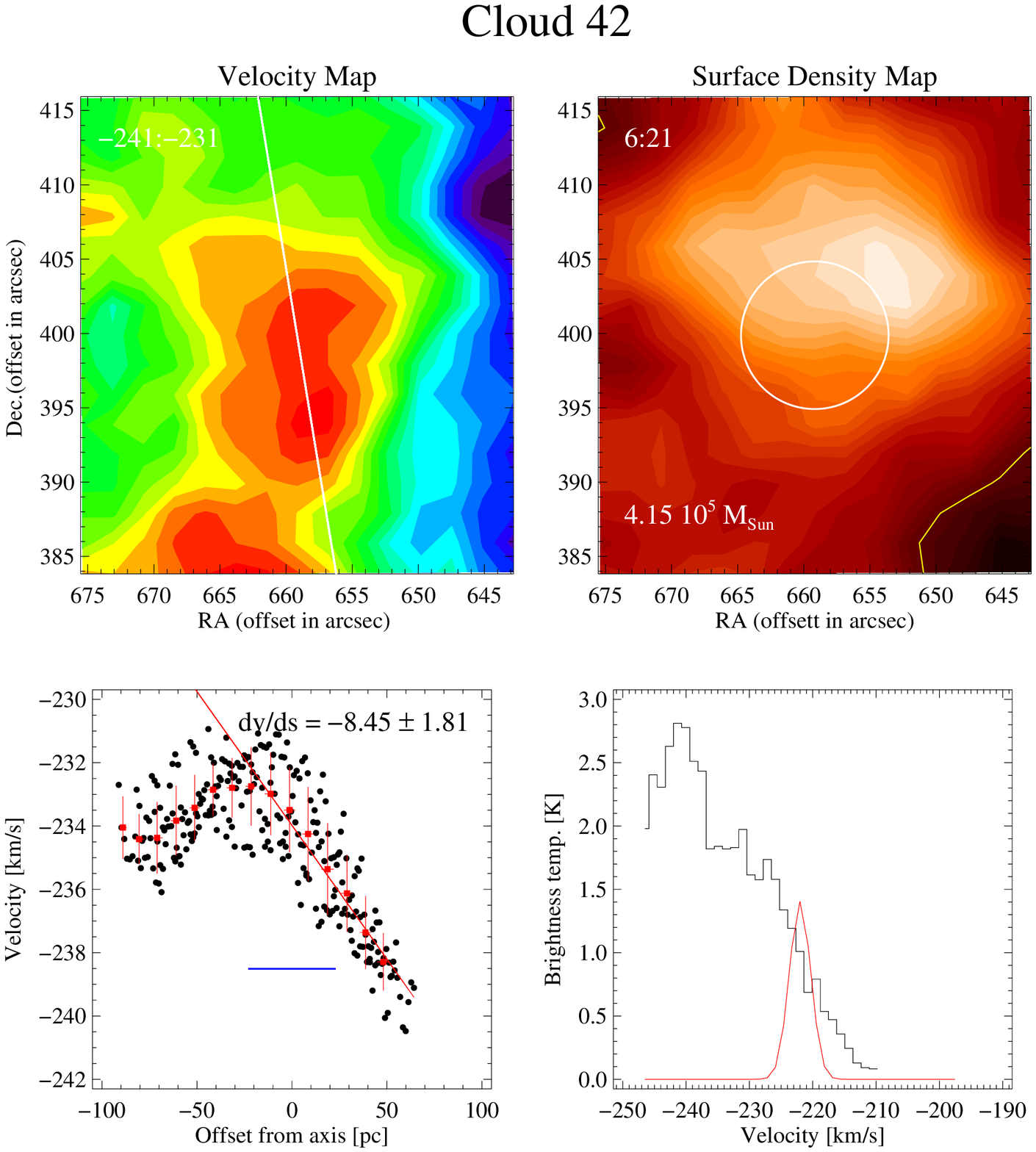}
\caption{See Figure \ref{fig:f22}.}
\end{figure*}

\newpage
\vspace*{2cm}
\begin{figure*}[htbp]
\includegraphics[scale=0.8]{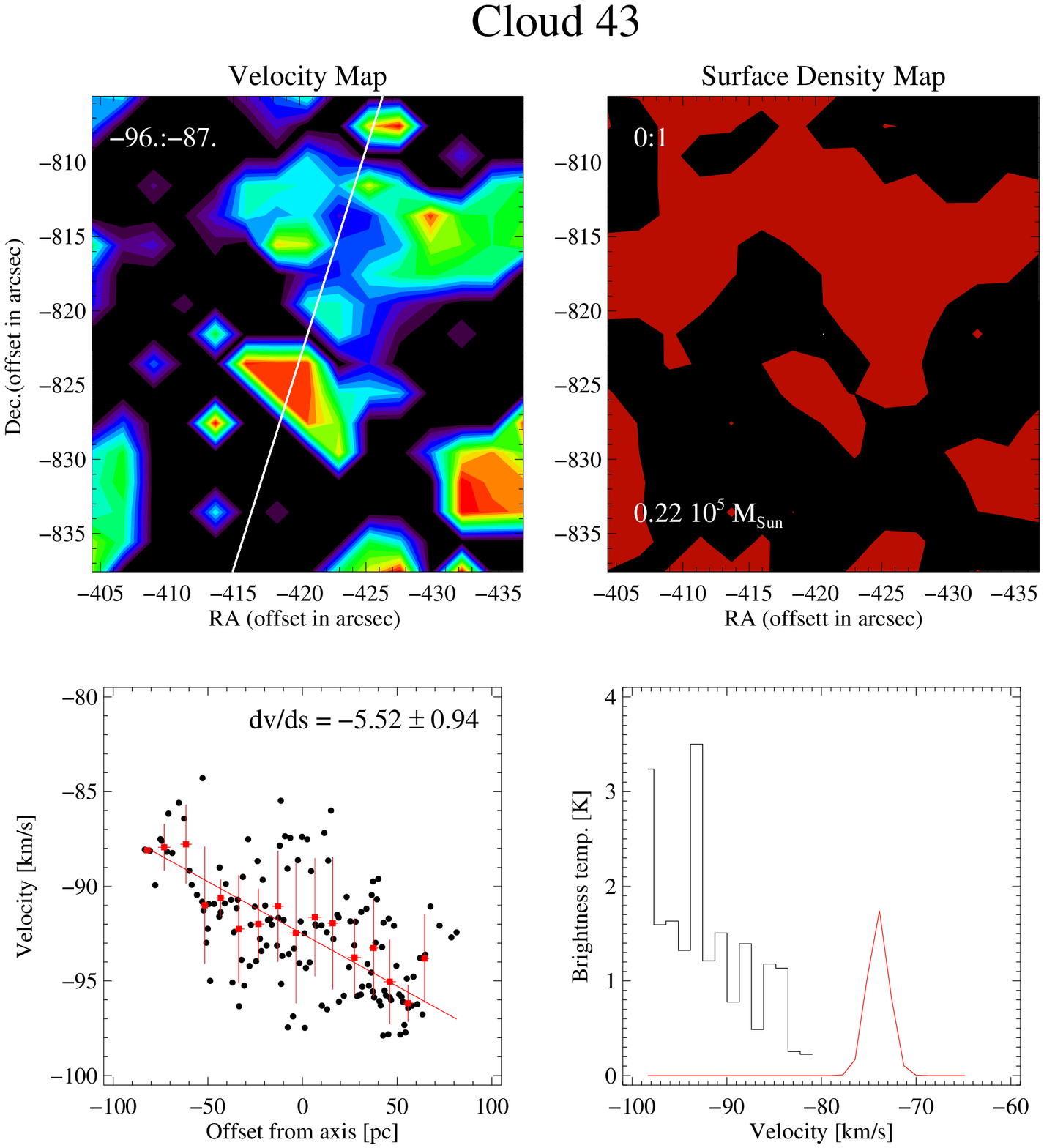}
\caption{See Figure \ref{fig:f22}.}
\end{figure*}

\newpage
\vspace*{2cm}
\begin{figure*}[htbp]
\includegraphics[scale=0.8]{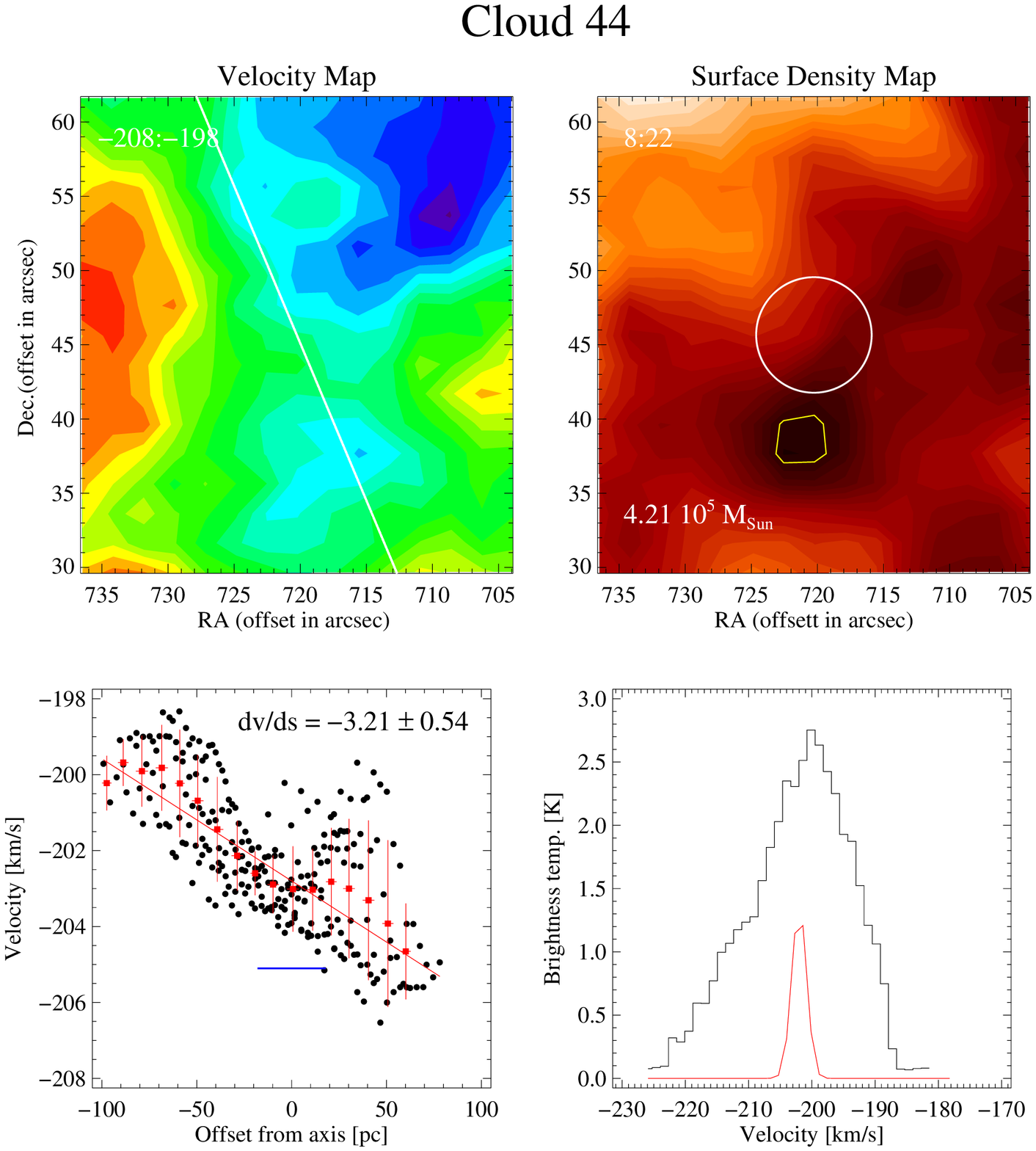}
\caption{See Figure \ref{fig:f22}.}
\end{figure*}

\end{document}